\documentclass{aa}  

\usepackage{graphicx}
\usepackage{txfonts}
\usepackage{natbib,twoopt}
\usepackage[breaklinks=true]{hyperref} 
\bibpunct{(}{)}{;}{a}{}{,}             
\makeatletter
  \newcommandtwoopt{\citeads}[3][][]{\href{http://adsabs.harvard.edu/abs/#3}%
    {\def\hyper@linkstart##1##2{}%
     \let\hyper@linkend\@empty\citealp[#1][#2]{#3}}}
  \newcommandtwoopt{\citepads}[3][][]{\href{http://adsabs.harvard.edu/abs/#3}%
    {\def\hyper@linkstart##1##2{}%
     \let\hyper@linkend\@empty\citep[#1][#2]{#3}}}
  \newcommandtwoopt{\citetads}[3][][]{\href{http://adsabs.harvard.edu/abs/#3}%
    {\def\hyper@linkstart##1##2{}%
     \let\hyper@linkend\@empty\citet[#1][#2]{#3}}}
  \newcommandtwoopt{\citeyearads}[3][][]%
    {\href{http://adsabs.harvard.edu/abs/#3}
    {\def\hyper@linkstart##1##2{}%
     \let\hyper@linkend\@empty\citeyear[#1][#2]{#3}}}
\makeatother

\hypersetup{linkcolor=red,citecolor=green,filecolor=cyan,urlcolor=magenta}

\usepackage{soul}

\usepackage{subfigure}

\usepackage{gensymb}

\usepackage{orcidlink}

\begin{document}

   \title{A new atmospheric characterization of the sub-stellar companion HR\,2562\,B with JWST/MIRI observations}

   \author{N. Godoy
          \inst{1}
          \orcidlink{0000-0003-0958-2150}
          \and
          E. Choquet
          \inst{1}
          \orcidlink{0000-0002-9173-0740}
          \and
          E. Serabyn
          \inst{2}
          \and
          C. Danielski
          \inst{3}
          \orcidlink{0000-0002-3729-2663}
          \and
          T. Stolker
          \inst{4}
          \and
          B. Charnay
          \inst{5}
          \orcidlink{0000-0003-0977-6545}
          \and
          S. Hinkley
          \inst{6}
          \orcidlink{0000-0001-8074-2562}
          \and
          P.~O. Lagage
          \inst{7}
          \and
          M.~E. Ressler
          \inst{2}
          \and
          P. Tremblin
          \inst{8}
          \orcidlink{0000-0001-6172-3403}
          \and
          A. Vigan
          \inst{1}
          \orcidlink{0000-0002-5902-7828}
          }

   \institute{ 
              Aix Marseille Universit\'e, CNRS, CNES, LAM, Marseille, France.\\
              \email{nicolas.godoy@lam.fr}
             \and 
              Jet Propulsion Laboratory, California Institute of Technology, Pasadena, CA 91109, USA.
             \and 
              INAF - Osservatorio Astrofisico di Arcetri, Largo E. Fermi 5, 50125, Firenze, Italy.
             \and 
              Leiden Observatory, Leiden University, Niels Bohrweg 2, 2333 CA Leiden, The Netherlands.
             \and 
              LESIA, Observatoire de Paris, Université PSL, CNRS, Sorbonne Université, Université Paris Cité, 5 place Jules Janssen, 92195 Meudon, France.
              \and 
              University of Exeter, Astrophysics Group, Physics Building, Stocker Road, Exeter, EX4 4QL, UK.
             \and 
              Université Paris-Saclay, Université Paris Cité, CEA, CNRS, AIM, 91191, Gif-sur-Yvette, France.
             \and 
              Université Paris-Saclay, UVSQ, CNRS, CEA, Maison de la Simulation, 91191, Gif-sur-Yvette, France.
             }

\date{Submitted to A\&A on March 12th, 2024. Accepted on June 23rd, 2024}

  \abstract
   {HR\,2562\,B is a planetary-mass companion at an angular separation of $0.56\arcsec$ ($19$\,au) from the host star, which is also a member of a select number of L/T transitional objects orbiting a young star. This companion gives us a great opportunity to contextualize and understand the evolution of young objects in the L/T transition. However, the main physical properties (e.g., $\mathrm{T_{eff}}$ and mass) of this companion have not been well constrained (34\% uncertainties on $\mathrm{T_{eff}}$, 22\% uncertainty for log(g)) using only near-infrared (NIR) observations.
   }
   { We aim to narrow down some of its physical parameters uncertainties (e.g., $\mathrm{T_{eff}}$: 1200K-1700K, log(g): 4-5) incorporating new observations in the Rayleigh-Jeans tail with the JWST/MIRI filters at $10.65$, $11.40$, and $15.50\,\mu m$, as well as to understand its context in terms of the L/T transition and chemical composition. 
   }
   {We processed the MIRI observations with reference star differential imaging (RDI) and detect the companion at high S/N (around $16$) in the three filters, allowing us to measure its flux and astrometry. We used two atmospheric models, \texttt{ATMO} and \texttt{Exo-REM}, to fit the spectral energy distribution using different combinations of mid-IR and near-IR datasets. We also studied the color-magnitude diagram using the \texttt{F1065C} and \texttt{F1140C} filters combined with field brown dwarfs to investigate the chemical composition in the atmosphere of HR\,2562\,B, as well as a qualitative comparison with the younger L/T transitional companion VHS\,1256\,b.}
   {We improved the precision on the temperature of HR\,2562\,B ($\mathrm{T_{eff}}$ = $1255$\,K) by a factor of $6\times$ compared to previous estimates ($\pm15$\,K vs $\pm100$\,K) using \texttt{ATMO}. The precision of its luminosity was also narrowed down to $-4.69\pm0.01$ dex. The surface gravity still presents a wider range of values (4.4 to 4.8 dex). While its mass was not narrowed down, we find the most probable values between $8\mathrm{M_{Jup}}$ ($3$-$\sigma$ lower limit from our atmospheric modeling) and $18.5\mathrm{M_{Jup}}$ (from the upper limit provided by astrometric studies). We report a sensitivity to objects of mass ranging between $2-5\mathrm{M_{Jup}}$ at $100$\,au, reaching the lower limit at \texttt{F1550C}. We also implemented a few improvements in the pipeline related to the background subtraction and stages 1 and 2. }
   {HR\,2562\,B has a mostly (or near) cloud-free atmosphere, with the \texttt{ATMO} model demonstrating a better fit to the observations. From the color-magnitude diagram, the most probable chemical species at MIR wavelengths are silicates (but with a weak absorption feature); however, follow-up spectroscopic observations are necessary to either confirm or reject this finding. The mass of HR\,2562\,B could be better constrained with new observations at $3-4\mu m$. Although HR\,2562\,B and VHS\,1256\,b have very similar physical properties, both are in different evolutionary states in the L/T transition, which makes HR\,2562\,B an excellent candidate to complement our knowledge of young objects in this transition. Considering the actual range of possible masses, HR\,2562\,B could be considered as a planetary-mass companion; hence, its name then ought to be rephrased as HR\,2562\,b.  }


   \maketitle
%

\section{Introduction} \label{sec:intro}

In the last decade, several sub-stellar objects (giant planets and brown dwarfs) have been discovered using direct imaging techniques with ground-based telescopes (e.g., 2MASSWJ 1207334-393254 or 2M1207, \citealt{Chauvin+2004}; HR\,8799\,bcde, \citealt{Marois+2008}; $\beta$\,Pictoris\,b, \citealt{Lagrange+2010}; 51\,Eridani\,b , \citealt{Macintosh+2015}; HIP\,79098(AB)\,b, \citealt{Janson+2019}). These telescopes (e.g., SCExAO, \citealt{Jovanovic+2015}; MagAO-X, \citealt{Males+2018}; SPHERE \citealt{Beuzit+2019}; GPI, \citealt{Macintosh+2014}) have a high angular resolution and efficient coronagraph in addition to benefiting from state-of-the-art systems for atmospheric correction (extreme adaptive optics system: ExAO; e.g., \citealt{Fusco+2013}), enabling studies of regions very close to the stars ($\sim15\,\mathrm{au}$ at $100$\,pc). However (mainly due to the sky thermal emission), such observations have mostly been made at near-infrared wavelengths (NIR, $<5\,\mu m$). The characterization of young sub-stellar objects entails atmospheric and dynamic studies to determine the orbital solution and main physical properties. From the atmospheric modeling, it is possible to know the effective temperature ($\mathrm{T_{eff}}$), surface gravity (log(g)), radius, mass, and even chemical composition (e.g., \citealt{Suarez+Metchev-2022}; \citealt{Miles+2023}), as well as the temperature-pressure profiles in the atmosphere (e.g., \citealt{Marley+2013}). All this information helps us to understand the current state and to trace the evolutionary path of these systems and make a comparison among the different theories of planetary formation (e.g., \citealt{Nielsen+2019}; \citealt{Vigan+2021}).

The NIR spectra of sub-stellar objects are characterized by carbon monoxide, methane, and water absorption features, which can be used to constrain turbulent mixing (e.g., \citealt{Konopacky+2013}). However, for some of these objects, parameters such as $\mathrm{T_{eff}}$ and log(g) (as well as radius and mass), may not be well constrained due to the limited spectral coverage and to the physical and chemical processes included in the various atmospheric models and clouds. On the other hand, some of the objects have a well-delimited age (e.g., $\beta$ Pic moving group with an age of $21\pm4$\,Myrs, \citealt{Gratton+2024}: 51\,Eri\,b, \citealt{Macintosh+2015}; AF\,Lep\,b, \citealt{Mesa+2023}; $\beta$\,Pic\,b, \citealt{Lagrange+2009}; Columba with an age of $36\pm8$\,Myrs, \citealt{Gratton+2024}: HR\,8799\,abcd, \citealt{Marois+2008}; $\kappa$\,And\,b, \citealt{Carson+2013}), which significantly help constraining these physical parameters. Also, observations made with different instruments can disagree in terms of recorded spectra and derived physical parameters of sub-stellar objects when using atmospheric models (e.g., \citealt{Sutlieff+2021}). Also, the derived physical parameters from these types of observations can be in disagreement with astrometric measurement when it comes to measuring the object's mass (e.g., \citealt{Chilcote+2017} and \citealt{Brandt+2021} for the mass of $\beta$\,Pictoris\,b). Observations in the mid-infrared (MIR) in the Rayleng-Jeans regime (e.g., \citealt{Danielski+2018} for JWST/MIRI coronographic observations and \citealt{Malin+2023} for MIRI/MRS) allow for better constraints on $\mathrm{T_{eff}}$ and log(g) (as well as the radius, \citealt{Carter+2022}) when combining with NIR observations. For this reason, observations made from the ground and space (NIR), as well as from space (MIR) allow us to improve the accuracy and precision of the atmospheric characterization of young sub-stellar objects.

Young brown dwarfs (BD) in the L-to-T transition present a great challenge to characterizing their atmospheres (e.g., \citealt{Suarez+2021}; \citealt{Sutlieff+2021}; \citealt{Miles+2023}). This comes from the fact that in this transition the passage of optically thick to thin clouds occurs, which may involve chemical disequilibrium processes, different cloud compositions at different atmospheric altitudes, differences in turbulent mixing, and different particle sedimentation. After the first observation and characterization of a young object in this transition in the near and MIR with high signal-to-noise ratio (S/N) and medium spectral resolution (VHS\,1256–1257\,b, \citealt{Miles+2023}) using the \textit{James Webb} Space Telescope (JWST, \citealt{Rigby+2023}), it is clear that the atmospheric models match the overall continuum spectrum shape well, but not some absorption features present in the spectral energy distribution (SED, e.g., $3.3\mu m$ $CH_{4}$ feature and silicates). This can be interpreted as much more complex physical-chemical processes occurring in the atmospheres of this type of source, at the top of the clouds that remain visible to cooler temperatures in the presence of lower gravity (e.g., \citealt{Marley+2012}), which need to be investigated, considered, and incorporated into theoretical models. 

The young brown dwarfs in the L-to-T transition differ in terms of certain physical characteristics from their older counterparts. For example, they are redder and cooler than field BDs, with ages greater than 1\,Gyr (e.g., \citealt{Metchev+Hillenbarand_2006}). The main reason for such a difference is that a low surface gravity with relatively high temperatures allows for a prevalence of clouds in the upper layers, which captures particles high in the atmosphere. This causes a decrease of emission at shorter wavelengths and an increase at longer wavelengths, causing the object to appear redder (e.g., \citealt{Curri+2011}; \citealt{Marley+2012}). Conversely, for the older ones, as the atmosphere cools, these particles condense in deeper layers, making them appear bluer. The presence of different chemical species (silicates, ammonia, methane, water, carbon monoxide, etc.) which have significant signatures in the mid-IR, marks different states in this transitional process. Therefore, a more refined characterization of these transitional objects can help us to understand better not only the physical-chemical changes during the transition but also their evolutionary path (young versus old sources). One of these L/T transition objects is HR\,2562\,B, which is part of a very small list of young L/T transition substellar objects orbiting a star. This companion, discovered by \cite{Konopacky+2016}, has a considerable uncertainty in the age estimates and in combination with discrepancies between ground-based observations(see \citealt{Sutlieff+2021}) generates discrepancies in the estimated physical parameters such as the $\mathrm{T_{eff}}$, log(g), and even the mass when comparing with astrometric studies (see Section\,\ref{sec:system}). Here, we present the first JWST coronagraphic observations (\citealt{Boccaletti+2015}; \citealt{Danielski+2018}) of HR\,2562\,B with the mid-IR instrument MIRI (\citealt{Rieke+2015}; \citealt{Wright+2015}) at $10.65$\,$\mu m$, $11.40$\,$\mu m$, and $15.50$\,$\mu m$.

The paper is organized as follows. In Section\,\ref{sec:system}, we summarize the main properties and context of the HR\,2562 system. Section\,\ref{sec:obs} describes the observations, strategy, data reduction, and post-processing and the astrophysical measurements retrieval. In Section\,\ref{sec:res}, we present our results and analysis. Section\,\ref{sec:dis} focuses on the mass and main characteristics of HR\,2562\,B and we compare this object to the younger VHS\,1256\,b object, recently observed with JWST/MIRI. Finally, in Section\,\ref{sec:sac} we summarize our results and conclusions.

\section{The HR\,2562 system}\label{sec:system}

HR\,2562 (or HD\,50571) is a young, F5V \citep{Gray+2006} spectral type star at a distance of $33.92$ pc (\citealt{Gaia+2016}; \citealt{Gaia-EDR3}). In Table\,\ref{table:star_parameters} we summarize all the main stellar parameters known to date. It has a debris disk detected by \cite{Moor+2006} using IRAS and Spitzer observations. \cite{Chen+2014} studied the debris disk SED, finding that the best-fit solution corresponds to two dust populations, one a warm disk located at 1.1\,au ($\sim400K$, $9.7\times10^{-7}\,\mathrm{M_{Moon}}$) and a cold disk at 340\,au ($45K$, $0.56\,\mathrm{M_{Moon}}$). By using Herschel observations in which the outer disk is marginally resolved, \cite{Moor+2015} constrained the inner edge of the disk between $18$ and $70$ au, and the outer radius at $187\pm20$ au, with an inclination of $78\pm6.3\degree$ (although in \citealt{Moor+2015} their best SED fit indicates a single component, unlike \citealt{Chen+2014}). Subsequently, \cite{Zhang+2023} constrained the location of the inner dust disk radius between $45-55$ au ($\sim1.7\arcsec$), based on disk-resolved ALMA observations and dynamical constraints, and the outer disk edge at $\sim260$ au ($\sim8.8\arcsec$) using only ALMA data. Also, they obtained better constraints the inclination with best fit $85\degree$ and lower limit of $79\degree$. We note that the disk marginally affects the stellar flux below $\sim20\mu m$ (see Fig.\,4 in \citealt{Chen+2014} and Fig.\,14 in \citealt{Zhang+2023}).

It is challenging to constrain the age of this system because it is not associated with a moving group or star-forming region with very good age estimates. \cite{Asiain+1999} classified HR\,2562 as a member of the ``B3'' subgroup Local Association, estimating an age of $300\pm120$ Myr, a value in good agreement with the one derived by \cite{Rhee+2007} using space motion, lithium nondetection, and X-ray luminosity. Subsequently, \cite{Mesa+2018} delimit the age between $200$ and $750$ Myr using lithium detection, with $450$ Myr as the representative age of the star. In the rest of this paper, we adopt a uniform age distribution between 250 and 750 Myr (see Table\,\ref{table:star_parameters}).

\begin{table}[]
\centering
\caption{\label{table:star_parameters}Main properties and stellar parameters of HR\,2562.}
\begin{tabular}{l c l}
\hline \hline
Parameter & Value & Ref \\
\hline
Spectral type &  F5V  & (1) \\
Right ascension (J2000) & 06:50:01.0151 & (2,3) \\
Declination (J2000) & -60:14:56.921 & (2,3) \\
Parallax (mas) & $29.4738 \pm 0.0185$ & (2,3) \\
Distance (pc) & $33.92 \pm 0.02$ & (2,3) \\
$\mu_{\alpha}\,(mas\,yr^{-1})$ & $4.830 \pm 0.025$ & (2,3) \\
$\mu_{\delta}\,(mas\,yr^{-1})$ & $108.527 \pm 0.023$ & (2,3) \\
Age (Myr) & $[250 - 750]$ & (4) \\
Mass ($M_{\odot}$) & $1.368 \pm 0.018$ & (4) \\
Radius ($R_{\odot}$) & $1.334 \pm 0.027$ & (4) \\
$\mathrm{T_{eff}}$ (Kelvin) & $6597 \pm 81$ & (5) \\
log(g) (dex) & $4.3 \pm 0.2$ & (4) \\
$\left[ \mathrm{Fe/H} \right]$ & $0.10 \pm 0.06$ & (4) \\
$A_{V}$ (mag) & 0.07 & (6) \\
$B - V$ (mag) & $0.454\pm 0.004$ & (7) \\
$V - I$ (mag) & $0.53 \pm 0.01$ & (7) \\
$B$ (mag) & $6.634 \pm 0.014$ & (7) \\
$H_{p}$ (mag) & $6.209 \pm 0.001$ & (8) \\
$GB_{p}$ (mag) & $6.222 \pm 0.003$ & (2,3) \\
$V$ (mag) & $6.15 \pm 0.01$ & (7) \\
$G$ (mag) & $6.004 \pm 0.003$& (2,3) \\
$GR_{p}$ (mag) & $5.629 \pm 0.004$ & (2,3) \\
$G_{rvs}$ (mag) & $5.502 \pm 0.006$ & (2,3) \\
$J$ (mag) & $5.305 \pm 0.020$ & (9) \\
$H$ (mag) & $5.128 \pm 0.029$ & (9) \\
$Ks$ (mag) & $5.020 \pm 0.016$ & (9) \\
$W1$ (mag) & $5.003 \pm 0.217$ & (10) \\
$W2$ (mag) & $4.794 \pm 0.108$ & (10) \\
$12\mu m$ (mag) & $4.712 \pm 0.065$ & (11,12) \\
$W3$ (mag) & $5.053 \pm 0.015$ & (10) \\
$25\mu m$ (mag) & $4.815 \pm 0.195$ & (11,12) \\
$W4$ (mag) & $4.990 \pm 0.029$ & (10) \\
$60\mu m$ (mag) & $1.925 \pm 0.239$ & (11,12) \\
$\mathrm{m_{F1065C}}$ (mag) & $4.995 \pm 0.056 $ & This paper  \\
$\mathrm{m_{F1140C}}$ (mag) & $4.994 \pm 0.056 $ & This paper  \\
$\mathrm{m_{F1550C}}$ (mag) & $4.996 \pm 0.055 $ & This paper  \\
$\mathrm{F_{F1065C}}$ (mJy) & $338 \pm 17$ & This paper \\
$\mathrm{F_{F1140C}}$ (mJy) & $298 \pm 15$ & This paper \\
$\mathrm{F_{F1550C}}$ (mJy) & $160 \pm 8$ & This paper \\
\hline
\end{tabular}
\tablefoot{References: ($1$) \cite{Gray+2006}; (2) \cite{Gaia-EDR3}; (3) \cite{Gaia+2016}; (4) \cite{Mesa+2018}; (5) \cite{Casagrande+2011}; (6) \cite{Morales+2006}; (7) \texttt{TYCHO}, \cite{TYCHO}; (8) \texttt{HIPPARCOS}, \cite{HIPPARCOS+0}; (9) \texttt{2MASS}, \cite{2MASS+0}; (10) \texttt{WISE}, \cite{WISE}; (11) \texttt{IRAS}, \cite{IRAS+1}; (12) \texttt{IRAS}, \cite{IRAS+2}.}
\end{table}

\begin{figure}
\centering
\includegraphics[width=8.5cm]{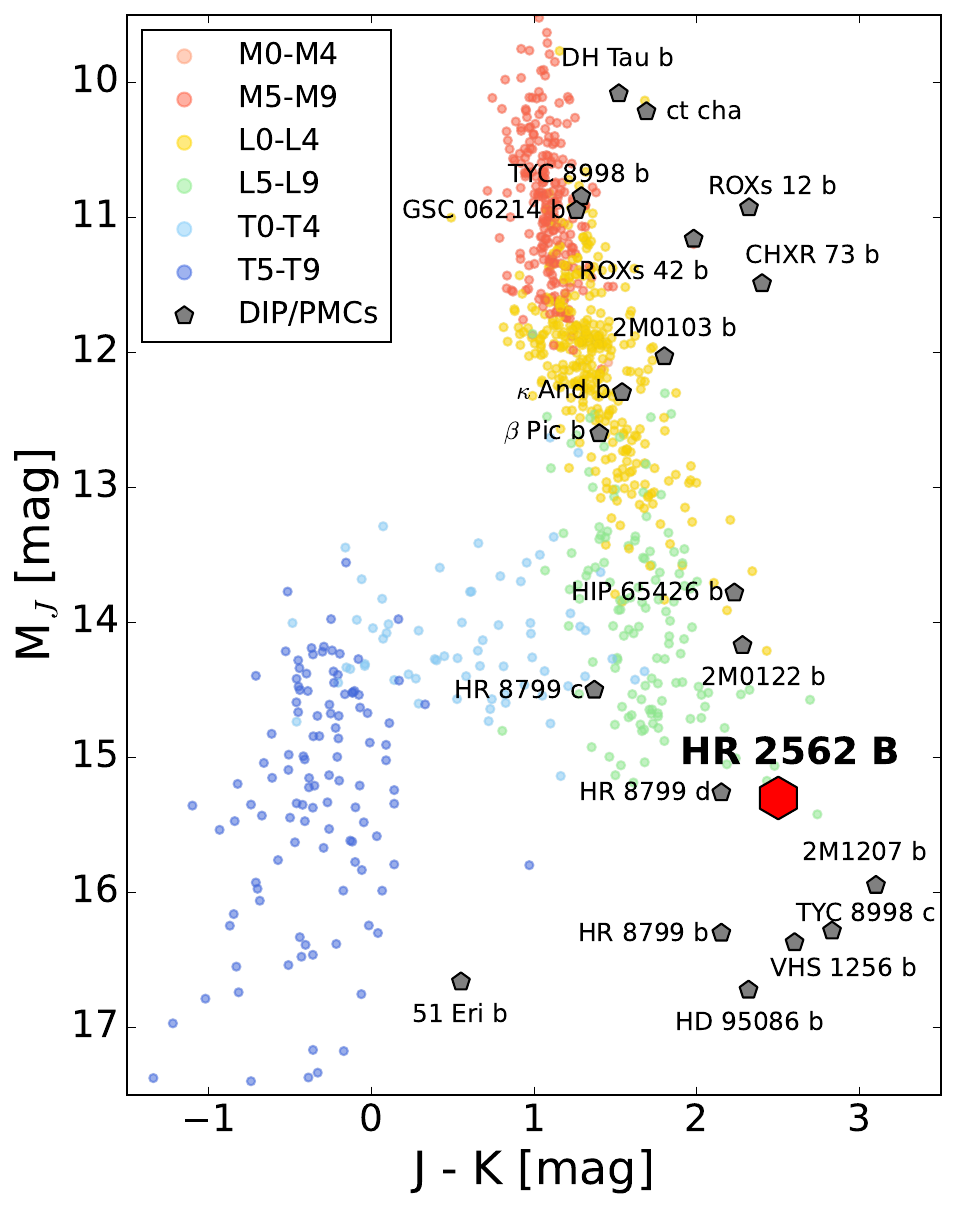}
\caption{ Color-magnitude diagram showing the position of HR\,2562\,B (red hexagon, \citealt{Konopacky+2016}) relative to the population of low-mass stars and brown dwarfs (colored circles), as obtained from \cite{Best+2021}. The different colors highlight the different spectral types. The gray pentagons correspond to selected directly imaged planets (DIP) and planetary-mass companions (PMCs).}
\label{fig:CMD_JK}%
\end{figure}

The sub-stellar companion, HR\,2562\,B, was discovered by \cite{Konopacky+2016} and classified as L$7\pm3$ spectral type in the L/T transition using the Gemini Planet Imager (GPI, \citealt{Macintosh+2014}) in the context of the ``Gemini Planet Imager Exoplanet Survey'' observations. The main parameters derived by \cite{Konopacky+2016} are the effective temperature of $1200\pm100$K, log(g) of $4.70\pm0.32$\,dex, a radius of $1.11\pm0.11\,R_{Jup}$, and a mass of $30\pm15\mathrm{M_{Jup}}$ at a projected separation of $20.3\pm0.3$ au ($0.618\pm0. 004\arcsec$). Subsequently, \cite{Mesa+2018} supplemented the observations with  the \textit{Spectro-Polarimetric High-contrast Exoplanet
REsearch}   (SPHERE, \citealt{Beuzit+2019}) instrument's \textit{Integral Field Spectrograph} (IFS, \citealt{Claudi+2008}) data in the YJ bands, as well as IRDIS (\citealt{Dohlen+2008}) H broad-band observations. \cite{Mesa+2018} constrained the atmospheric parameters (e.g., effective temperature and mass) and achieved a good agreement with the previous estimates. In addition, from these SPHERE observations, the companion was classified as T2–T3 spectral-type object. Figure\,\ref{fig:CMD_JK} shows the position of HR\,2562\,B in the color-magnitude diagram in which the spectral types of a population of brown dwarfs as obtained from \cite{Best+2021}\footnote{Photometry from the ultracool dwarfs catalog: \url{http://bit.ly/UltracoolSheet}.} are highlighted. The most recent analysis of its atmospheric properties was made by \cite{Sutlieff+2021}, using narrow-band MagAO/Clio2  (\citealt{Close+2010, Close+2013}; \citealt{Sivanandam+2006}; \citealt{Morzinski+2014}) observations at $3.9\mu m$ with the vector Apodizing Phase Plate coronagraph (vAPP, \citealt{Otten+2017}). They reported consistent mass estimates ($29\pm15\,\mathrm{M_{Jup}}$) but a wider range on effective temperature and surface gravity ($1200-1700$\,K and $4.0-5.0$\,dex, respectively) depending on the bandpass considered in the spectral modeling using SPHERE and GPI data.

\cite{Maire+2018} provided the first dynamical and astrometric study of HR\,2562\,B using SPHERE and GPI data, giving a first constraint on its orbital parameters. They refined the semi-major axis to $30^{+11}_{-8}$\,au and the inclination to $87^{+1}_{-2}\deg$\footnote{The uncertainties in the semi-major axis and inclination correspond to the lower and upper limits.}, confirming that the orbit is coplanar with the disk. The most important parameter is the eccentricity since it has a direct impact on the dynamical mass estimates. Assuming the companion carves the cavity within the disk and using N-body simulations, \cite{Maire+2018} found that the eccentricities between $\sim0.2$ and 0.3 ($e>0.15$, discarding the zero-eccentricity scenario) are compatible with the disk geometry for an orbital period of 100 and 200 yrs, respectively. Subsequently, the dynamical mass was calculated using new GPI epochs by \cite{Zhang+2023}, obtaining a probable mass of about $10\,\mathrm{M_{Jup}}$ and a $3$-$\sigma$ upper limit of $18.5\,\mathrm{M_{Jup}}$, as well as an orbit that is coplanar with the disk (in agreement with the conclusion presented in \citealt{Maire+2018}). This mass constraint is in agreement (i.e., within the uncertainties at $3\sigma$) with the prediction from the spectroscopic analyses (\citealt{Konopacky+2016}; \citealt{Mesa+2018}; \citealt{Sutlieff+2021}), but still with a significant difference in the best mass estimates (i.e., $10\,\mathrm{M_{Jup}}$ vs $29\,\mathrm{M_{Jup}}$) that would intrinsically change the nature of the object from BD to a planetary body. Therefore, this upper limit in the companion mass can help to constrain the atmospheric modeling using spectra to accurately narrow down the atmospheric and physical parameters.

With a spectral type between L4  and T3 (\citealt{Konopacky+2016}; \citealt{Mesa+2018}), this object is one of the few known transitional planetary-mass companions (see Figure\,\ref{fig:CMD_JK}), placing it as one of the few transitional planetary-mass companions. A discrepancy of $35\%$ ($\Delta$H=\,0.35\,mag) is observed between the GPI and SPHERE H-band data (\citealt{Mesa+2018}; \citealt{Sutlieff+2021}). This may be due to a different flux calibration, different procedures in the data reduction, post-processing, and companion extraction, or to an actual variability of the object. Such a variability could be due to a cloudy fast rotational surface (\citealt{Radigan+2014}; \citealt{Metchev+2015}). However, the variability amplitudes for substellar objects typically range from 0.01 to 0.3 mag (e.g., $\sim0.1\Delta$ mag for a $\sim120$\,Myrs object of $\sim20\,\mathrm{M_{Jup}}$, \citealt{Sutlieff+2023}), making this option more difficult to consider, but not entirely impossible. Indeed, the variability for L/T transitional objects is expected to be more prominent considering the distribution of the nonuniform clouds in the atmosphere (e.g., \citealt{Metchev+2015}, \citealt{Miles+2023}). Also, young dwarfs are found to be more variable than old field dwarfs (e.g., \citealt{Vos+2022}). In their spectral analysis, \cite{GRAVIY+2020} chose to re-scale the GPI H-band photometry by a factor of $\sim0.9$ (obtained from the modeling), attributing the discrepancy to an instrumental, data processing, or calibration bias. Although, the effect of the discrepancy on the values of mass, effective temperature, and radius are within $3$-$\sigma$, the uncertainty in age plays a more important role when comparing the data with evolutionary models such as \texttt{ATMO} (\citealt{Tremblin+2015}; \citealt{Tremblin+2016}; \citealt{Phillips+2020}). Observations at thermal infrared wavelengths can help to constrain these companion's parameters.

\section{Observations and data reduction}\label{sec:obs}

\subsection{Observations and strategy}

The observations of the HR\,2562 system were carried out under the guaranteed time observation (GTO) program 1241 (PI: M. Ressler), which has the goal of constraining the effective temperature, mass, and presence of ammonia of four known sub-stellar companions using the JWST/MIRI instrument. The observations were taken on March 10 2023 within five hours of execution time. We used the four-quadrant phase masks coronagraph (4QPMs; \citealt{Rouan+2000}) with the narrow-band filters \texttt{F1065C}, \texttt{F1140C}, and \texttt{F1550C}. The observing strategy and exposure parameters were optimized using the MIRI Coronagraphic Simulation pipeline described in \cite{Danielski+2018}. Both the observational strategy and setup are explained below.

Observations were designed to minimize both the effects of flux-lost in the post-processing and the execution time. All the observations (using the filters \texttt{F1065C}, \texttt{F1140C}, and \texttt{F1550C}) of HR\,2562 were made with a single telescope orientation, with a telescope orientation constraint to place HR\,2562\,B in the bottom right side of the detector, at $45\degree\pm5\degree$ from the 4QPM spider arms (i.e., TA4 aperture PA $\sim79\degree$). This position was selected to minimize as much as possible the effect of the coronography transmission on the companion. Given the closeness of the companion to the star ($\sim0.6$\arcsec, \citealt{Konopacky+2016}) and the limited rotation capability of the telescope about its optical axis ($\sim10\degree$), using multiple telescope orients to subtract the starlight (angular differential imaging, \citealt{Marois+2006}) would induce excessive flux-subtraction of the companion, and it would hinder the detection of HR\,2562\,B. We thus chose to use only reference differential imaging (RDI, \citealt{Smith+Terrile-1984}) processing techniques, in order to minimize overheads associated with telescope rolls. From the \textit{Hubble} Space Telescope, we know that RDI is often the best strategy to image extrasolar objects thanks to the much better wavefront stability of space observatories compared to ground-based telescopes (\citealt{Lafreniere+2009}; \citealt{Soummer+2011}; \citealt{Schneider+2014}; \citealt{Hagan+2018}). This is also in agreement with the JWST predicted performance with the sub-pixel dithering strategy referred to as ``small grid dither'' (\citealt{Lajoie+2016}). Recent results presented by \cite{Carter+2022} show an outstanding RDI performance of JWST/MIRI, demonstrating the great stability of the instrument and that our choice for the RDI strategy was the best for this program.

We selected HD\,49518 as the reference star (K4III spectral type, \citealt{Houk+1975}; $J=4.768$\,mag, $J-K=0.994$\,mag, \citealt{Cutri+2003}). We chose this star based on the \textit{JWST User Documentation}\footnote{\url{https://jwst-docs.stsci.edu/jwst-mid-infrared-instrument/miri-observing-strategies/miri-coronagraphic-recommended-strategies}} and given the following: \textbf{1)} a star angularly close to our target ($1.6\degree$) to minimize wavefront drifts; \textbf{2)} a star bright enough to guarantee reasonable S/N at shorter exposure times considering the grid dither strategy adopted for the reference star (i.e., $5\times$ more exposures than for the science target); \textbf{3)} since at the Rayleigh-Jeans tail, the PSF chromatic mismatch is more relaxed (differences given the spectral type), we chose this giant star (class III), which is overluminous, compared to main-sequence stars and sufficiently bright despite being $330$\,pc away; \textbf{4)} a target without known companions (including no-binary stars from Washington Double star Catalog, \citealt{WDSC} ), and checking the Gaia Early Data Release 3 (\citealt{Gaia-EDR3}) parameter values against multiplicity following the recommendations of \cite{Fabricius+2021}\footnote{The values of the parameters for HR\,2562 are: renormalized unit weight error for astrometry, \texttt{ruwe}$=1.002$; the amplitude of the image parameters determination goodness-of-fit versus the position angle of the scan, \texttt{ipd\_gof\_harmonic\_amplitude}$=9.17\,10^{-3}$; and the percent of successful image parameters determination windows with more than one peak, \texttt{ipd\_frac\_multi\_peak}$=0.0$.}. The relatively large distance of the reference star ($330$\,pc) also helps with the last point, since a putative companion or circumstellar disk would likely be unresolved from the star, and thus limit the risk of contaminating the science field after starlight subtraction. At that distance, the median binary separation ($50$\,au, \citealt{Moe-Kratter+2021}; \citealt{Huang+2018}; \citealt{Andrews+2021}) corresponds to $1.3$\,pixel only, and the median disk radius ($60$\,au, \citealt{Wyatt-2005}) corresponds to only $1.6$\,pixels, neither of which significantly affects the coronagraphic PSF. To get a broader library of PSFs, each reference observation was repeated at five dither positions, following the small-grid dither patterns \texttt{5-point-small-grid} in a cross (i.e., ``$\times$'') configuration with a dither of $\sim10$\,mas ($0.01$ pixels). The dither-patter technique (\citealt{Rajan+2015}; \citealt{Schneider+2017}) has the advantage to allow for the creation of a library that reflects different alignments between the star and the coronagraph given the 10-15\,mas pointing accuracy of the telescope, which facilitates the subtraction of the coronographic PSF science star. This approach was first described by \cite{Lajoie+2016} with MIRI simulations before launch, and then proven by \cite{Boccaletti+2022} during JWST commissioning, and by \cite{Carter+2022} with the analysis of HIP\,65426 as part of the ERS program (\citealt{Hinkley+2022}). Given the contrast and separation of HR\,2562\,B, our simulations showed that a PCA-reduction using a library with five dither positions would be enough to detect the companion while relaxing observational constraints on the reference star, compared to the nine-point grid pattern. Furthermore, we decided to use the five dither positions instead of the nine dithers to save the precious time of the JWST; at the same time we were able to save more time for this GTO program.

Given the presence of the ``glow stick'' stray light in all the observations made with MIRI, discovered during commissioning (\citealt{Boccaletti+2022}), we added background observations for both the science target and the reference star to allow the subtraction of this feature. We selected regions about $7\arcmin$ from the two targets free of any known sources in the 2MASS, WISE, and GALEX databases. We dithered the background observations by $\sim17.7\arcsec$ to minimize the risk of having the glowstick signature contaminated by unknown background sources (e.g., galaxies). The observing setups are shown in Table\,\ref{table:obs_setup} for both stars HR\,2562 and HD\,49518.

Regarding the order of execution, we started with the background observations for HR\,2562; then we observed the star HR\,2562, followed by the reference star. The order was selected to minimize the wavefront variation at the shortest wavelength; finally, we observed the background for the reference star. 

\begin{table*}[]
\caption{Observational setup.}
\centering
\begin{tabular}{ c c c c c c c c c c c c c }
\hline\hline
Target  & Filter & $\lambda_{\mathrm{cen}}$\tablefootmark{a} & BW\tablefootmark{a} & $\lambda_{\mathrm{eff}}$\tablefootmark{b} & $\mathrm{W_{eff}}$\tablefootmark{b} & FWHM\tablefootmark{b}  & $\mathrm{N_{groups}}$ & $\mathrm{N_{int}}$ & $\mathrm{t_{exp}}$ & $\mathrm{N_{dither}}$ & $\mathrm{t_{total}}$ & Total\\
        &    & [$\mu m$] & [$\mu m$] & [$\mu m$] & [$\mu m$] & [$\mu m$] & & & [sec] & & [sec] & integration\\
\hline
HR\,2562   & F1065C & 10.575 & 0.75 & 10.554 & 0.567 & 0.568 & 98   & 3 & 23.48 & 1 & 70.94 & 3 \\
           & F1140C & 11.30  & 0.80 & 11.301 & 0.604 & 0.587 & 98   & 3 & 23.48 & 1 & 70.94 & 3 \\ 
           & F1550C & 15.50  & 0.90 & 15.508 & 0.704 & 0.734 & 1250 & 4 & 299.6 & 1 & 1199.12 & 4 \\
\hline
HD\,49518  & F1065C & 10.575 & 0.75 & 10.554 & 0.567 & 0.568 & 29   & 3 & 06.95 & 5 & 21.33 & 15 \\
           & F1140C & 11.30  & 0.80 & 11.301 & 0.604 & 0.587 & 29   & 3 & 06.95 & 5 & 21.33 & 15 \\
           & F1550C & 15.50  & 0.90 & 15.508 & 0.704 & 0.734 & 496  & 3 & 118.88 & 5 & 357.12 & 15 \\
\hline
\end{tabular}
\tablefoot{
\tablefoottext{a}{Central wavelength ($\lambda_{\mathrm{cen}}$) and bandwidth (BW) were taken from \cite{Boccaletti+2022}. Note that these definitions and values reported by \cite{Boccaletti+2022} are different than the ones used in the VSO Filter Profile Service (\citealt{Rodrigo+2012}; \citealt{Rodrigo+2020}).}
\tablefoottext{b}{Effective wavelength ($\lambda_{\mathrm{eff}}$), effective width ($\mathrm{W_{eff}}$), and full width at half maximum (FWHM) were taken from VSO Filter Profile Service (\citealt{Rodrigo+2012}; \citealt{Rodrigo+2020}). }
}
\label{table:obs_setup}
\end{table*}

\subsection{Data reduction and pre-processing}\label{sec:cosmetics}

We reduced the data using the \texttt{JWST}\footnote{\url{https://jwst-pipeline.readthedocs.io/}}\footnote{\url{https://jwst-pipeline.readthedocs.io/en/latest/jwst/jump/description.html}} pipeline routines (version 1.9.6, \citealt{Bushouse+2022}; CRDS version 11.16.21), with the \texttt{spaceKLIP}\footnote{\url{https://github.com/kammerje/spaceKLIP/tree/dev_v1/jk}} pipeline (version 0.1, \citealt{Kammerer+2022}). We modified some minor aspects in the pipeline to optimize the pre-processing and cosmetics calibrations. First, we ran \texttt{spaceKLIP} with the same setups presented in \cite{Carter+2022}. Given the poor reduction (persistent artifacts, strong cosmic rays residuals, persistently hot or bad pixels) obtained when using these standard parameters, we modified part of the routine and input parameters to improve the quality of the reduction. We divided these modifications into two groups, the first one related to the reduction itself (from the \texttt{*uncal*} to \texttt{*rateints*}, and \texttt{*calints*} files related to the ramp fitting and calibration steps), and the second group related to the removal of the background. After the stage 1 step, a number of bad pixels and cosmic rays are left uncorrected regardless of the chosen bad pixel parameter (e.g., ramp-fit threshold). One possible explanation for this is that some cosmic rays were very energetic and spread out, making their optimal extraction difficult or even impossible with the actual performance of the pipeline. Another possible explanation is that the cosmic rays reach the detector at the beginning of the observations, carrying a wrong correction in the ramp fit. To correct them, we define a custom mask that tags them as bad pixels, which is used at the very beginning of the stage 1 routine processing the uncalibrated files (i.e., applying an interpolation in those bad pixels). We note that we do this only to pixels affected by cosmic rays or groups of pixels that are not well corrected in the pipeline, so we were able to exclude hot or isolated bad pixels in this step. This procedure has an impact mostly at $15\mu m$, since the frames were more hit by cosmic rays due to longer exposure times (see Table\,\ref{table:obs_setup}). The correction of agglomerations of pixels hit by cosmic rays creates local areas with different noise statistics, which moderately affects the final contrast limits.

At stage 1, the ramp-fit is a crucial step that allows us to correct all types of artifacts, read-out patterns, bad pixel registration, and cosmic rays in our observations. The so-called ``jumps'' in detection with nondestructive read ramps (\citealt{Anderson+Gordon-2011}) are produced for a cosmic ray that reaches the detector and suddenly increases the number of counts registered by a pixel, for example. A threshold value is used to identify these jumps in the ramp groups, and it is based on the estimated signal and noise registered. We selected a threshold of 7 as it showed the best compromise between the detection of ``bad pixels'' (hot pixels, cosmic rays, etc.), and the over-detection of good pixels (pixels with low fluctuation at the noise level that can be wrongly selected). When obtaining the \texttt{*calints.fits} files from stage 2 we performed an additional correction of remaining bad pixels and isolated mask pixels using different techniques from the \texttt{VIP} package (\citealt{Gomez-Gonzalez+2017}; \citealt{Christiaens+2023}). We applied the \texttt{cube\_correct\_nan} function to stage 2 products (\texttt{*calints}, before background subtraction), then we identified and corrected the most prominent outliers using $\sigma$-clipping with the function \texttt{cube\_fix\_badpix\_isolated}. Then, we proceeded to create the median background and identify the remaining outliers using spatial filtering, subtracting the smoothed frames from the original one, and using the function \texttt{find\_outliers} to identify the outliers. We used the same functions to replace those outliers. After applying the background subtraction (see below), we iterated a last time the $\sigma$-clipping procedure with \texttt{cube\_fix\_badpix\_isolatedin} in two steps using first $\sigma=20$ in size of $15$\,pixels, and then using $\sigma=7$ in size of $5$\,pixels. We preferred this combination to first identify persistent bad pixels within the ``glow stick'' and coronagraphic stellar PSF.  

Regarding stage 2 itself, we keep all the default parameters for the correction and proper calibration of the frames (as is explained in \citealt{Carter+2022}). In particular, we used the default flat correction instead of discarding or not using it (e.g., \citealt{Boccaletti+2023}). Also, we do not use Gaussian blur. For the routines of \texttt{spaceKLIP}, we use a $\sigma$=3 for the $\sigma$-clipping to identify bad pixels with 25 iterations, a threshold of 3 to identify outliers in a smoothed version of each integration. The flux calibration is also done by the JWST stage 2 pipeline, with the default parameters set in \texttt{spaceKLIP}.

\begin{figure}
\centering
\includegraphics[width=4.2cm]{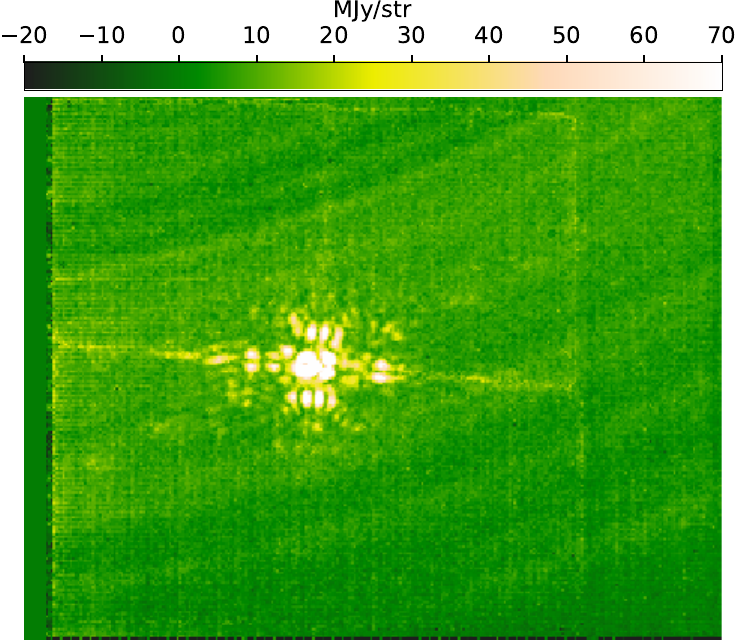}
\includegraphics[width=4.2cm]{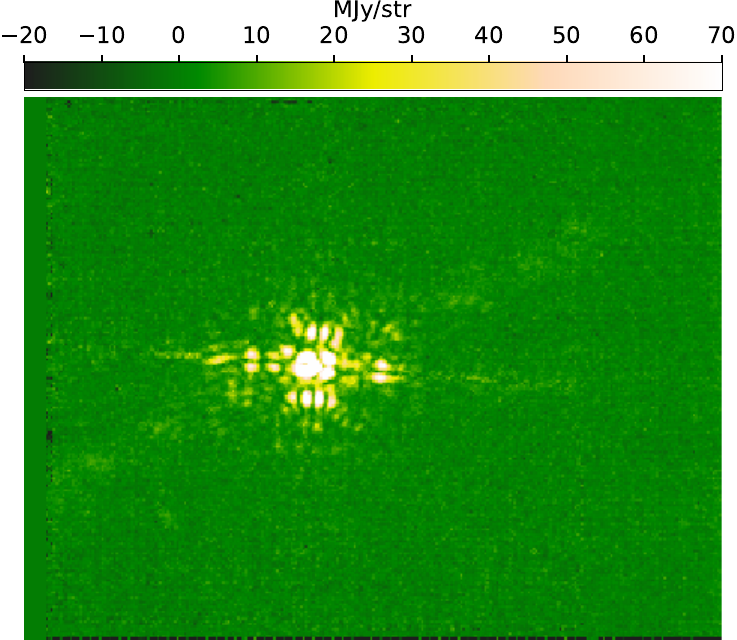}
\caption{First science frame of HR\,2562, \texttt{F1140C} filter using the two different background subtraction approach. \textit{Left}: Removal of the background using the median between all the available background frames at $\sim 11\mu m$ for HR\,2562. \textit{Right}: Removal of the background but keeping the cube structure, i.e., the first frame of each background observation was combined and subtracted from the first science frame.  }
\label{fig:Bkg_sub_example}%
\end{figure}

Regarding the background subtraction itself on all \texttt{*calints.fits} files corrected, we proceeded as follows. After verifying that there are no astrophysical sources in the field of view, we combined the background integrations into one single background frame per target and per filter (six in total). However, when performing the background subtraction we realized the presence of structures in the first integration of each sequence (as already mentioned in \citealt{Carter+2022}). Therefore, we took a different approach and decided to keep the cube structure (observing sequence for both sciences and backgrounds), instead of combining all integrations into a single one. This approach has two points to consider. First, it helps us to optimally remove the structure present in the first integration of each sequence, so we can keep the first integration in the sequence and not just remove it (as was done by \citealt{Carter+2022}). This allows us to maximize the S/N of the companion in the final image by using all the science frames. Figure\,\ref{fig:Bkg_sub_example} shows the first integration of HR\,2562 in the \texttt{F1140C} filter after removing the corresponding background using the combination of all the integrations (median combined frames, left image), and preserving the cube structure (observing sequence, a cube of integrations, right image). We see that using the background individual integrations improves significantly the calibration of the first science frame. Secondly, with this method, we only have a couple of frames to combine, obtained from the dithered background field. Our approach results in a similar noise level in each combined background frame ($3.70$ and $7.67$\,MJy/sr for the background of the science and reference stars), compared when all the integrations are combined into a single one ($3.85$ and $8.59$\,MJy/sr, respectively). We noted that each of the integrations in the sequence has a slightly different background level, so our procedure can remove more efficiently not only the structures shown in Fig.\,\ref{fig:Bkg_sub_example} but also these differences in the background level. We highlight this point since a more accurate background removal means less dispersion (noise) when combining the frames and, consequently, slightly better contrast limits.

\begin{figure}
\centering
\includegraphics[width=4.2cm]{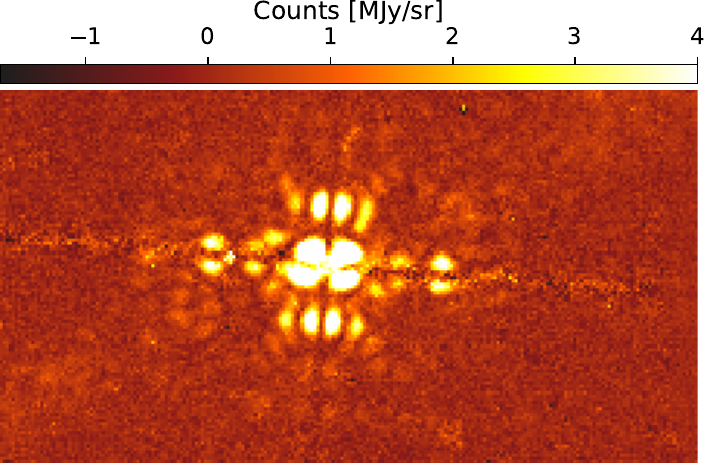}
\includegraphics[width=4.2cm]{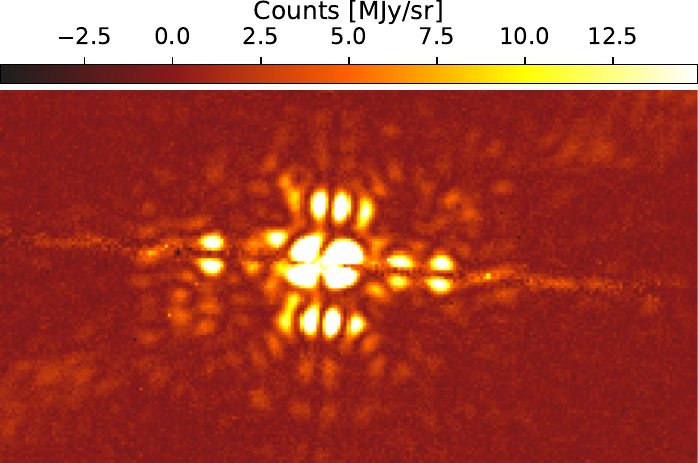}
\caption{Example of the normal procedure for the background subtraction at \texttt{F1550C} filter. Left: Third integration of HR\,2562. Right: Third integration of the reference star. The images were cropped to better see the effect of background subtraction around the coronographic PSF. }
\label{fig:Bkg_sub_sci-ref}
\end{figure}

\begin{figure}
\centering
\includegraphics[width=4.2cm]{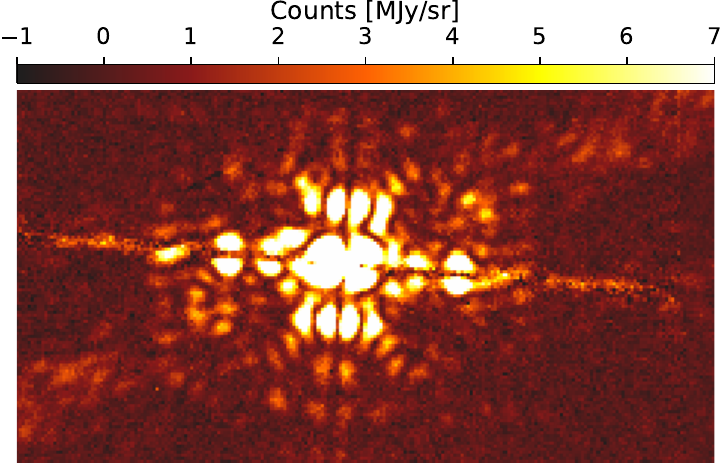}
\includegraphics[width=4.2cm]{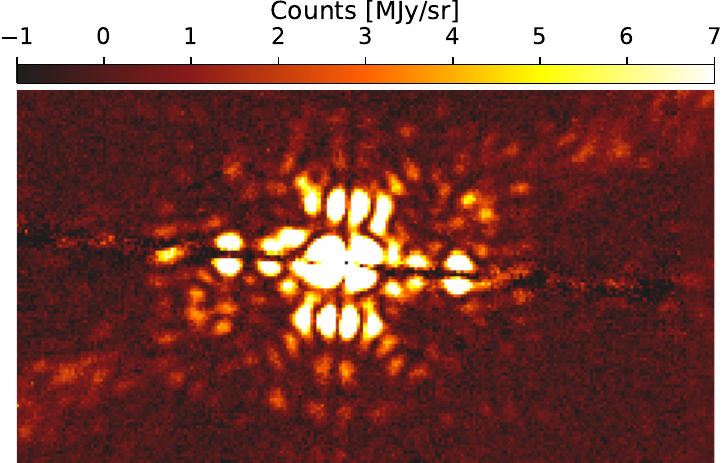}
\caption{Example of background subtraction applied in the reference star at \texttt{F1550C}. Left: Third integration using a normal procedure of background subtraction. Right: Same integration using an optimization for the background subtraction. The images were cropped to better see the effect of background subtraction around the coronographic PSF.}
\label{fig:Bkg_sub_bad-good}
\end{figure}

In the case of the \texttt{F1550C} filter, the ``glow stick'' looks undersubtracted in the reference integrations compared to the science ones, which generates an over-subtraction in the post-processing when using RDI. Figure\,\ref{fig:Bkg_sub_sci-ref} shows the third integrations for HR\,2562 (left image) and the reference star (right image), where is visible the undersubtraction of the ``glow stick'' for the reference star. We tried to solve this issue using different approaches, for example, adjusting the background level to better remove the ``glow stick,'' also first removing the median background and then removing the ``glow stick'' by re-scaling it, we tried re-scaling the ``glow stick’’ without removing the background, or even all the previous procedures focusing now on different combinations and sizes of regions dominated by the ``glow stick’’, among others. However, regardless of the background removal approach used, the improvements were negligible when using RDI. Figure\,\ref{fig:Bkg_sub_bad-good} shows an example of the normal background subtraction (left image) and our procedure to remove more efficiently the ``glow stick'' (right image) for the third integration of the reference star. One possible explanation is the time window between the science and background and reference and background observations. Indeed, for the science-background observations, the average $\Delta$time is 12\,minutes, while for the reference-background is 28\,minutes. A difference of $\sim75$ minutes between the reference star and its background observations at 15$\mu m$ can mean that the background observations become less representative of our reference star observations. In addition, the time difference between the science and reference stars observations is around $3$\,hours for 15$\mu m$ and $2$\,hours for $10$ and $11$\,$\mu m$. This could mean that it exists a temporal evolution of the ``glow stick'', which could have a greater impact at longer wavelengths. This is out of the scope of this study but will be investigated in the future. Another possibility is that we are biased when subtracting the background due to the flat correction (\citealt{Boccaletti+2023}). A less representative flat can generate an overcorrection of the coronographic transmission and ``glow stick,'' generating this mismatching between the science and reference star. However, we do not apply this approach since our target is less affected by the glow stick feature, compared to the HR\,8799 planetary system presented in \cite{Boccaletti+2023}. Instead, we just applied the correction for the background subtraction based on the background removal and re-scaling of the ``glow stick’’ mentioned above.

\subsection{Post-processing, contrast, and photometry}\label{sec:raw_contrast}

\begin{figure}
\centering
\includegraphics[width=8.2cm]{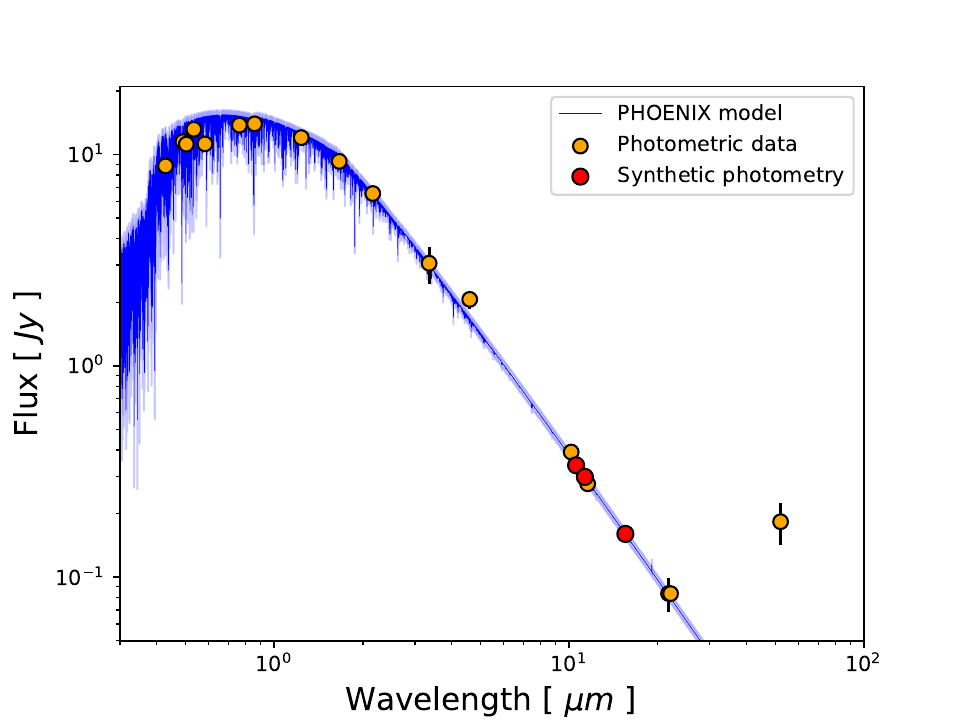}
\caption{Spectral energy distribution of the star HR\,2562. Orange dots correspond to the photometric data from Table\,\ref{table:star_parameters}. Red dots correspond to our synthetic photometry using the \texttt{F1065C}, \texttt{F1140C}, and \texttt{F1550C} bandpasses. The blue line corresponds to the mean between all the realizations of \texttt{PHOENIX} models with the respective uncertainties.}
\label{fig:Stellar_SED}%
\end{figure}

The \texttt{spaceKLIP} pipeline has incorporated \texttt{KLIP} (\citealt{Soummer+2012}) with the python routine \texttt{pyKLIP} (\citealt{Wang+2015}), which we used to optimally subtract the starlight with principal component analysis. \texttt{spaceKLIP} uses the stellar spectrum and magnitude to calibrate the contrast and extract the companion photometry (see the text below for more details). We used the main stellar parameters from Table\,\ref{table:star_parameters} to generate a stellar spectrum model which is used as inputs for \texttt{spaceKLIP}. \cite{Mesa+2018} provided great accuracy in the stellar properties such as the effective temperature (from spectral measurements), surface gravity, mass, radius, and metalicity (see Table\,\ref{table:star_parameters}). We used the \texttt{PHOENIX} atmospheric models as used by \texttt{spaceKLIP}, and the parameters and their uncertainties listed in Table\,\ref{table:star_parameters} to generate a synthetic spectrum of the star. To validate the stellar model, we collected archival photometry data from \texttt{Vizier} (\citealt{Vizier-data}), and used the \texttt{VOSA}\footnote{\url{http://svo2.cab.inta-csic.es/theory/vosa/}} tools (\citealt{Bayo+2008}) to estimate the dilution factor ($\mathrm{M_d}$=$\mathrm{R}^2_{\mathrm{star}}/\mathrm{Distance}^2$) needed to re-scale the theoretical models. We proceed to generate 1\,000 synthetic spectra considering all the parameters and uncertainties using a Monte Carlo approach and assuming a Gaussian distribution centered at each parameter and sigma equal to the parameter uncertainty. At the same time, we calculated the synthetic photometry at each MIRI filter (\texttt{F1065C}, \texttt{F1140C}, and \texttt{F1550C}). From the synthetic spectra, we compute the mean stellar spectrum that is used in \texttt{spaceKLIP} for contrast calibration, as well as the standard deviation as a representation of the uncertainty. We computed the synthetic photometry in the same manner. Figure\,\ref{fig:Stellar_SED} shows the spectral energy distribution of the HR\,2562 star, the archival photometry in orange, the synthetic photometry in red, and the solid line meaning synthetic spectrum, and related uncertainties.

\begin{figure}[htb]
    \centering 
 \begin{subfigure}{}
  \includegraphics[width=2.1cm]{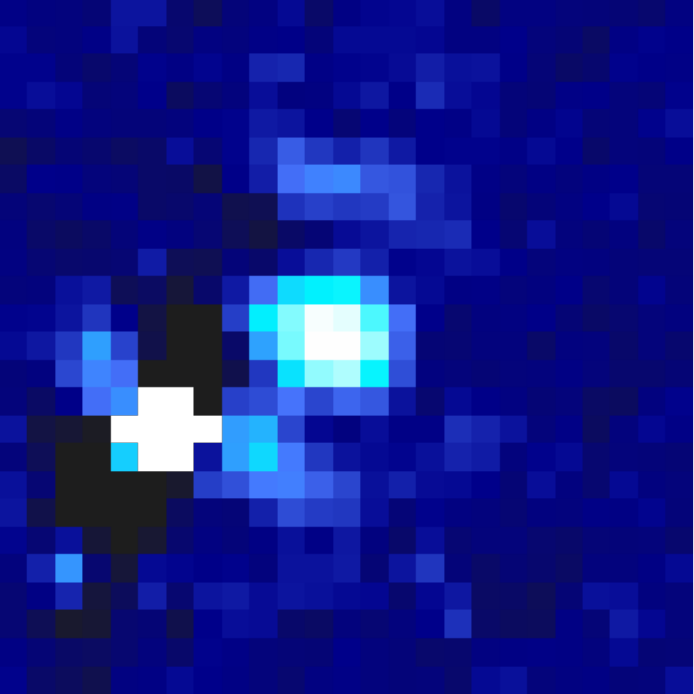}
 \end{subfigure}\hfil 
 \begin{subfigure}{}
  \includegraphics[width=2.1cm]{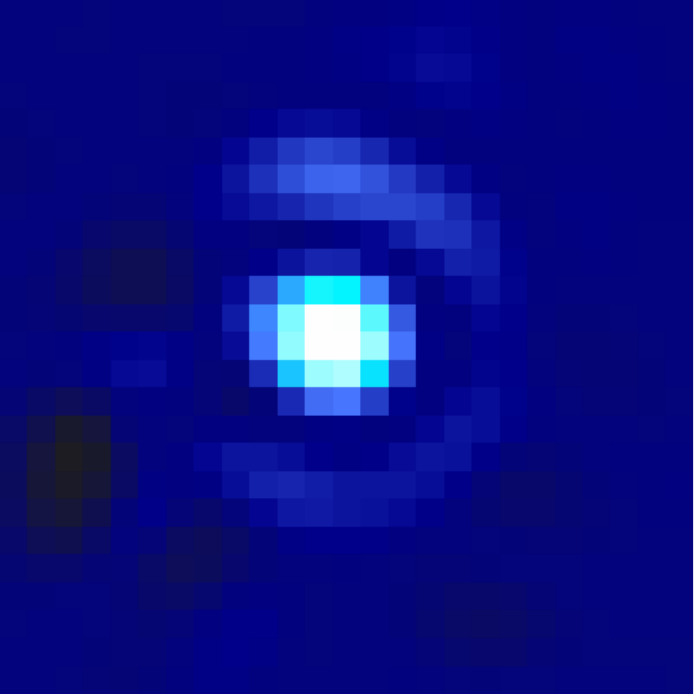}
 \end{subfigure}\hfil 
 \begin{subfigure}{}
  \includegraphics[width=2.6cm]{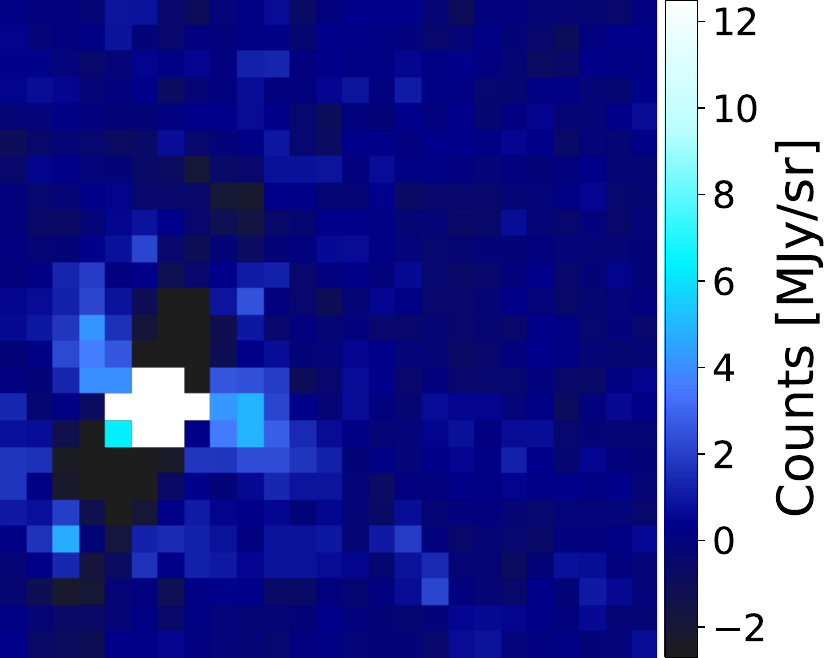}
 \end{subfigure}
 \medskip
     \centering 
 \begin{subfigure}{}
  \includegraphics[width=2.1cm]{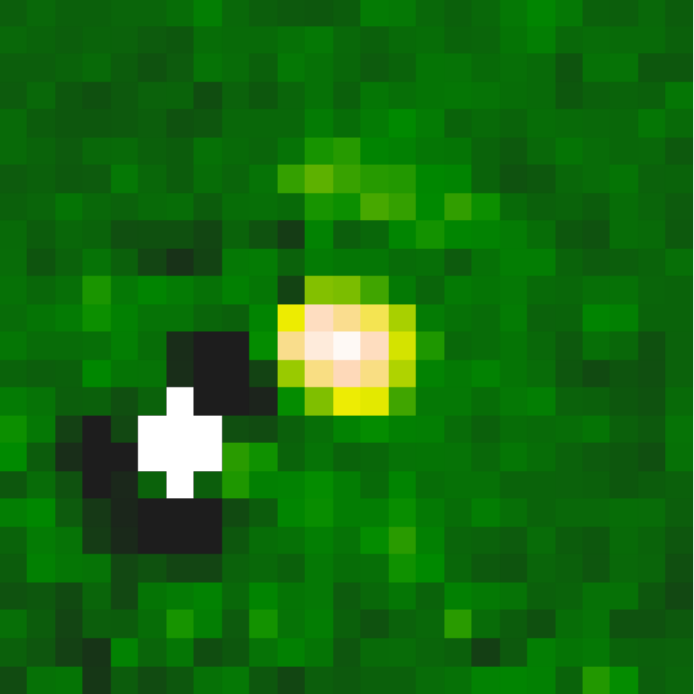}
 \end{subfigure}\hfil 
 \begin{subfigure}{}
  \includegraphics[width=2.1cm]{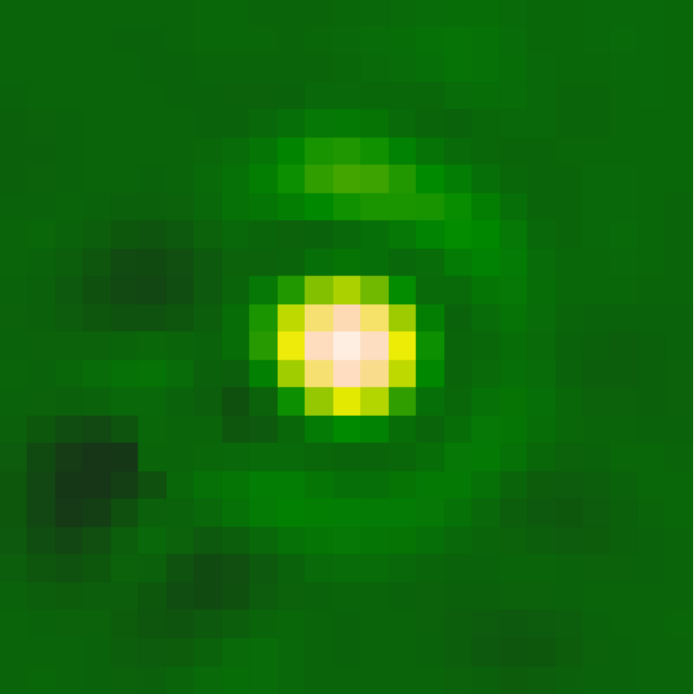}
 \end{subfigure}\hfil
 \begin{subfigure}{}
  \includegraphics[width=2.6cm]{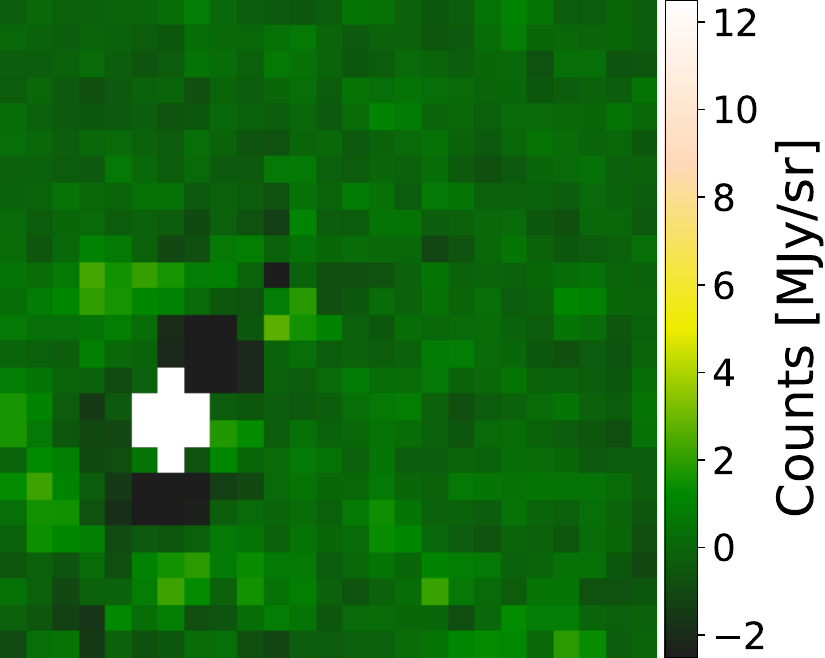}
 \end{subfigure}
 \medskip
     \centering 
 \begin{subfigure}{}
  \includegraphics[width=2.1cm]{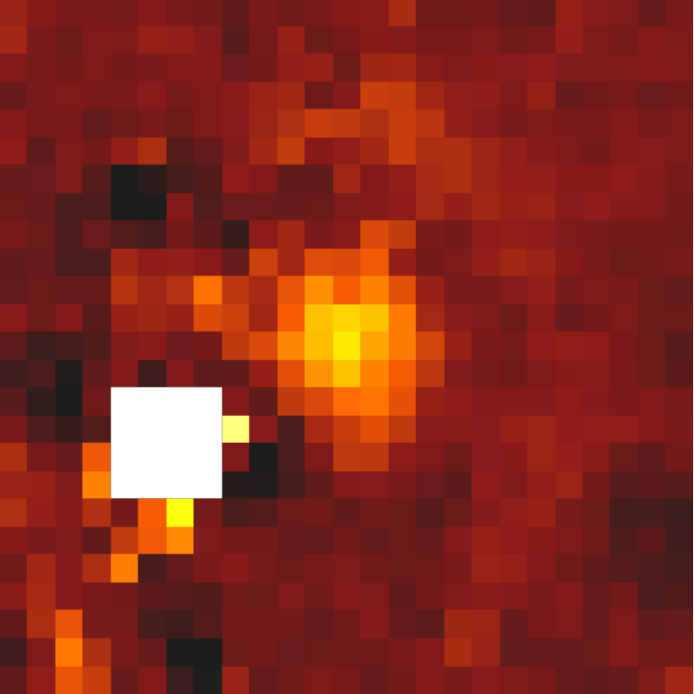}
 \end{subfigure}\hfil
 \begin{subfigure}{}
  \includegraphics[width=2.1cm]{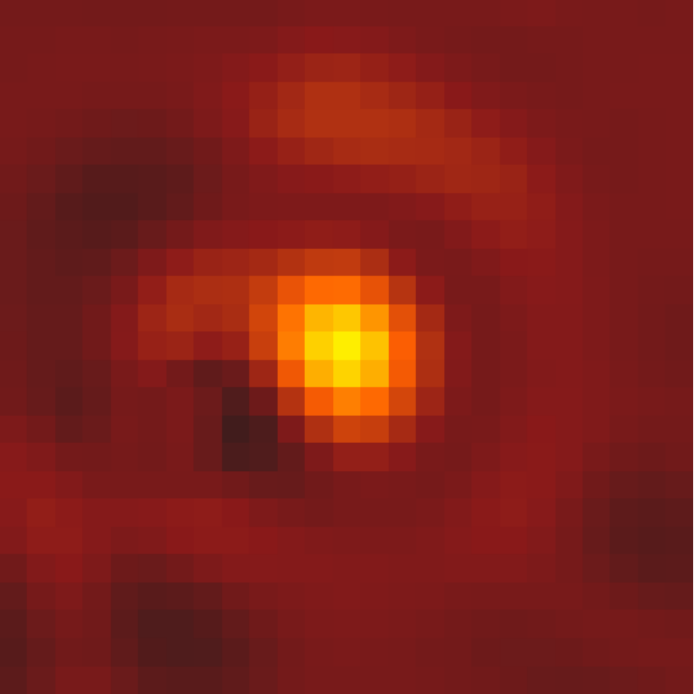}
 \end{subfigure}\hfil
 \begin{subfigure}{}
  \includegraphics[width=2.6cm]{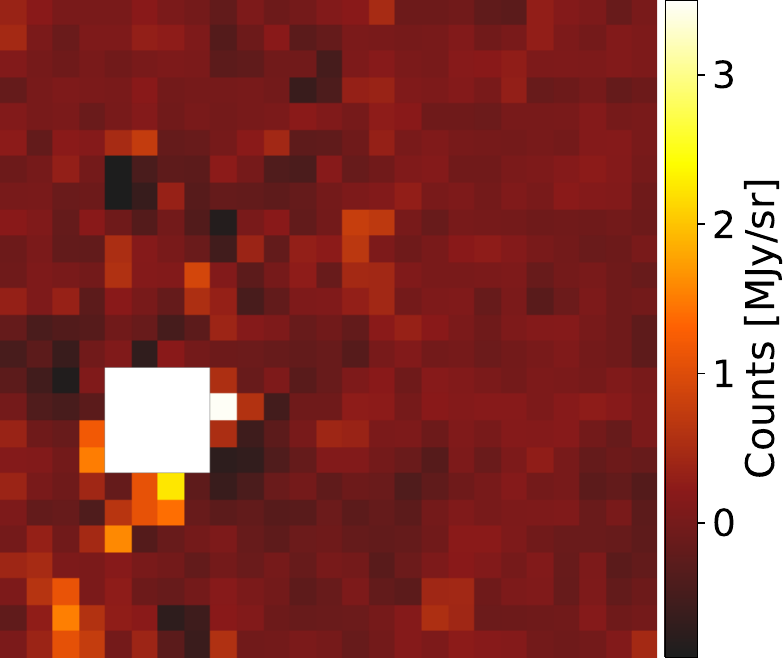}
 \end{subfigure}
\caption{ Modeling and extraction of the companion HR\,2562\,B. 
From \textit{top} to \textit{bottom}: Filters \texttt{F1065C}, \texttt{F1140C}, and \texttt{F1550C}. \textit{Left}: Science image. \textit{Middle}: Best \texttt{spaceKLIP} model of the companion using \texttt{webb\_psf}, corresponding to 6, 7, and 7 components for the filters \texttt{F1065C}, \texttt{F1140C}, and \texttt{F1550C}, respectively. \textit{Right}: residuals after subtracting the model from the science data. The field of view corresponds to $25\times25$ pixels ($\sim2.75\arcsec\times2.75\arcsec$). The images in each row have the same color scale.}
\label{fig:Extraction_comp}
\end{figure}

Concerning the starlight subtraction, \texttt{spaceKLIP}  provides several parameters to optimize the starlight subtraction. We used a reduction zone that covers the entire image (i.e., we did not use annuli mask), to compute the contrast curve and extract the companion properties. We explored the full range of \texttt{KLIP} components provided by the reference star dithers (i.e., $15$ components per filter), and selected the numbers that maximize the S/N of the companion, minimize the residuals in the companion extraction, and minimize the uncertainties and reach stable values for the flux and position. For each component after applying RDI, we extracted the companion photometry and astrometry following the routines presented and described in \texttt{spaceKLIP} and \cite{Carter+2022}. The \texttt{spaceKLIP} pipeline uses \texttt{webb\_psf} to generate a library of PSF models to fit the companion and extract the flux and position in each of the filters. Figure\,\ref{fig:Extraction_comp} shows the best-fit companion modeling and extraction in the three MIRI filters. Using routines from the \texttt{VIP} package we estimated the S/N of the companion at each number of components, and we measured the noise in the residual frames (after subtracting the best model of the companion). From the modeling for the companion extraction, we obtained the flux, contrast, position of the companion, and their related uncertainties. Figure\,\ref{fig:Comp_stat} shows the S/N of the companion, root mean squares, fluxes, and relative positions as a function of the components used for each filter. 


\begin{figure*}[htb]
    \centering
 \begin{subfigure}{}
  \includegraphics[width=4.8cm]{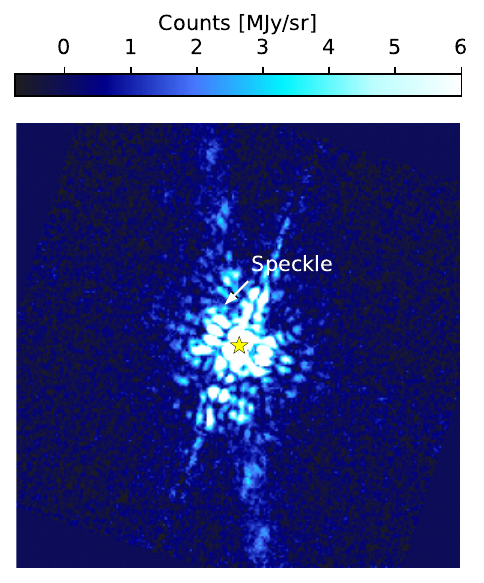}
 \end{subfigure}\hfil 
 \begin{subfigure}{}
  \includegraphics[width=4.8cm]{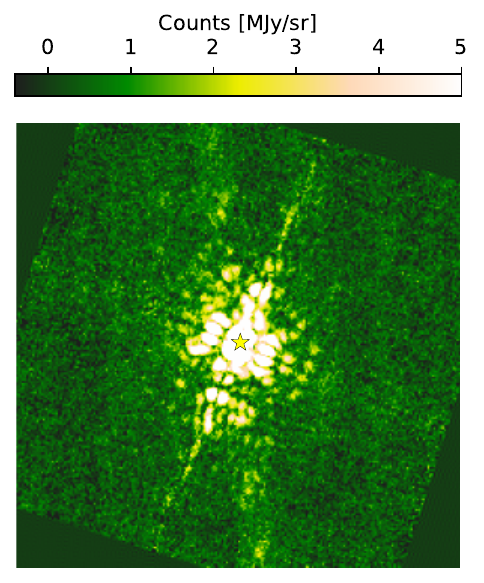}
 \end{subfigure}\hfil 
 \begin{subfigure}{}
  \includegraphics[width=4.8cm]{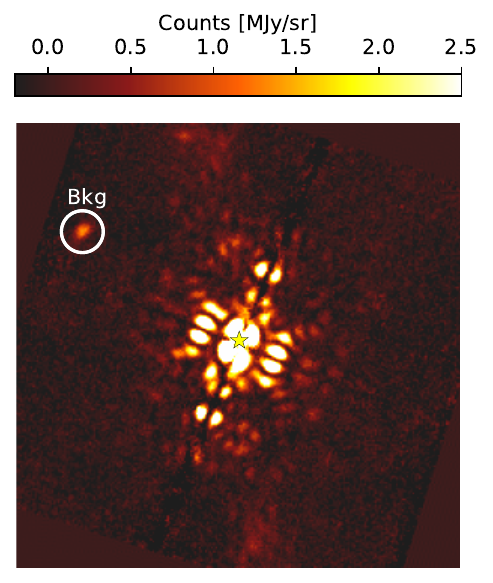}
 \end{subfigure}
 \medskip
 \begin{subfigure}{}
  \includegraphics[width=4.8cm]{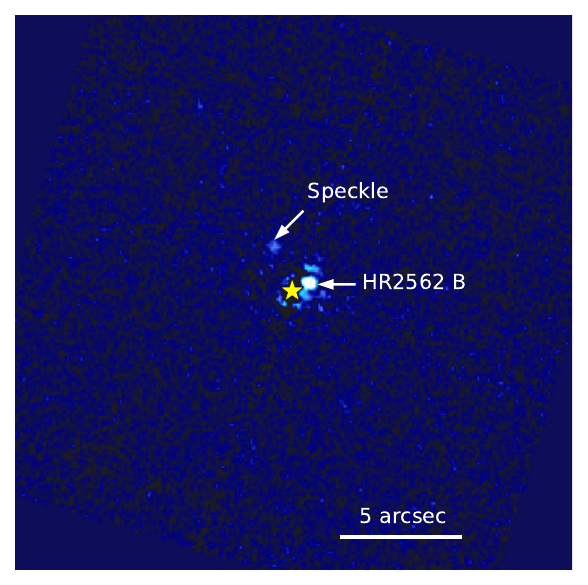}
 \end{subfigure}\hfil 
 \begin{subfigure}{}
  \includegraphics[width=4.8cm]{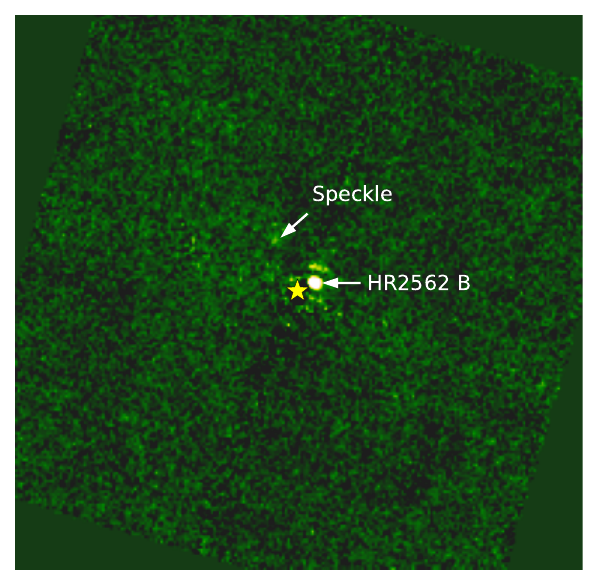}
 \end{subfigure}\hfil 
 \begin{subfigure}{}
  \includegraphics[width=4.8cm]{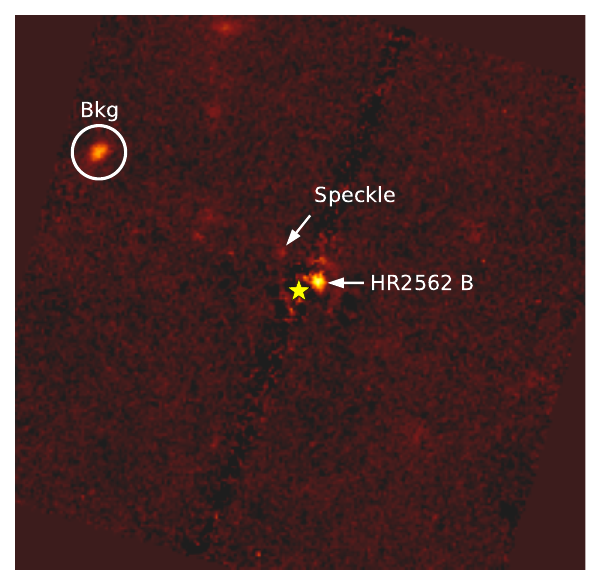}
 \end{subfigure}
\caption{Pre- and post-processing frames of HR\,2562. \textit{Top}: Background subtracted and stacked science frames. The stacking is only for visualization purposes. \textit{Bottom}: Starlight-subtracted image using RDI and \texttt{KLIP} at the optimal number of components. From left to right: \texttt{F1065C}, \texttt{F1140C}, and \texttt{F1550C} filters. The top white arrow at \texttt{F1065C}, \texttt{F1140C}, and \texttt{F1550C} filters marks the most prominent residual speckle. The horizontal white arrow highlights the position of HR\,2562\,B. The white circle at $15\mu m$ filter F1550C marks the background source. Each of the columns (i.e., filters) have the same color scale. The yellow star in each subplot marks the position of the star HR\,2562. North is up and east to left.}
\label{fig:images}
\end{figure*}

From this analysis, we selected $6$, $7$, and $7$ components as the best choice, respectively, for filters \texttt{F1065}, \texttt{F1140}, and \texttt{F1550}. Figure\,\ref{fig:images} shows the best final reduction of pre- and post-starlight subtraction using the \texttt{KLIP} approach with the optimal number of components for each filter. We note that even with the stellar suppression, few artifacts and speckles persist in all the filters. We highlighted in the \texttt{F1065C} images with an arrow the most notorious artifact in the images with the ``speckle'' label. Comparing its position in the three images we identify it as a residual speckle as it moves radially by $\sim4.5$ pixels (see Fig.\,\ref{fig:Astr_c3}). This comes from the fact that we used the 5-point strategy instead of the 9-point one, giving us a slightly lower diversity of frames and, therefore, a ``less optimal'' stellar suppression. However, the simulations done for this program, show that the 5 dither pattern is enough to detect the companion with sufficient S/N (which is the main goal of this observation). In Appendix\,\ref{Apx:speckle} we present a more detailed analysis of this source, concluding that it is effectively a speckle. However, as can be seen directly in the post-processed images, the performance of the RDI-KLIP reduction is reasonably good, and all the persistent speckles can be identified using the color and astrometric information (e.g., radial drift that depends on wavelength). We also note that there are some persistent artifacts in the post-processed images at $15\mu m$. These come from the imperfect corrections of cosmic rays, and persistent bad pixels, since we only masked the most prominent ones. However, these artifacts do not have an important impact on our companion analysis. Although all these artifacts are at $15\mu m$, one of them labeled ``bkg'' in Figure\,\ref{fig:images}, is indeed a real astrophysical source and is clearly visible even in all the raw frames individually. More details about this source are presented in Appendix\,\ref{Apx:c2_source}. From our analysis, we concluded that this source is a background galaxy. Table\,\ref{table:Fluxes} shows the angular separation and position angle for HR\,2562\,B and the ``bkg'' source, as well as the fluxes in the three bands. Note that the system is inclined in almost an edge-on configuration (\citealt{Zhang+2023}), however, the debris disk poorly/marginally irradiates at wavelengths below $20\mu m$ (see \citealt{Chen+2014}; \citealt{Zhang+2023}). Considering the values from the SED fitting and the disk dimension presented in \cite{Chen+2014} and \cite{Zhang+2023} and the stellar fluxes in Table\,\ref{table:Fluxes}, we estimate a disk contribution below $0.01$\,mJy for a PSF-size area.

\begin{figure}
\centering
\includegraphics[width=7.5cm]{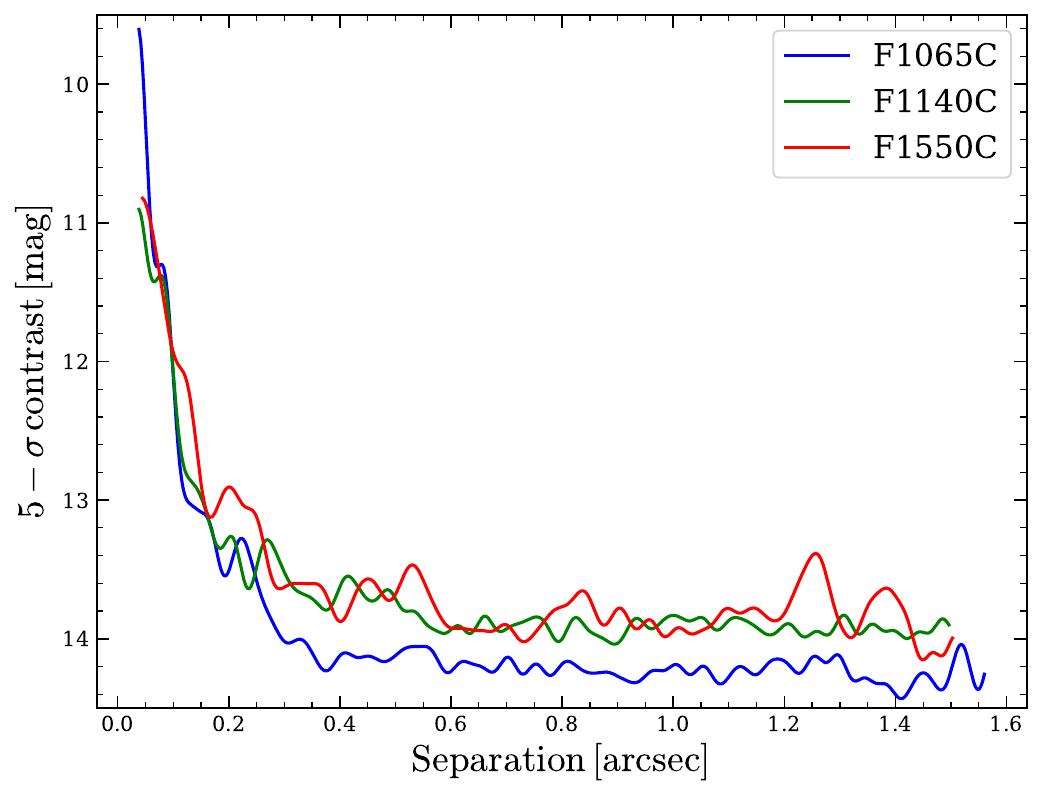}
\caption{Smoothed contrast limits for all JWST/MIRI observations ($10\mu m$, $11\mu m$, and $15\mu m$).  }
\label{fig:Contrast_mag}%
\end{figure}

We also estimated the contrast limits for each filter. We adopt a conservative (but more aggressive) approach and set the number of components to $15$ to have the optimal performance of RDI at all angular separations. We subtracted the companion and masked the galaxy in all post-processing images, to obtain an unbiased contrast limit at inner/outer angular separations. The contrast limit is computed using \texttt{spaceKLIP} (\texttt{raw\_contrast} and \texttt{cal\_contrast} functions), and considers the low number statistics (\citealt{Mawet+2014}). The contrast limit is corrected from the coronagraph transmission and from over-subtraction induced by KLIP, calculated by injecting and recovering calibration point sources in the raw and klip-subtracted images. We smoothed the frames using a Gaussian kernel to reduce the local noise fluctuation in the contrast, and also we smoothed the contrast itself using the Fourier spectral smoothing method to have a more uniform and noiseless curve as a function of angular separation. The final contrast limits are shown in Figure\,\ref{fig:Contrast_mag}.

\begin{table*}[]
\centering
\caption{Physical properties measured with MIRI coronagraphic observations.}
\begin{tabular}{ l c c c c c c }
\hline\hline
Target  & $10\mu m$ flux & $11\mu m$ flux & $15\mu m$ flux & PA      &   Angular Sep. \\
        &     [mJy]      &     [mJy]      &     [mJy]      &  [deg]  & [\arcsec] \\
\hline
HR\,2562    & $338\pm17$ & $298\pm15$ & $160\pm8$ & --- & --- \\
HR\,2562\,B & $0.162\pm0.026$ & $0.140\pm0.022$ & $0.067\pm0.010$ & $-59.45\pm0.23$ & $0.750\pm0.026$ \\ 
Bkg source  & $<0.014\pm0.001$ & $<0.016\pm0.001$ & $0.010\pm0.005$ & $55.29\pm0.03$ & $10.099\pm0.056$ \\
\hline
\end{tabular}
\tablefoot{Each of the flux refers to the apparent flux at each filter (i.e., \texttt{F1065C}, \texttt{F1140C}, and \texttt{F1550C}). The angular separation and position angle were estimated by combining all three filters at each best number of components previously determined. The angular separation was estimated assuming a pixel scale of $0.11\arcsec$. }
\label{table:Fluxes}
\end{table*}

\subsection{Archival data}

We used public data from GPI (\citealt{Konopacky+2016}), SPHERE (\citealt{Mesa+2018}), and MagAO (\citealt{Sutlieff+2021}) for our spectral fit analysis.

\begin{figure}
\centering
\includegraphics[width=9cm]{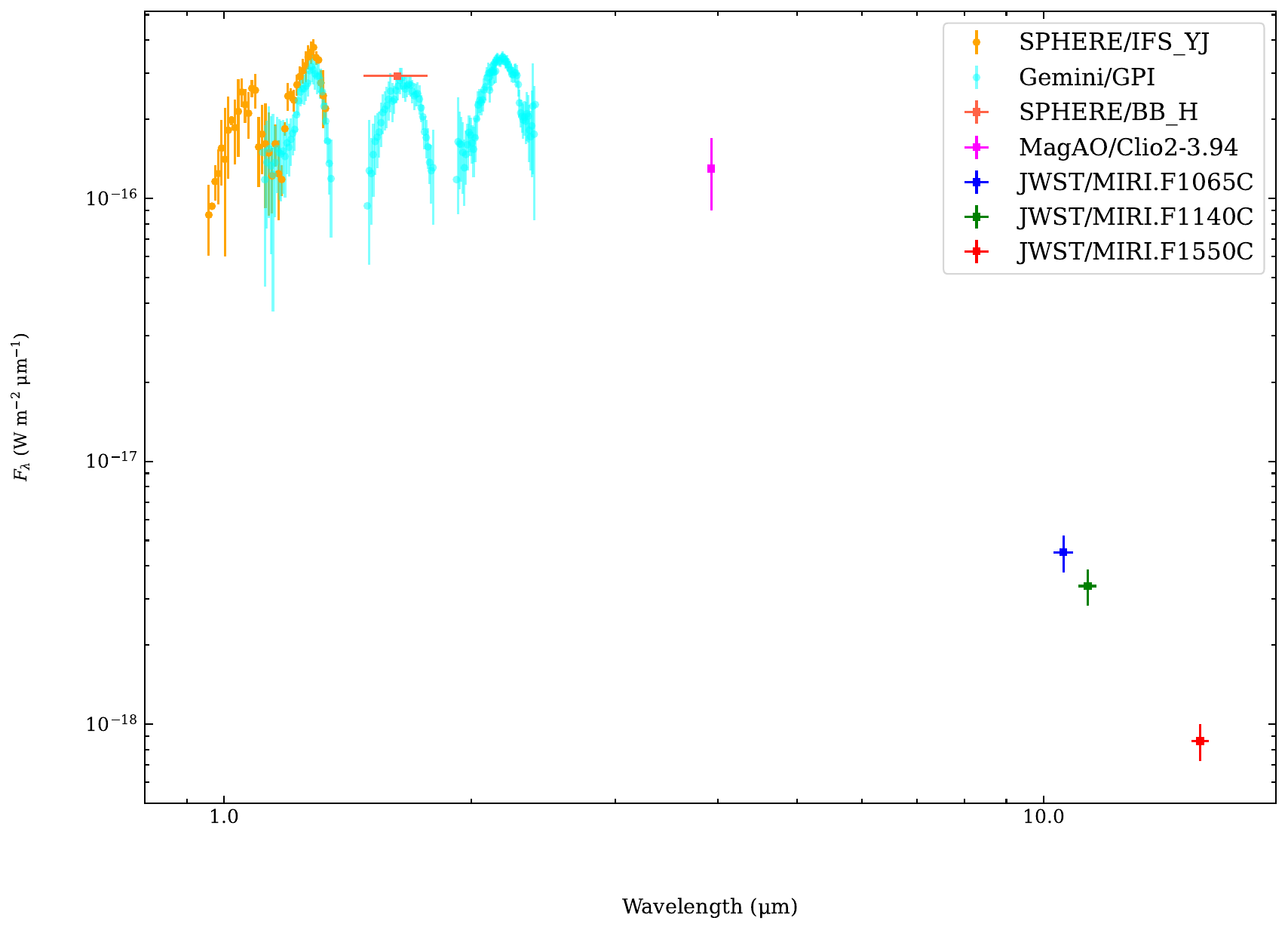}
\caption{Spectral energy distribution of HR\,2562\,B. Each of the points corresponds to different observations (SPHERE, GPI, MagAO, and JWST) identified by different colors as shown in the legend. Errorbar corresponds to 1-$\sigma$ uncertainty.  }
\label{fig:Full_sed_only}%
\end{figure}

The first detection of HR\,2562\,B was done with GPI (\citealt{Konopacky+2016}) using the H-band spectroscopic mode, followed by K1-K2 bands observations days later. A follow-up campaign was carried out a month after using the K2 and J-band spectroscopy to confirm the co-moving nature and perform spectroscopic analysis. \cite{Mesa+2018} used the SPHERE instrument to get observations with the integral field spectrograph (IFS; \citealt{Claudi+2008}) in Y and J bands ($0.95-1.35\mu m$), and the dual-band spectrograph (IRDIS; \citealt{Dohlen+2008}) with the H broadband filter. MagAO observations (\citealt{Sutlieff+2021}) were made using the vAPP observing mode with an ultra-narrow band filter at $3.9\mu m$ with the MagAO (\citealt{Close+2012}; \citealt{Morzinski+2014}) system on the 6.5-m \textit{Magellan} Clay telescope at LCO, Chile. Figure\,\ref{fig:Full_sed_only} shows the GPI, SPHERE, and MagAO data, along with our MIRI MIR observations.

\section{Analysis and results}\label{sec:res}

 \subsection{Mass limits and sensitivity}

\begin{figure*}
\centering
\includegraphics[width=5.75cm]{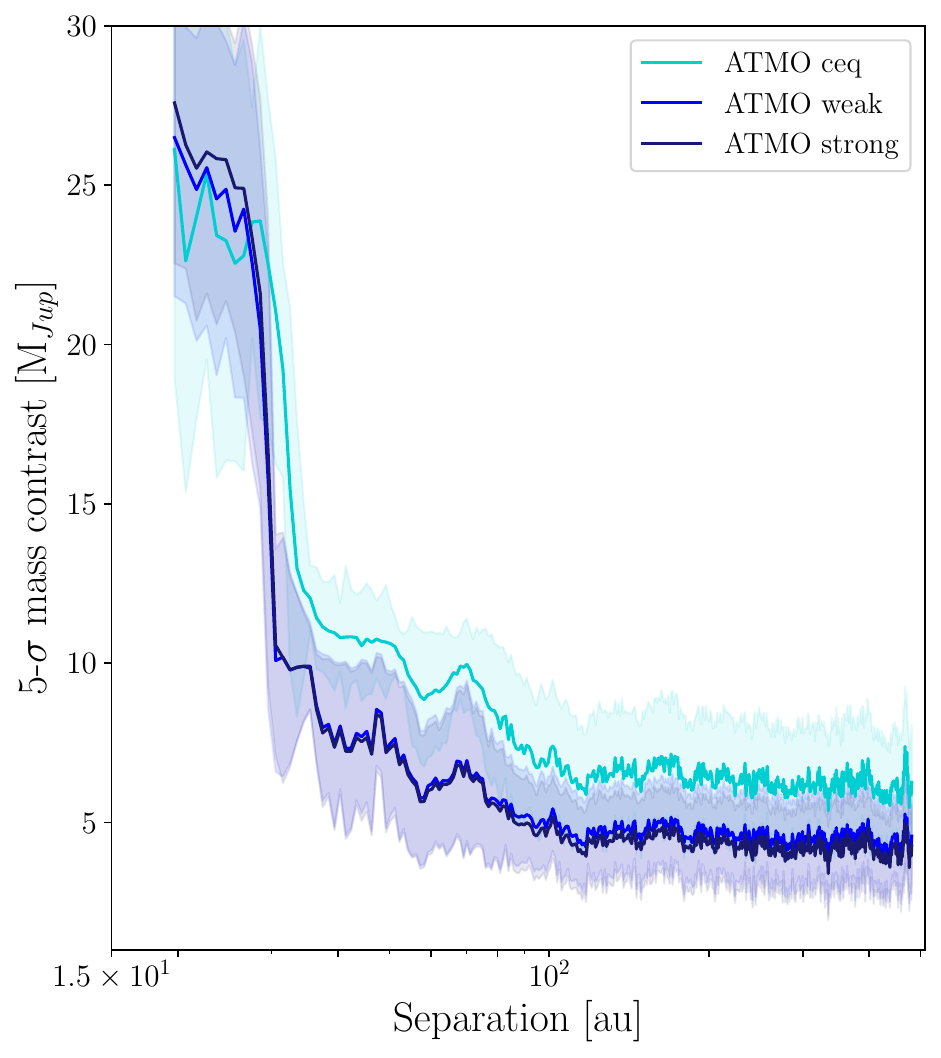}
\includegraphics[width=5.75cm]{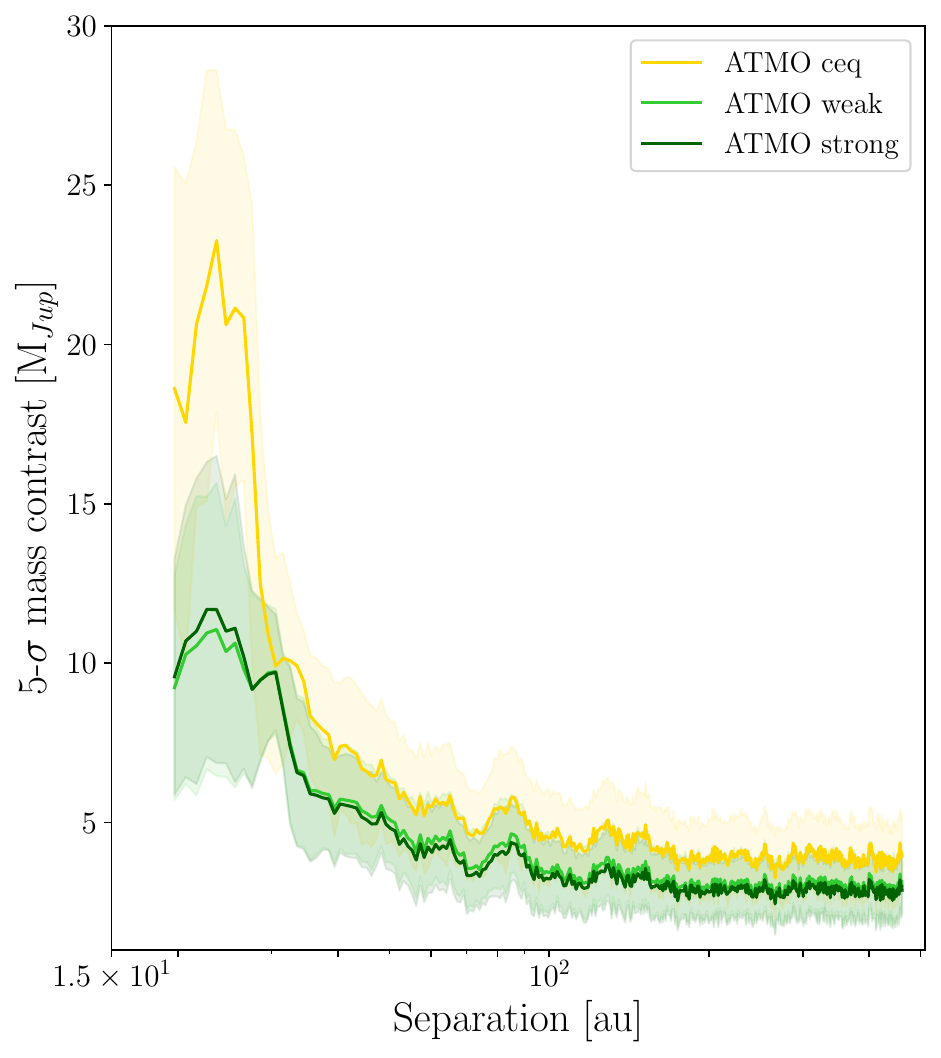}
\includegraphics[width=5.75cm]{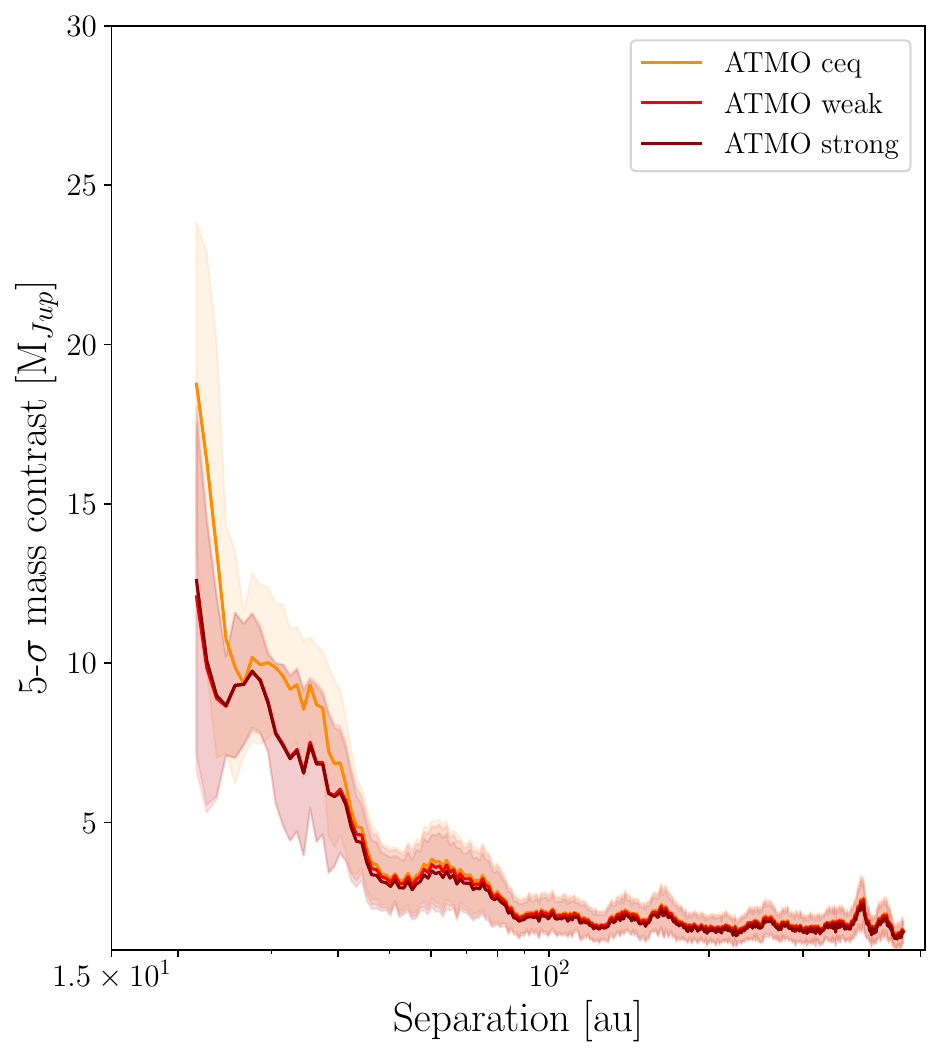}
\caption{ Mass sensitivity as a function of projected separation. From \textit{left} to \textit{right}: Mass limit at $10\mu m$ (\texttt{F1065C}), $11\mu m$ (\texttt{F1140C}), and $15\mu m$ (\texttt{F1550C}). The different curves in each subplot correspond to the different \texttt{ATMO} models used: Chemical equilibrium, weak chemical disequilibrium, and strong chemical disequilibrium. The shaded areas correspond to $1\sigma$ uncertainty.}
\label{fig:Mass_contrast-ATMOceq}%
\end{figure*}

\begin{figure}
\centering
\includegraphics[width=5.75cm]{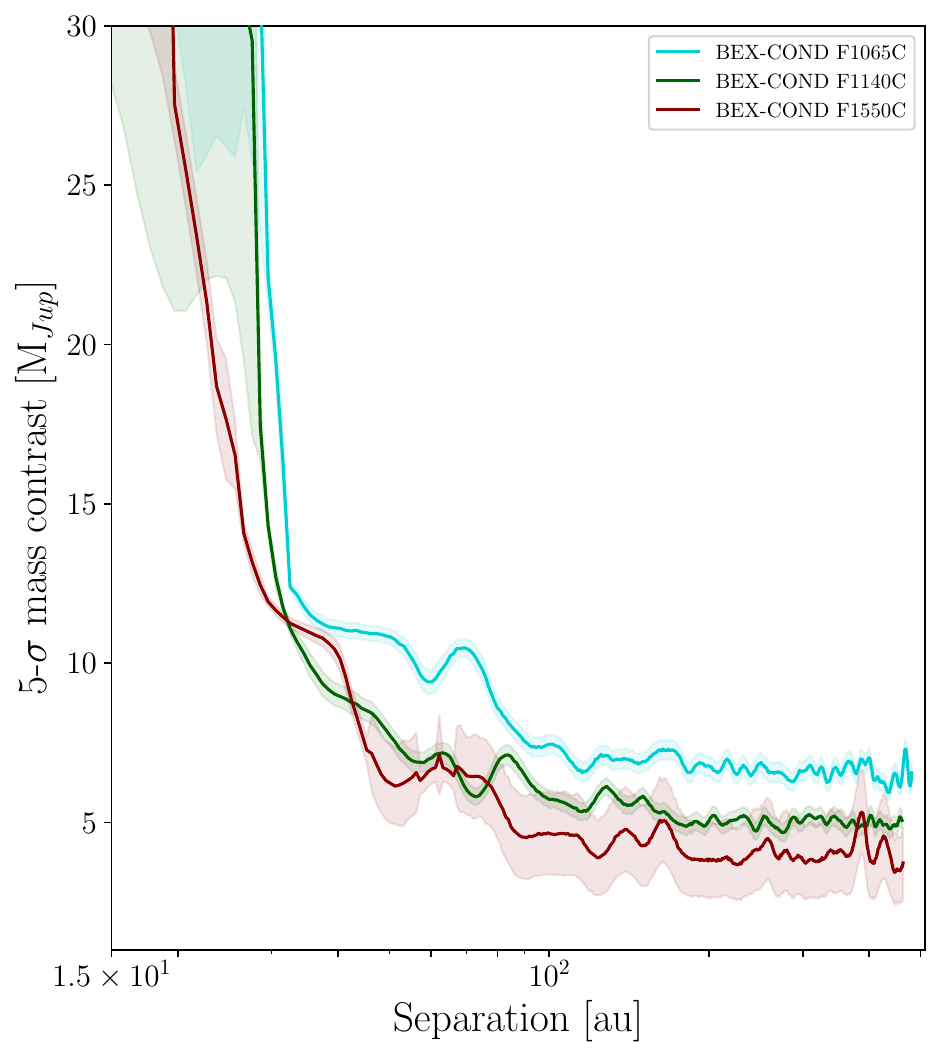}
\caption{Same as Figure\,\ref{fig:Mass_contrast-ATMOceq} but using the \texttt{BEX-COND} atmospheric models for \texttt{F1065C} (cyan), \texttt{F1140C} (dark green), and \texttt{F1550C} (dark red). }
\label{fig:Mass_contrast-BEX-COND}%
\end{figure}

We used the synthetic stellar magnitudes (Section\,\ref{sec:raw_contrast}), the parallax and the age (Table\,\ref{table:star_parameters}), with the contrast limits obtained in Section\,\ref{sec:raw_contrast}, to compute the substellar companion mass limits to which our MIRI data are sensitive. We compare the detection limits obtained with two different families of evolutionary models, \texttt{ATMO} and \texttt{BEX-COND}. The \texttt{ATMO} models (\citealt{Phillips+2020}) assume a non-significant contribution of clouds in the atmosphere of the sub-stellar object, but instead use fingering convection models to reproduce the observed spectrum. \texttt{ATMO} is based on three possible scenarios: atmospheric chemical equilibrium, weak chemical imbalance, and strong chemical imbalance. Figure\,\ref{fig:Mass_contrast-ATMOceq} shows the mass detection limit as a function of the projected separation in astronomical units for the three ATMO models and for the three filters ($10$, $11$, and $15$ $\mu m$). We studied the three models and we have not found significant differences in our results, except for the case of chemical equilibrium where the mass limit is slightly higher for the filters at $10$ and $11\mu m$. Our observations are sensitive to objects with masses, in average, of $17\pm3\,\mathrm{M_{Jup}}$ at a separation of $26$\,au, corresponding approximately to the location of HR\,2562\,B ($27\pm5\,\mathrm{M_{Jup}}$ for \texttt{F1065C}, $12\pm5\,\mathrm{M_{Jup}}$ for \texttt{F1140C}, and $13\pm5\,\mathrm{M_{Jup}}$ for \texttt{F1550C})\footnote{The uncertainties are estimated from simple error propagation, see Fig\,\ref{fig:PMD-ATMO_uncert_ceq}}. For larger separations (i.e., $100$\,au), the mass sensitivity is $6\pm2\mathrm{M_{Jup}}$, $4\pm1\mathrm{M_{Jup}}$, and $3\pm1\mathrm{M_{Jup}}$ for \texttt{F1065C}, \texttt{F1140C}, and \texttt{F1550C}, respectively. Similarly, we used the evolutionary models of \texttt{COND} (\citealt{Allard+2001}) coupled\footnote{We did not combine both models/codes, but merged to create an extended grid to move from high to low planetary mass regimen.} with \texttt{BEX} (\citealt{Linder+2019}) to extend the lower-mass bound, in the same way as shown in \cite{Vigan+2021} (BEX-COND evolutionary track). The main difference with the \texttt{ATMO} models is that \texttt{COND-BEX} uses clouds in the atmosphere to reproduce the shapes and emission of the atmosphere of sub-stellar objects. Similarly as for \texttt{ATMO}, our observations, on average, are sensitive to objects of mass $31\pm 4\,\mathrm{M_{Jup}}$ at $26$ au using \texttt{COND-BEX} evolutionary models. For larger separations, at $100$\,au, the mass sensitivity is $7\pm 1\,\mathrm{M_{Jup}}$ in average. Figures\,\ref{fig:Mass_contrast-ATMOceq} and \ref{fig:Mass_contrast-BEX-COND} show the detection limits in mass versus projected separation in au for \texttt{ATMO} and \texttt{COND-BEX}, respectively. Mass sensitivity varies depending on the filter, reaching the deepest limit at $15\mu m$. The range of masses detectable at $100$ au is comparable and compatible with both models.

\begin{figure*}
\centering
\includegraphics[width=5.75cm]{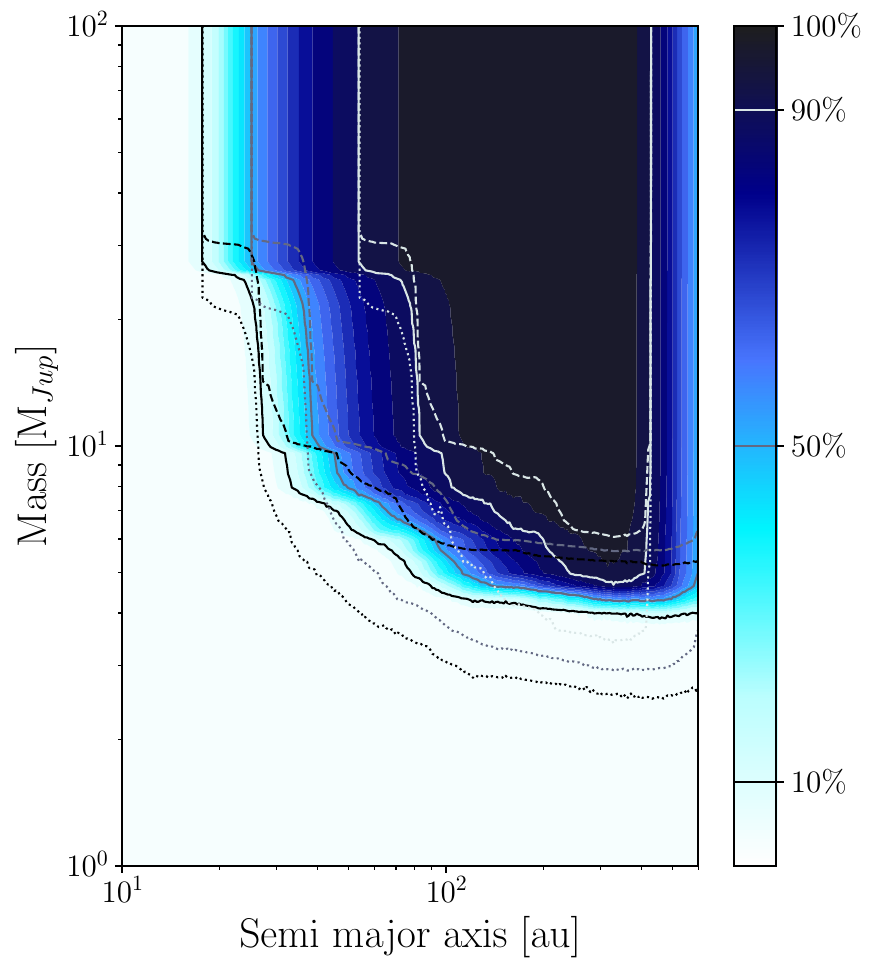}
\includegraphics[width=5.75cm]{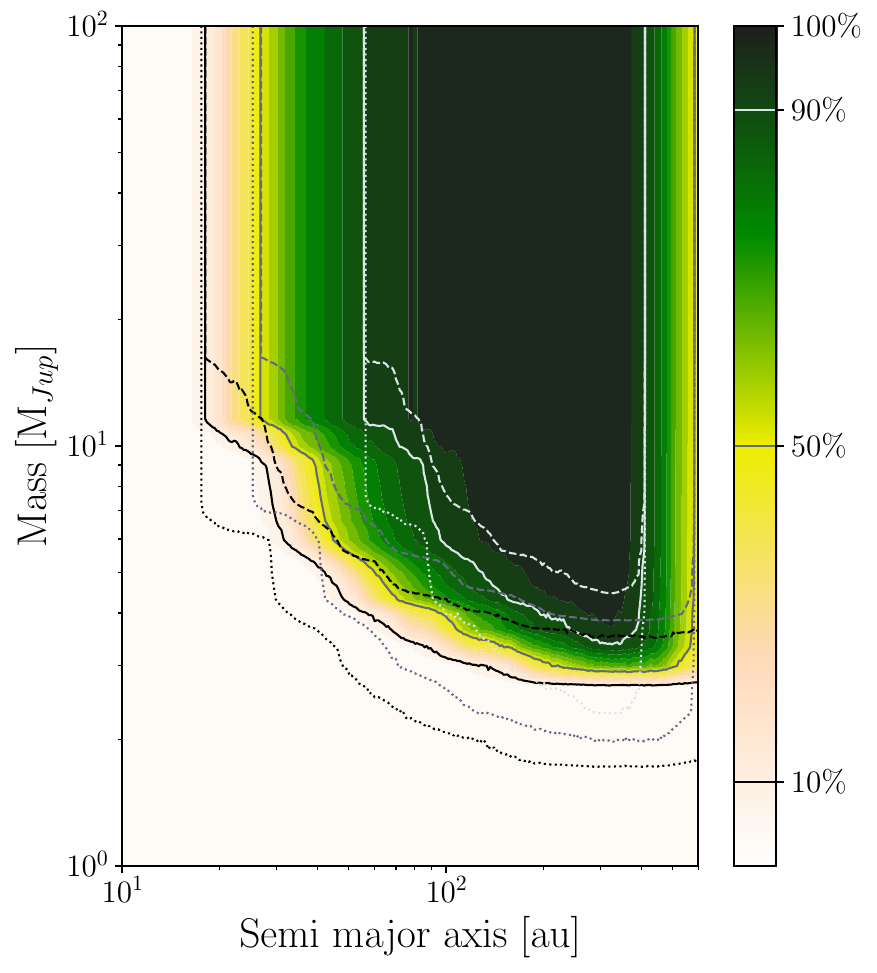}
\includegraphics[width=5.75cm]{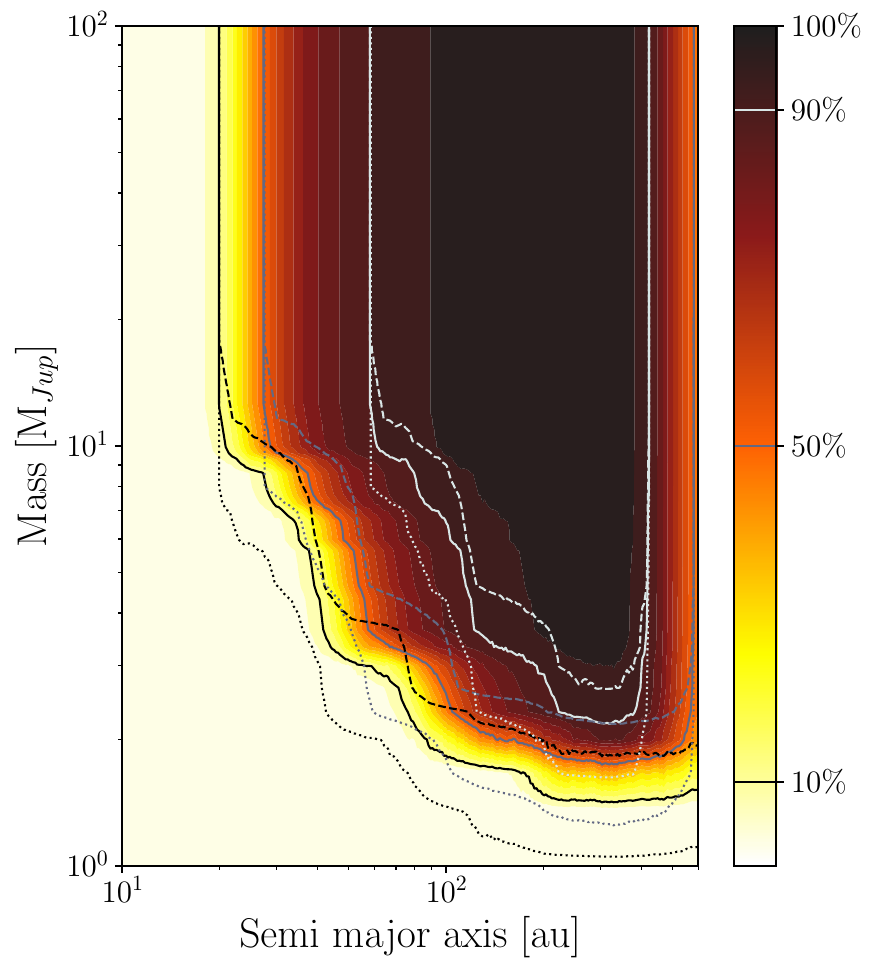}
\caption{Sensitivity of JWST/MIRI observations using \texttt{ATMO-strong} (strong chemical disequilibrium) evolutionary models. From \textit{left} to \textit{right}: $10\mu m$ filter \texttt{F1065C}, $11\mu m$ \texttt{F1140C}, and $15\mu m$ \texttt{F1550C}. The color bar in each plot means the detection probability and the solid lines highlight the 10\%, 50\%, and 90\% detection thresholds. The dotted and dashed lines correspond to $1\sigma$ uncertainties, respectively. }
\label{fig:PMD-ATMOceq}%
\end{figure*}

We determined the sensitivity maps (or probability detection map) using the ``Exoplanet Detection Map Calculator'' (\texttt{Exo-DMC}\footnote{\url{https://github.com/mbonav/Exo_DMC}}, \citealt{Exo-DMC}) code. We simulated $2.5\times 10^8$ orbits with parameters between $10$ and $600$ au, and a range of masses between $1$ and $120\,M_{Jup}$. We adopted a conservative approach, not using the known inclination of the disk and of HR\,2562\,B as prior. Figure\,\ref{fig:PMD-ATMOceq} shows the sensitivity maps using the contrast limits at $10$, $11$, and $15\,\mu m$ obtained with \texttt{ATMO} and strong chemical disequilibrium, while Figure\,\ref{fig:PMD-BEX-COND} shows the maps obtained with \texttt{BEX-COND}. In Appendix\,\ref{apx:PMD} we show the \texttt{ATMO} sensitivity mass for the chemical equilibrium case.

\begin{figure*}
\centering
\includegraphics[width=5.75cm]{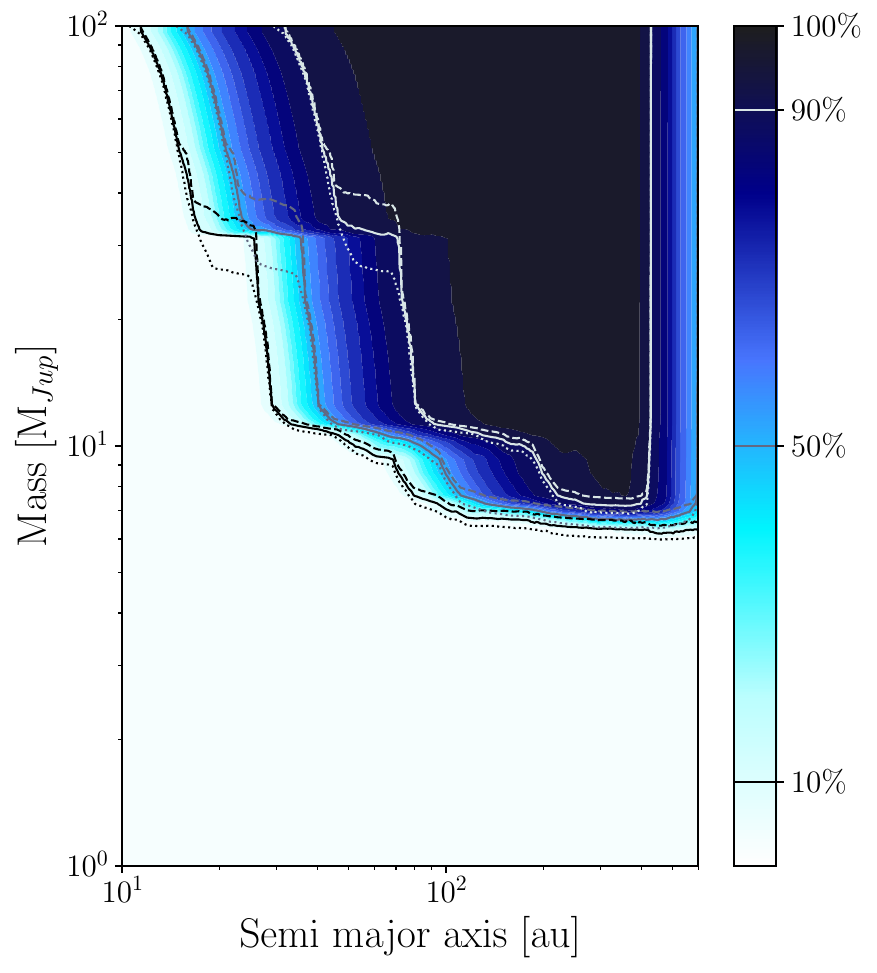}
\includegraphics[width=5.75cm]{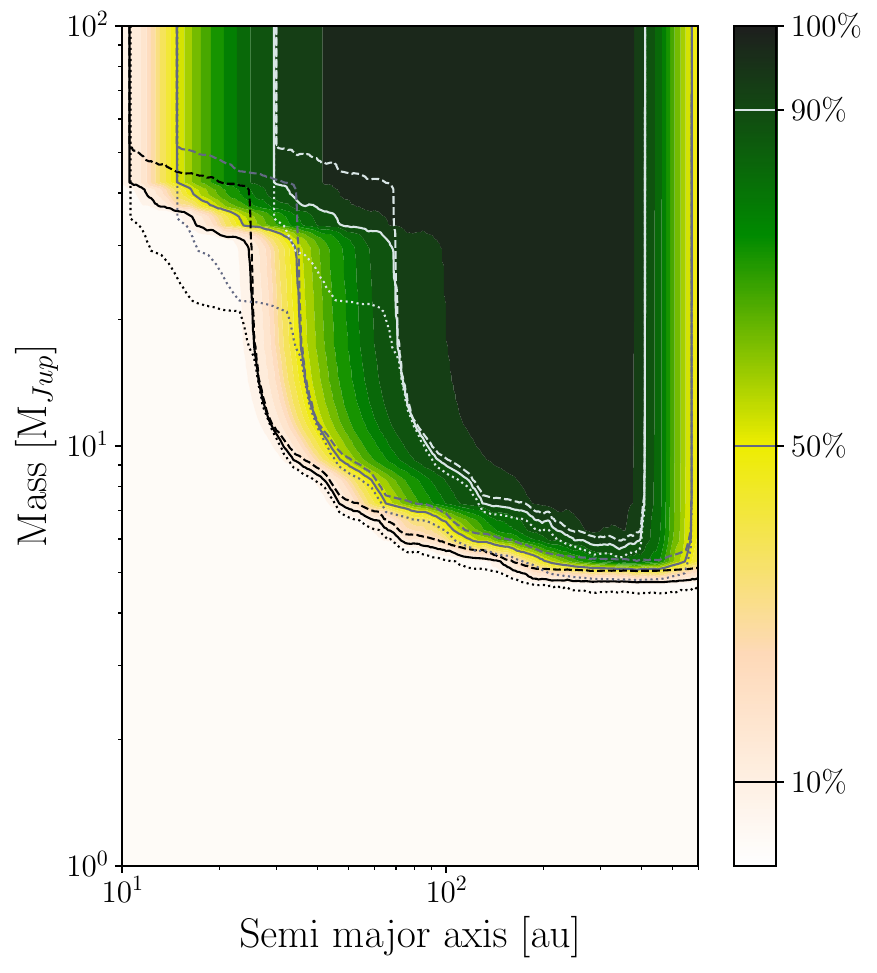}
\includegraphics[width=5.75cm]{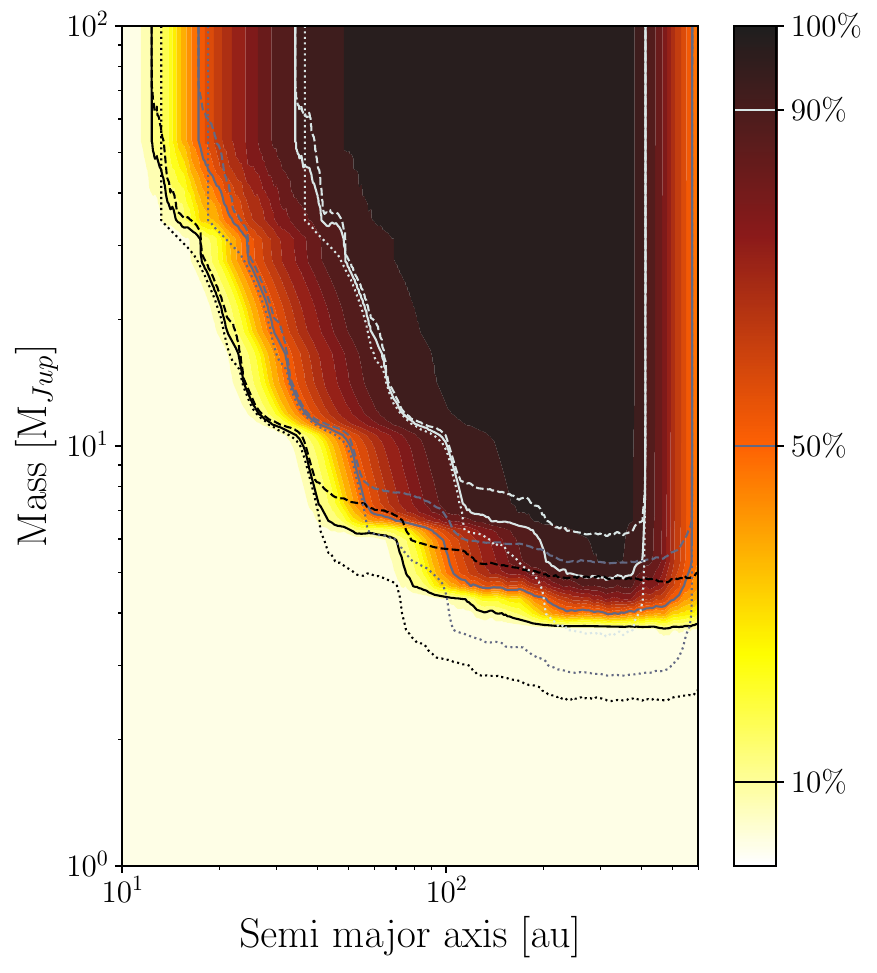}
\caption{Same as Figure\,\ref{fig:PMD-ATMOceq} but using the \texttt{BEX-COND} evolutionary models. }
\label{fig:PMD-BEX-COND}%
\end{figure*}

One of the factors that most influence the magnitude-mass transformation is age. Despite \cite{Mesa+2018} efforts, the age of HR\,2562 is not well constrained and has a much wider estimation/uncertainty range (250 to 750\,Myr) compared to, for example, VHS\,1256–1257 (VHS\,1256, $140\pm20$\,Myrs, \citealt{Dupuy+2023}). The uncertainty on the age of the system is reflected on the mass detection limits as noticeable uncertainties and error bars (see, e.g., Figure\,\ref{fig:Mass_contrast-ATMOceq}). We propagated the uncertainties using Monte Carlo simulations to reflect their impact on the sensitivity maps. Figure\,\ref{fig:PMD-ATMOceq} and \ref{fig:PMD-BEX-COND} shows the sensitivity maps, where the dotted and dashed lines mark the $1$-$\sigma$ uncertainty of each solid contour line (corresponding to different detection probability values). These uncertainties change the probability ranges, making it possible for us to observe objects down to $10-15\,\mathrm{M_{Jup}}$ at separations of $\sim 25\,\mathrm{au}$, in agreement with our estimates for HR\,2562\,B.

The sensitivity maps show that our observations can reach a $50\%$ probability of detection for a $10\,\mathrm{M_{Jup}}$ at $50$\,au. However, in regions closer to the star ($<30$\,au) we quickly lose sensitivity, reaching the order of $20-30\,\mathrm{M_{Jup}}$ (at 50\% sensitivity), and $10-15\,\mathrm{M_{Jup}}$ in the detection limits at $20\,\mathrm{au}$. This is particularly important given that HR\,2562\,B is at a projected distance of $23\,\mathrm{au}$ (semi-major axis of $24$-$30\,\mathrm{au}$, depending on the eccentricity), and whose mass has been restricted to no more than $18.5\,\mathrm{M_{Jup}}$ from multi-epoch dynamical constraints (\citealt{Zhang+2023}). However, the detection of this object is in good agreement with these evolutionary models given the large uncertainties in the probability map detection (see Appendix\,\ref{apx:PMD}).

\subsection{Color-magnitude diagram from MIRI observations}\label{sec:bands}

One advantage of using color-magnitude diagrams is to identify sub-groups of populations (e.g., Y-dwarfs, AGB stars), to infer some physical properties. The choice of filters (wavelengths) is essential to make this distinction. The filters \texttt{F1065C} and \texttt{F1140C} were designed to distinguish the objects with ammonia absorption in their atmospheres (e.g., T-dwarfs), but we can also try to infer the presence of other substances such as methane and silicates. The presence of these substances is directly linked to the pressure-temperature profile, the chemical (dis)equilibrium, and the presence of clouds in the upper atmosphere. For example, silicates are characterized by small-size particles ($\leq 1\mu m$), and form clouds that show absorption at $10-12\,\mu m$ in the upper atmosphere (e.g., \citealt{Cusing+2006}). Silicates are often present in objects with low temperatures such as T and L dwarfs, with the strongest absorption for L4-L6 objects, corresponding to $\mathrm{T_{eff}}$ between $1480$-$1690$\,K (\citealt{Filippazzo+2015}). Silicates produce much redder colors on these objects (\citealt{Suarez+2023}), while at lower temperatures ($\mathrm{T_{eff}}<\sim1250$\,K, \citealt{Suarez+Metchev-2022}) the silicates condense to deeper altitudes, making these dwarfs bluer, a color-change representative of the L/T transition. The presence of silicates is not observable for brown dwarfs later than T0 due to this condensation temperature threshold. Also, the appearance of the $\sim8\mu m$ methane absorption (which overlaps with the silicate absorption feasure) at the L/T transition (\citealt{Cusing+2006}; \citealt{Suarez+Metchev-2022}) could also influence the non-detection of silicates for T0 or later objects.

On the other hand, the presence of ammonia is also expected at lower temperatures where its dissociation is less effective, corresponding to spectral types early-Y (\citealt{Cushing+2011}) and T2-T9 and even as early as T1.5 (\citealt{Suarez+Metchev-2022}). However, the ammonia could be also found at higher temperatures. Indeed, the composition, pressure-temperature profile, mixing processes, and metalicity can play an important role in the presence of ammonia (\citealt{Lodders+Fegley-2002}). For example, ammonia has been measured in atmospheres of about $1000$\,K (\citealt{Tannock+2022}). Ammonia has the potential to be detected up to $\mathrm{T_{eff}}$ = $1200$\,K when the atmosphere is in non-chemical equilibrium. We refer to \cite{Danielski+2018} for more details on this topic, applied to MIRI coronagraphic observations. Detection and non-detection of ammonia in a range of L-T dwarfs can thus help to understand the chemical equilibrium and species in the atmosphere. The ammonia double absorption feature falls within the \texttt{F1065C} range, namely, between $10.2\mu m$ and $10.8\mu m$.

Methane, on the other hand, can only exist as a molecule at temperatures below $1400$\,K since the chemical reaction between methane and carbon monoxide is less efficient at these temperatures (\citealt{Lodders+2002}). The equilibrium of a chemical reaction depends strongly on the pressure-temperature profile and concentration of each molecule in the reaction equation. For this reason, lower temperatures help to the prevalence of methane in the chemical (des)equilibrium reaction. Also, the presence of methane is common in T-dwarf (\citealt{Cushing+2005}), between L8 and T9 (\citealt{Lodders+Fegley-2002}). The absorption at MIR wavelengths is predominant between $\sim7\mu m$ and $\sim9\mu m$, but can affect the shape of the SED even at $10\mu m$. Although methane has been observed in other sub-stellar objects, determining its presence at the T/L transition may provide us with a unique opportunity to understand the evolution of planetary atmospheres, composition, and chemical disequilibrium as a function of effective temperature and surface gravity at early ages. 

Although the best way to study the presence of these substances is through spectroscopy (e.g., \citealt{Miles+2023}), medium band filter imaging is also very useful for identifying objects that may have any of these substances. These chemical components can be identified by comparing the photometry in the \texttt{F1065C} and \texttt{F1140C} MIRI filters, which cover part or all of the aforementioned absorption features. For example, in a color-magnitude diagram using \texttt{F1065C} and \texttt{F1140C} MIRI filters, much bluer colors are expected for substances that survive at lower temperatures (negative values in the \texttt{F1140C}-\texttt{F1065C} color-magnitude).

\begin{figure*}
\centering
\includegraphics[width=5.8cm]{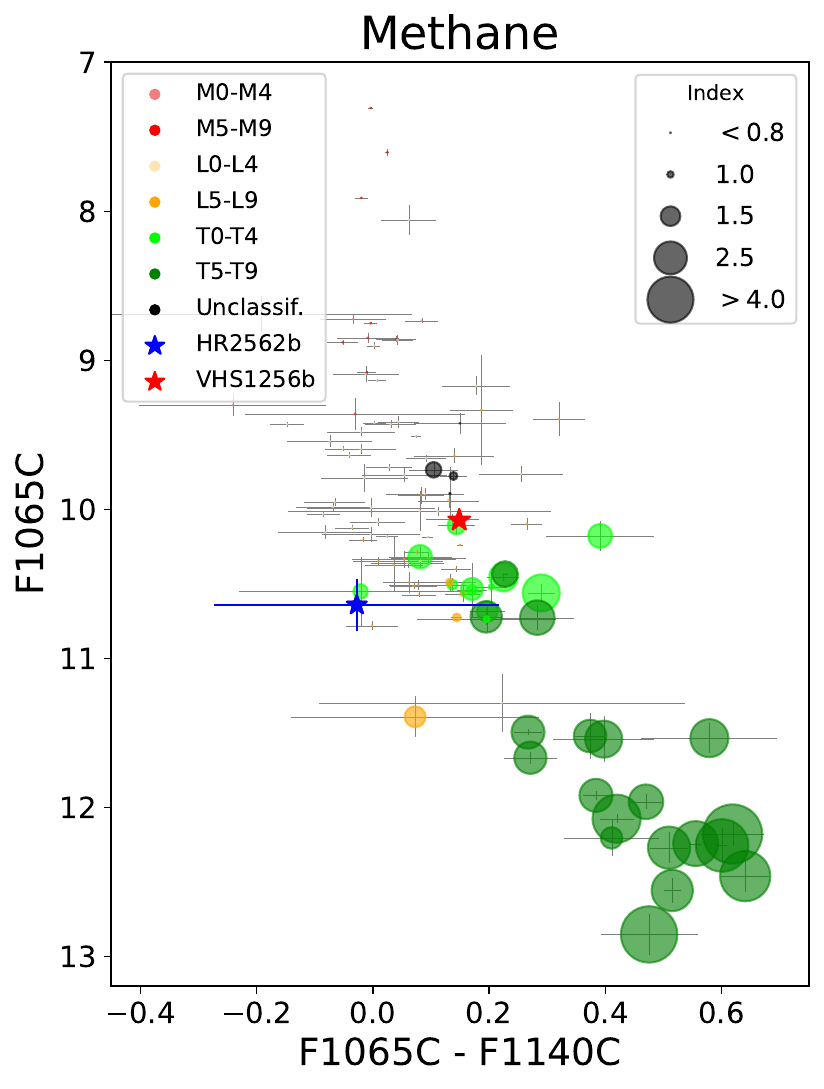}
\includegraphics[width=5.8cm]{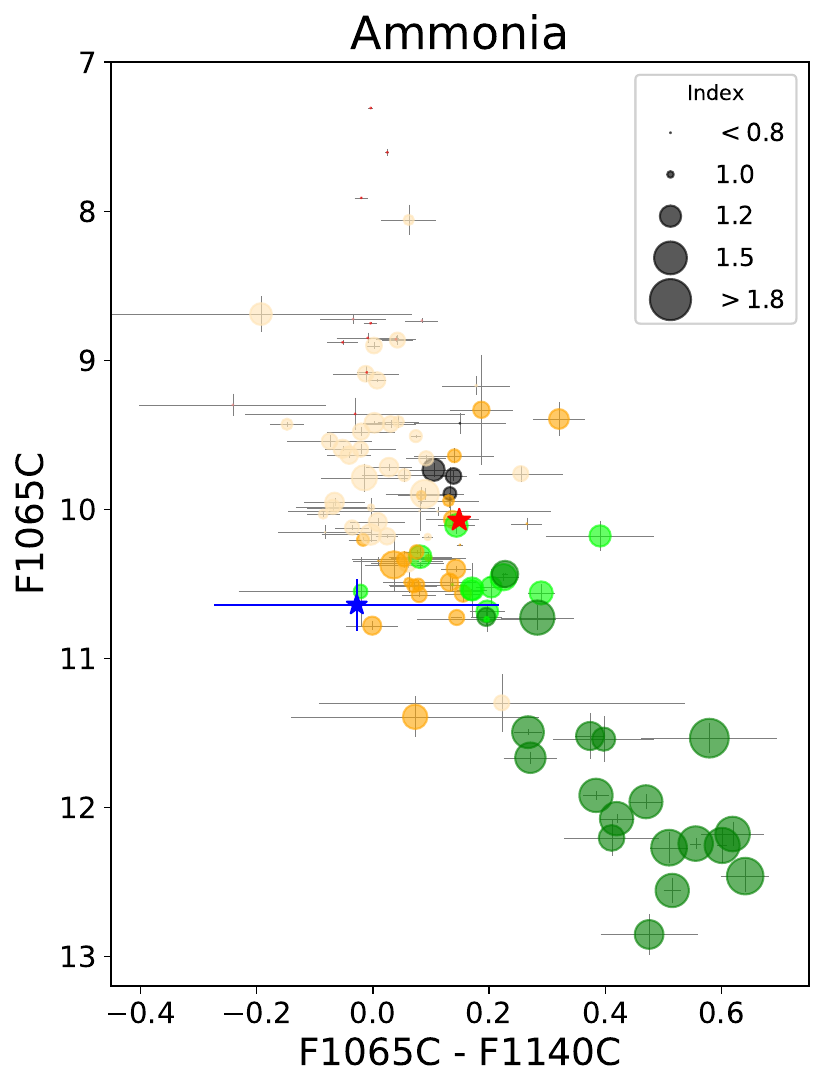}
\includegraphics[width=5.8cm]{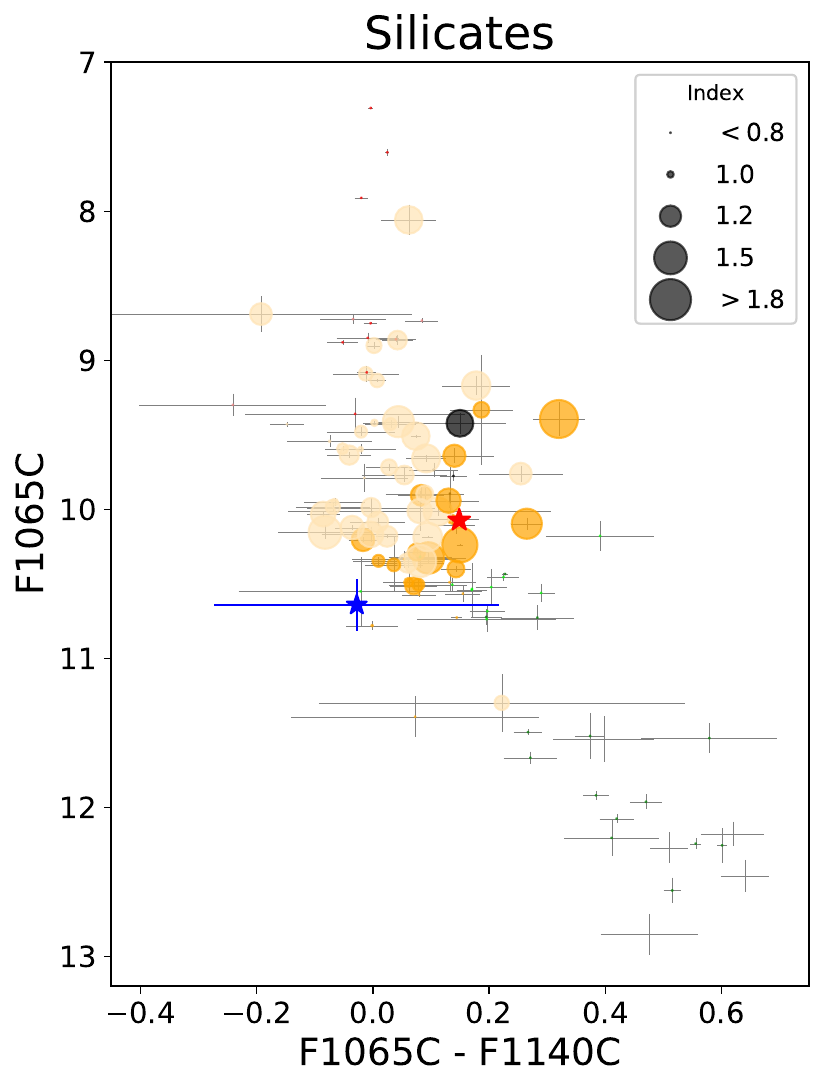}
\caption{ Color-magnitude diagram using the $10\mu m$ and $11\mu m$ filters from MIRI. The dots correspond to the photometry obtained from the Spitzer spectra sample (\citealt{Suarez+Metchev-2022}), while the colors refer to the spectral type. The subplots correspond to the same color-magnitude diagram but show the methane (\textit{left}), ammonia (\textit{middle}), and silicates (\textit{right}) spectral indices, as defined in \cite{Suarez+Metchev-2022}, related to the depth of the absorption feature of each component. The circle sizes in each subplot refer to the value of the spectral index, as defined by \cite{Suarez+Metchev-2022}. The gray dots in each sub-panel correspond to species non-detection (i.e., plotted with index=0, non-highlighted sources). The black dots refer to unclassified spectral types. The blue star corresponds to HR\,2562\,B, and the red star to VHS\,1256\,b from \cite{Miles+2023}. Note that the Y-axis (i.e., \texttt{F1065C}) refers to the absolute magnitude.}
\label{fig:CMD_index}%
\end{figure*}

In order to investigate the presence of ammonia, silicates, and methane in the atmosphere of HR\,2562\,B, we compared its photometry in the \texttt{F1065C} and \texttt{F1140C} filters with those of the field dwarf population with known and quantified abundances of these species. To do so, we used the measured \texttt{Spitzer IRS} spectra of $113$ field dwarfs for which the depth of the absorption features have been quantified by a dedicated species spectral index by \cite{Suarez+Metchev-2022}, and we computed the integrated photometry in the MIRI \texttt{F1065C} and \texttt{F1140C} filters. Figure\,\ref{fig:CMD_index} shows the corresponding color-magnitude diagram, in which we have over-plotted our measured photometry for HR\,2562\,B (blue star). The different plotted objects correspond to integrated photometry using the spectra taken from \cite{Suarez+Metchev-2022}. Each of these plots highlights the sub-population with strong absorption features for each species and inter-compare HR\,2562\,B within these samples. In each of the figures, the size of the colored circle represents the abundance of the corresponding species, as it is represented by the spectral index.

We tried to understand if the color shown by HR\,2562\,B (blue star in Figure\,\ref{fig:CMD_index}) is indicative of any of these substances. However, unlike VHS\,1256\,b (red star) which is located clearly in the silicates trend, HR\,2562\,B is located at a transition point between silicates and ammonia. Note that we can only directly distinguish silicates and ammonia from these filters, and indirectly the methane. Considering the uncertainties, the three species could be present in the atmosphere (but in different amounts/concentrations). A spectroscopic observation and analysis are key to knowing which species dominate/are present in the atmosphere. From our modeling (see Section\,\ref{sec:modeling}), an effective temperature of $1250\pm25\,K$ put HR\,2562\,B in the limit between the regimes where silicates ($\mathrm{T_{eff}}\gtrsim 1250$\,K), ammonia ($<1200\,K$), and methane ($\mathrm{T_{eff}}\lesssim 1400$\,K) are expected. Note that the presence of ammonia is possible in the atmosphere, however, the depth of the absorption expected at these temperatures is not strong enough to produce a deep shape in the spectrum at $10\mu m$, as can be seen in Figure\,\ref{fig:CMD_index} and Fig.\, 4 and 6 in \cite{Suarez+Metchev-2022}. However, for slightly lower temperatures, ammonia starts to affect more significantly the shape of the spectra. The detection of ammonia can help to investigate the environmental conditions that help to the prevalence of this species at high temperatures. In the case of silicates, we can compare the case of VHS\,1256\,b, which has different values for the $\mathrm{T_{eff}}$ in the literature (e.g., $1100\,K$, \citealt{Miles+2023}; $1240\,K$, \citealt{Hoch+2022}; $1380\,K$, \citealt{Petrus+2023}). From \cite{Miles+2023} we know that VHS\,1256\,b has silicates at these wavelengths, meaning that the silicates could survive even at slightly lower temperatures than $1250\,K$ which generates tensions with the theory. It is possible that other unknown or not considered physicochemical processes in the atmosphere could allow the survival of silicates at lower temperatures (for example, different pressure-temperature profiles coming from differential convective layers in the atmosphere). For the species to be present we expect HR\,2562\,B to present notorious absorption features. However, methane slightly affects the shape of the SED around $10\mu m$ due to a broad absorption feature at $7$-$9\mu m$ (for low temperature objects), thus having an indirect impact on the MIRI colors, while ammonia has more direct impact on the MIRI color with a strong absorption feature at $10.6\mu m$. Both species can contribute to producing redder colors, although more strongly for ammonia. At high temperatures ($\sim1200\,K$), and as in the color-magnitude diagram, the methane does not have a strong influence on the color of the sub-stellar objects. Another possibility could be the presence of water (lower flux than in the models, see Figure\,\ref{fig:ATMO_best_model}). We explored this option in Appendix\,\ref{apx:CMD_w}, in which Figure\,\ref{fig:CMD_water} shows the same color-magnitude diagram as in Figure\,\ref{fig:CMD_index} but coded with the water spectral index. We conclude the presence of water does not have an important impact on the \texttt{F1065C-F1140C} color, since water vapor is present in the L and T spectral types with a low spectral index when comparing with silicates in the range L1-L7, ammonia for $>$T5, and methane for $>$L7 (\citealt{Suarez+Metchev-2022}). This means that the color \texttt{F1065C-F1140C} is dominated by the other chemical species, even if water vapor generates characteristic absorptions in the SED, these are indistinguishable between spectral types at these wavelengths. We deduced that silicates are more favorable to be detected in HR\,2562\,B but with a low spectral index (i.e., weak silicate absorption), and ammonia but with a low chance at these wavelengths, given the low surface gravity and effective temperature. With a low surface gravity, the deep atmosphere is hotter which is prone to form $\mathrm{N}_{2}$ rather than $\mathrm{NH}_{3}$. For example, a log(g) of 4.4 generates a weaker absorption of ammonia than a log(g) of 5.0 (for different fixed temperatures using \texttt{ATMO}). However, the key parameter is the effective temperature, for which higher temperatures (1200-1400K) produce weak or null absorption signatures. Since HR\,2562\,B is at the limit of detectability for any species, medium-resolution spectroscopy in the $6-20\mu m$ range is required to precisely identify signatures for these species.

As mentioned in previous studies (e.g., \citealt{Konopacky+2016}), HR\,2562\,B is a transitional object between T and L spectral types, but it is different from VHS\,1256\,b, which shows clear silicate absorption and a different SED (\citealt{Miles+2023}). Although both objects belong to this transition between T and L spectra, they are in different evolution stages. In fact, \cite{Miles+2023} present similar temperatures for VHS\,1256\,b, in addition to having a better constrained age ($140\pm20\,\mathrm{Myrs}$), and a mass that is estimated to be less than $20\,\mathrm{M_{Jup}}$ (comparable to HR\,2562\,B, $\leq18.5\,\mathrm{M_{Jup}}$). Comparing the values of log(g), luminosity, and radius, we can affirm that both objects are similar, but in different evolutionary stages/transitions, which is reflected in their positions in the color-magnitude diagram. 

Another explanation could be the different viewing angles both objects. \cite{Suarez+2023} investigated the silicates index as a function of the brown dwarf's inclination, finding that edge-on objects present a stronger absorption, and the face-on a weak absorption. This could explain the different locations of VHS\,1256\,b and HR\,2562\,B in the color-magnitude diagram (Fig.\,\ref{fig:CMD_index}). The inclination of the rotation axis of VHS\,1256\,b is $\sim$ 90\degree (edge-on, \citealt{Zhou+2020}), which means the silicates produce a stronger absorption (as was shown in \citealt{Miles+2023}). However, from \cite{Zhang+2023}, the inclination of the orbit of HR\,2562\,B is $\sim$83\degree. The weaker silicate feature in HR\,2562\,B may indicate that the object is seen closer to pole-on than VHS\,1256\,b, and may thus have been perturbed at some point in its evolution, causing a misalignment between its orbit and rotational axis. A detailed investigation of HR\,2562\,B atmosphere would be particularly important since it would help us get a much broader view of the L/T transition. In addition, understanding the differences between the two objects could help us understand the evolution mechanisms in brown dwarf atmospheres. The main difference between both objects could lie in the sedimentation of the silicates, which could be more advanced in the case of HR\,2562\,B, for example. Therefore HR\,2562\,B is an excellent candidate for spectroscopic observations (e.g., with JWST/MIRI-MRS to detect or discard the silicate absorption), that will be necessary to understand the main differences with VHS\,1256\,b, chemical composition, and pressure-temperature profiles.

\subsection{HR\,2562\,B atmospheric characterization }\label{sec:modeling}

We used two types of models for the atmospheric characterization of HR\,2562\,B: \texttt{ATMO} (\citealt{Tremblin+2015}; \citealt{Tremblin+2016}; \citealt{Phillips+2020}) and \texttt{Exo-REM} (\citealt{Baudino+2015}; \citealt{Charnay+2018}; \citealt{Blain+2021}). Briefly, \texttt{ATMO} assumes cloudless atmospheric models, in which the adiabatic convective processes and disequilibrium chemistry reproduce the reddening observed in the spectrum. On the other hand, \texttt{Exo-REM} is a model that includes clouds of different compositions, so the spectral appearance is influenced by the Pressure-Temperature profile and metalicity, among others (see \citealt{Miles+2023} for more recent comparison between \texttt{ATMO} and \texttt{Exo-REM} with other models). These models are usually used under the premise that gas giant planets have clouds in the upper layers, which substantially modifies the shape of the SED. We used the open source code \texttt{species}\footnote{\url{https://github.com/tomasstolker/species}}\footnote{\url{https://species.readthedocs.io/en/latest/}} (\citealt{Stolker+2020}) to fit the HR\,2562\,B observations with \texttt{ATMO} and \texttt{Exo-REM} using the data presented in Figure\,\ref{fig:Full_sed_only} (Table\,\ref{table:Obs_summary}). In the next paragraphs, we will present our analysis using \texttt{ATMO} and \texttt{Exo-REM} separately.

\textit{ - \texttt{ATMO} fit procedure}

\begin{figure*}
\centering
\sidecaption
\includegraphics[width=12cm]{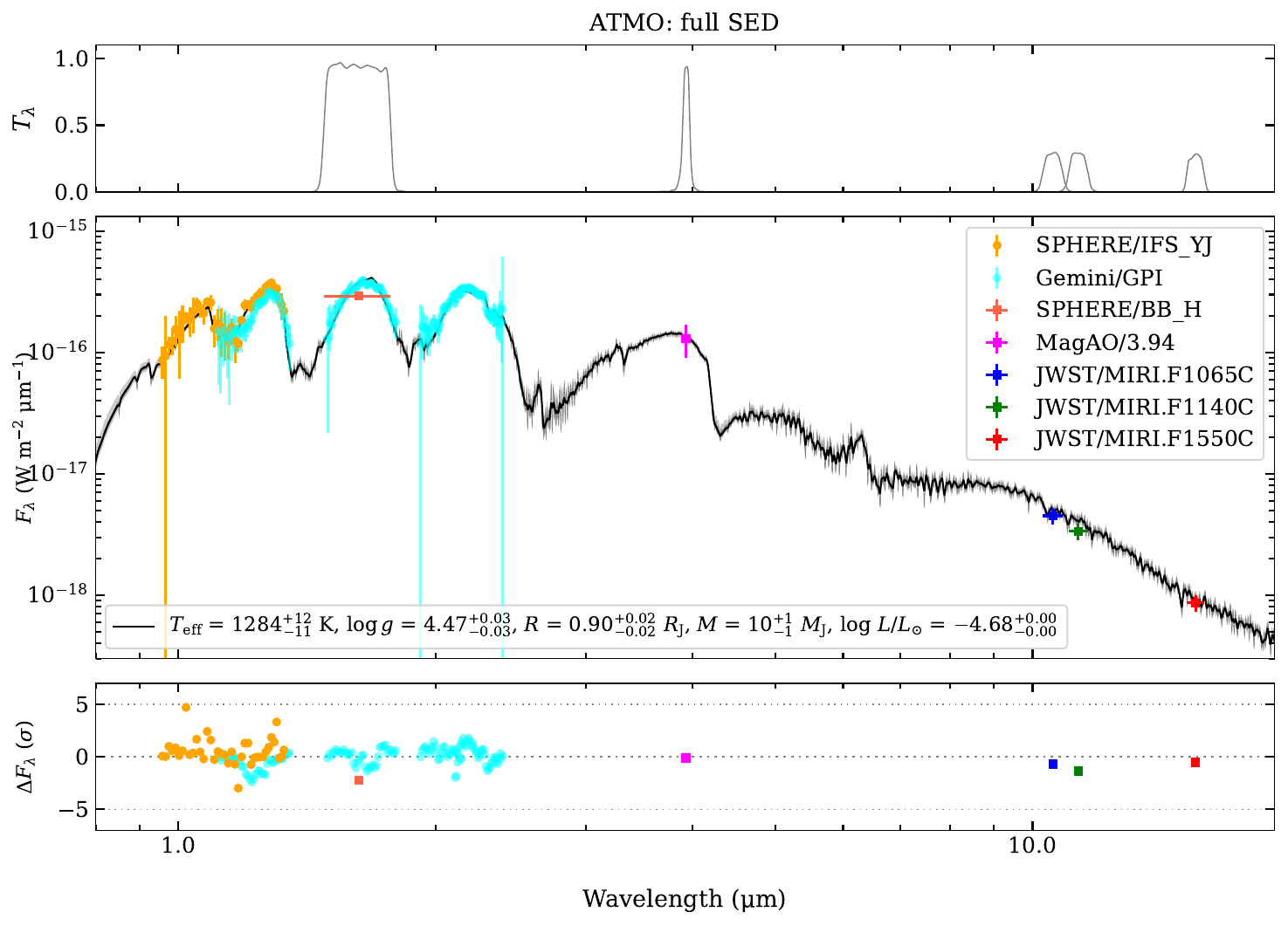}
\caption{Best fit model using \texttt{ATMO} and the entire set of data. \textit{Top panel}: Transmission filters. \textit{Middle panel}: SED fitting. \textit{Bottom panel}: Residuals from the fit. The black and gray lines in the middle panel correspond to the best fit from \texttt{ATMO} and the respective uncertainties. The different color symbols are the dataset. The uncertainty for the luminosity corresponds to $0.0049$.}
\label{fig:ATMO_best_model}%
\end{figure*}

In the case of \texttt{ATMO} (for which we used an \texttt{ATMO} grid computed by \citealt{Petrus+2023}), we combined different datasets to investigate the constraints on the model parameters provided by the MIRI data alone or jointly with NIR observations. These combinations and the best values found for the atmospheric and physical parameters are summarized in Table\,\ref{table:Phys_constr_ATMO}. The main parameters we fit are $\mathrm{T_{eff}}$, log(g), and radius, while the mass is calculated using the log(g) and radius. The (spectroscopic) mass is inferred using the equation $\mathrm{M}\sim 10^{\mathrm{log(g)}}\times\mathrm{R}^2$. In the same way, the luminosity is inferred using $\mathrm{T_{eff}}$ and radius: $\mathrm{L}\sim \mathrm{R}^2\times\mathrm{T_{eff}}^4$. Unlike mass (and indirectly log(g) and radius), all the other parameters were treated without restrictions (i.e., no priors used). \texttt{ATMO} also consider as free parameters the $\mathrm{[Fe/H]}$ and $\mathrm{C/O}$ ratio. However, we did not consider those in our analysis since the best-fit values converged to the edge of the grid meaning those could be no realistic or representative. To fit the data using the atmospheric models, \texttt{species} uses Bayesian inference (with \texttt{MultiNest}) to estimate the posterior distribution for each parameter and provides the best values and related uncertainties. For the particular case of \texttt{ATMO}, we used a Gaussian prior equal to the most probable mass obtained from astrometric studies (i.e., $10\,\mathrm{M_{Jup}}$, \citealt{Zhang+2023}). Figure\,\ref{fig:ATMO_best_model} shows the \texttt{ATMO} best-fit model.

\begin{table*}[]
\centering
\caption{Best parameters from the SED fit using \texttt{ATMO} models with different data combinations.}
  \begin{tabular}{ l c c c c c c c c c c }
    \hline \hline
  Data & Teff & Mass & Radius & log\,(g) & log(L/$L_{\odot}$) \\
       &  [K] & [M$_\mathrm{Jup}$] & [R$_\mathrm{Jup}$] &  &  \\ 
    \hline
MIRI               & $1555^{+289}_{-333}$ & $7^{+4}_{-5}$ & $0.71^{+0.13}_{-0.07}$ & $4.55^{+0.21}_{-0.41}$ & $-4.55^{+0.22}_{-0.28}$  \\
MagAO+MIRI         & $1599^{+242}_{-288}$  & $8^{+4}_{-4}$ & $0.70^{+0.11}_{-0.07}$ & $4.58^{+0.20}_{-0.36}$ & $-4.51^{+0.18}_{-0.23}$ \\
SPHERE+MIRI        & $1360^{+35}_{-35}$ & $5^{+5}_{-3}$ & $0.80^{+0.03}_{-0.02}$ & $4.31^{+0.28}_{-0.31}$ &$-4.68^{+0.03}_{-0.02}$\\
GPI+MIRI           & $1289^{+16}_{-19}$ & $10^{+2}_{-2}$ & $0.90^{+0.03}_{-0.02}$ & $4.48^{+0.06}_{-0.08}$ & $-4.67^{+0.01}_{-0.01}$ \\
SPHERE+MagAO +MIRI  & $1359^{+35}_{-34}$ & $5^{+5}_{-3}$ & $0.80^{+0.02}_{-0.02}$ & $4.30^{+0.27}_{-0.31}$ & $-4.69^{+0.03}_{-0.03}$ \\
GPI+MagAO+MIRI     & $1289^{+16}_{-19}$ & $10^{+2}_{-2}$ & $0.90^{+0.03}_{-0.02}$ &  $4.48^{+0.06}_{-0.08}$ & $-4.67^{+0.01}_{-0.01}$ \\
\hline
Full SED           & $1284^{+12}_{-11}$ & $10^{+1}_{-1}$ & $0.90^{+0.02}_{-0.02}$ &  $4.47^{+0.03}_{-0.03}$ & $-4.68^{+0.01}_{-0.01}$ \\
\hline
  \end{tabular}
\tablefoot{A mass Gaussian prior of mean $10\,\mathrm{M_{Jup}}$ was used in all the cases. The luminosity uncertainty was rounded to $0.01$.}
\label{table:Phys_constr_ATMO}
\end{table*}

\begin{figure*}
\centering
\sidecaption
\includegraphics[width=12cm]{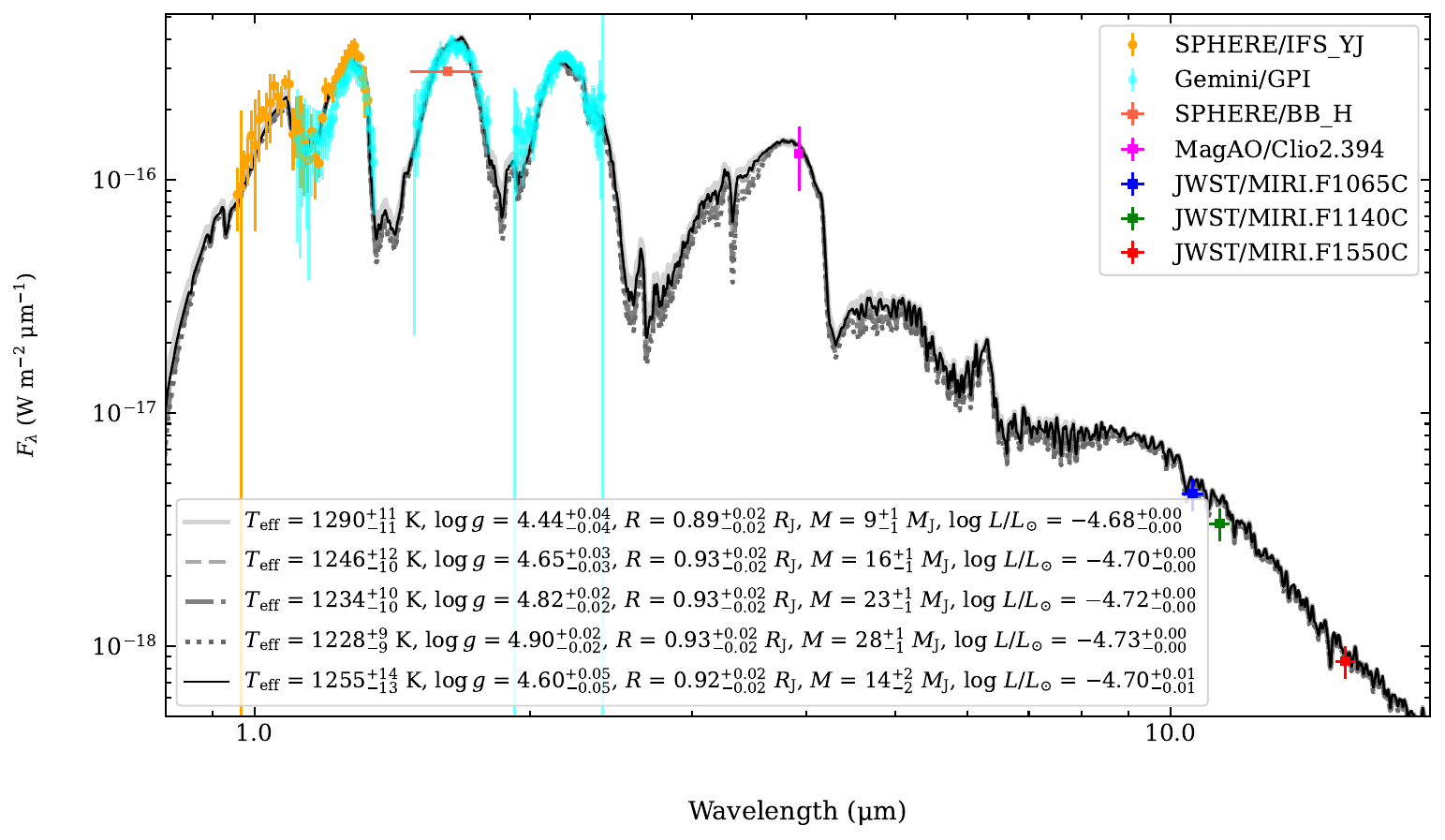}
\caption{SED fitting using the \texttt{ATMO} models with different prior in masses and the full dataset. The different black and gray lines styles correspond to the different best fit for each prior: solid light gray line uses a $10\,\mathrm{M_{Jup}}$ prior; gray dashed line a $18\,\mathrm{M_{Jup}}$ prior; dashed-dotted black line $25\,\mathrm{M_{Jup}}$ prior; dotted black line a $30\,\mathrm{M_{Jup}}$ prior; and the solid black line the no prior case. The colored symbols correspond to the observational datasets.  }
\label{fig:ATMO_best_model_priors}%
\end{figure*}

To understand the effect of the prior that we adopted for the mass, we repeated this fit using different priors of $10$, $18$, $25$, and $30$ Jupiter masses (assuming a Gaussian distribution and a small $\sigma <0.03$), and also considering the case ``without'' prior on the mass\footnote{We use a Gaussian mass prior of $35\,\mathrm{M_{Jup}}$ and $\sigma$=$15\,\mathrm{M_{Jup}}$ to avoid convergences issues since the best-mass converge to really low-masses (below $1\,\mathrm{M_{Jup}}$).}. The results of these fits are shown in Table\,\ref{table:Phys_constr_ATMO_prior}. Figure\,\ref{fig:ATMO_best_model_priors} shows the best fits for each of the priors. It is noteworthy that among all the priors, only the one with the greatest mass (i.e., $30\,\mathrm{M_{Jup}}$) presents a noticeable discrepancy with the data, but still within the uncertainties. It can be seen in the figure that there is a tendency to change in $\mathrm{T_{eff}}$ and log(g) as we increase the mass prior, even though these differences are not reflected in the fitted data (the most impacted value being the log(g)). Also, other trends are observed, for example when increasing the $\mathrm{T_{eff}}$ decreases the radius (and indirectly using the radius and log(g), the mass), and increase the luminosity with the $\mathrm{T_{eff}}$. We discuss these results in Section\,\ref{subsec:mass}.

\begin{table*}[]
\centering
\caption{Physical parameters from the best fit using \texttt{ATMO} models and the entire dataset, assuming different mass priors. }
  \begin{tabular}{ l c c c c c c c c c c }
    \hline \hline
  Mass & Teff & Mass & Radius & log\,(g) & log(L/$L_{\odot}$) \\
  Prior     &  [K] & [M$_\mathrm{Jup}$] & [R$_\mathrm{Jup}$] &  &  \\ 
    \hline
No prior     & $1255^{+14}_{-13}$ & $14^{+2}_{-2}$ & $0.92\pm0.02$ & $4.60^{+0.05}_{-0.05}$ & $-4.70\pm0.01$ \\
\hline
$10\,\mathrm{M_{Jup}}$  & $1290^{+11}_{-11}$ & $9^{+1}_{-1}$ & $0.89\pm0.02$ & $4.44^{+0.04}_{-0.04}$ & $-4.68\pm0.01$  \\
$18\,\mathrm{M_{Jup}}$  & $1246^{+12}_{-10}$  & $16^{+1}_{-1}$ & $0.93\pm0.02$ & $4.65^{+0.03}_{-0.03}$ & $-4.70\pm0.01$ \\
$25\,\mathrm{M_{Jup}}$  & $1234^{+10}_{-10}$ & $23^{+1}_{-1}$ & $0.93\pm0.02$ & $4.82^{+0.02}_{-0.02}$ &$-4.72\pm0.01$\\
$30\,\mathrm{M_{Jup}}$  & $1228^{+9}_{-9}$ & $28^{+1}_{-1}$ & $0.93\pm0.02$ & $4.90^{+0.02}_{-0.02}$ &$-4.73\pm0.01$\\
\hline
  \end{tabular}
\label{table:Phys_constr_ATMO_prior}
\end{table*}

\begin{figure*}
\centering
\includegraphics[width=4.8cm]{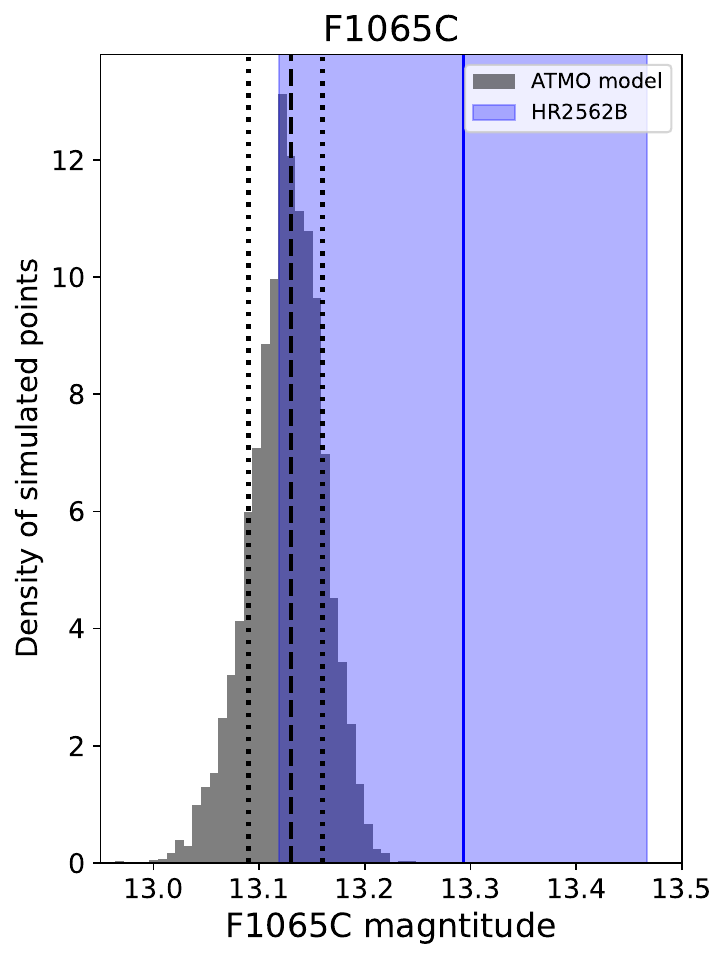}
\includegraphics[width=4.8cm]{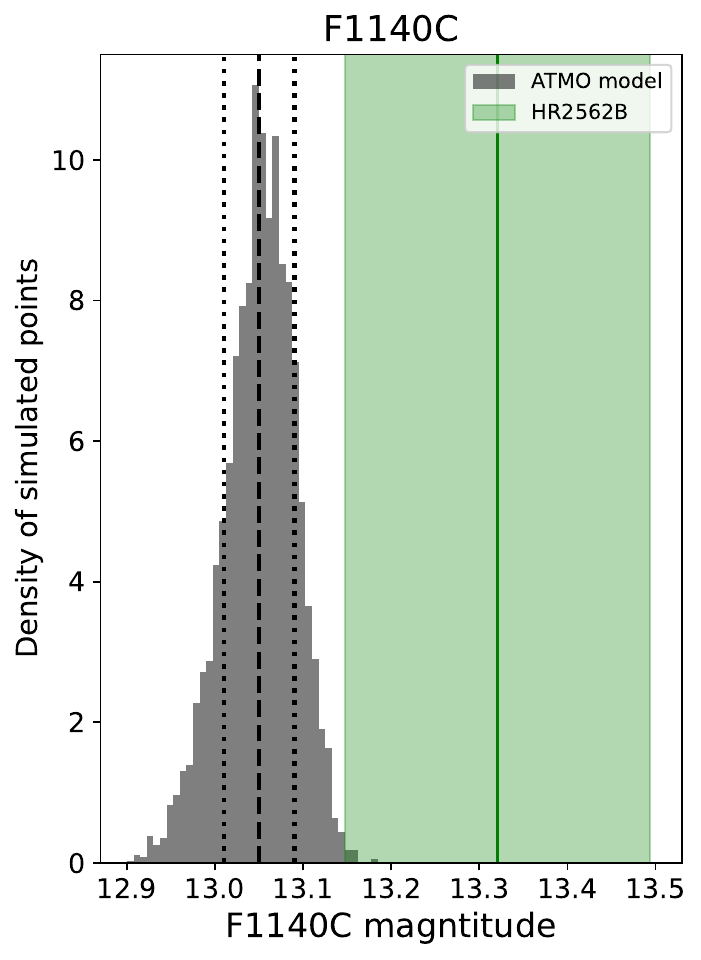}
\includegraphics[width=4.8cm]{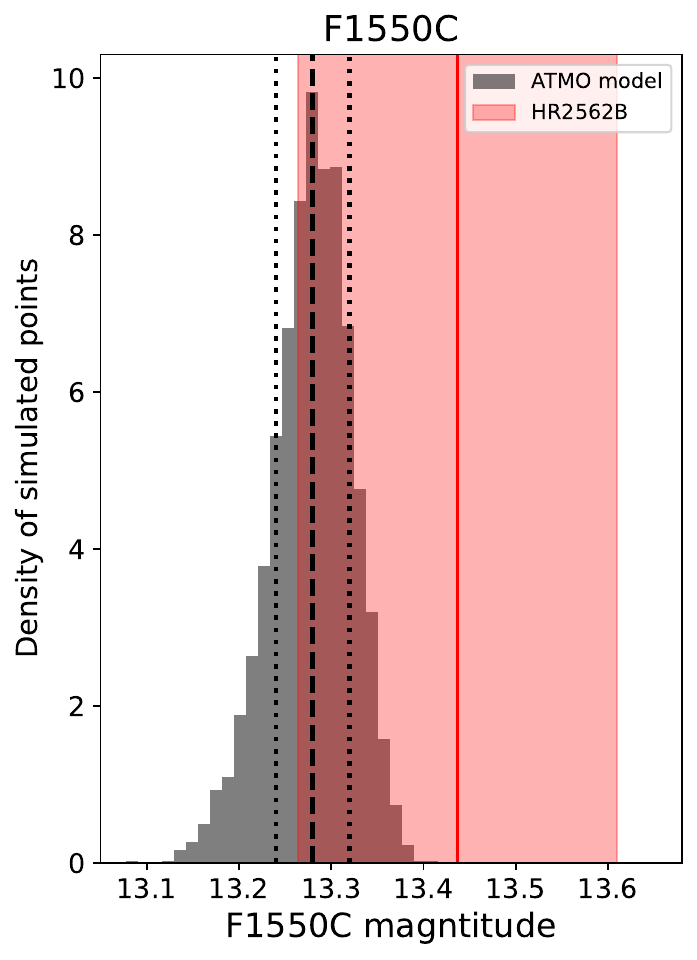}
\caption{Histogram of synthetic magnitudes obtained from the best-fit model from \texttt{ATMO} using MCMC ($10\,000$ points), which includes all the observations of HR\,2562\,B. From left to right: magnitudes at \texttt{F1065C}, \texttt{F1140C}, and \texttt{F1550C} filters. The colored vertical lines and filled, transparent area correspond to the magnitude and uncertainties of HR\,2562\,B at each filter. The dotted and dashed vertical black lines in each subplot correspond to $15.9\%$, $50\%$, and $81.1\%$ percentiles, respectively. }
\label{fig:hist_filt_ATMO}%
\end{figure*}

It should be noted that there is a mismatch with the \texttt{ATMO} models at $11\mu m$ (\texttt{F1140C}), the model overpredicting the companion flux compared to the measured flux. To investigate and quantify this, we proceed to generate $10\,000$ synthetic photometry at \texttt{F1065C}, \texttt{F1140C}, and \texttt{F1550C} MIRI filters using the best-fit models and the posterior distribution of the parameters we found using \texttt{ATMO} (which include all the observational datasets). Figure\,\ref{fig:hist_filt_ATMO} shows the histograms of these calculated magnitudes for each filter (gray bars), the measured HR\,2562\,B magnitudes (colored vertical line), and each uncertainty (colored and transparent boxes). The median of each distribution corresponds\footnote{The uncertainties correspond to the $15.9\%$ and $84.1\%$ percentiles of the distribution.} to $13.13^{+0.03}_{-0.04}$, $13.05\pm0.04$, and $13.28\pm0.04$ magnitudes for \texttt{F1065C}, \texttt{F1140C}, and \texttt{F1550C}, respectively. The HR\,2562\,B magnitudes at \texttt{F1065C} and \texttt{F1550C} are in good agreement with the best-fit model at $1\sigma$ level (left and right panels in Fig.\,\ref{fig:hist_filt_ATMO}), while the magnitude at \texttt{F1140C} (middle panel) shows a significant ($>$1-$\sigma$) deviation from the distribution at the same level. For example, $\Delta \mathrm{mag_{F1140C}}$=$0.26\pm0.18$mag, while $\Delta \mathrm{mag_{F1550C}}$=$0.16\pm0.17$mag meaning the difference is within $1\sigma$ for \texttt{F1550C} but not for \texttt{F1140C}. This difference could be attributable to the presence of silicate clouds (see Section\,\ref{sec:bands}). \texttt{ATMO} cannot reproduce these features since it does not include clouds with such compositions in the MIR. Notably, from \cite{Wang-2023} we could attribute the lack of flux at $11\mu m$ to the presence of silicates.

\begin{figure*}
\sidecaption
\includegraphics[width=12cm]{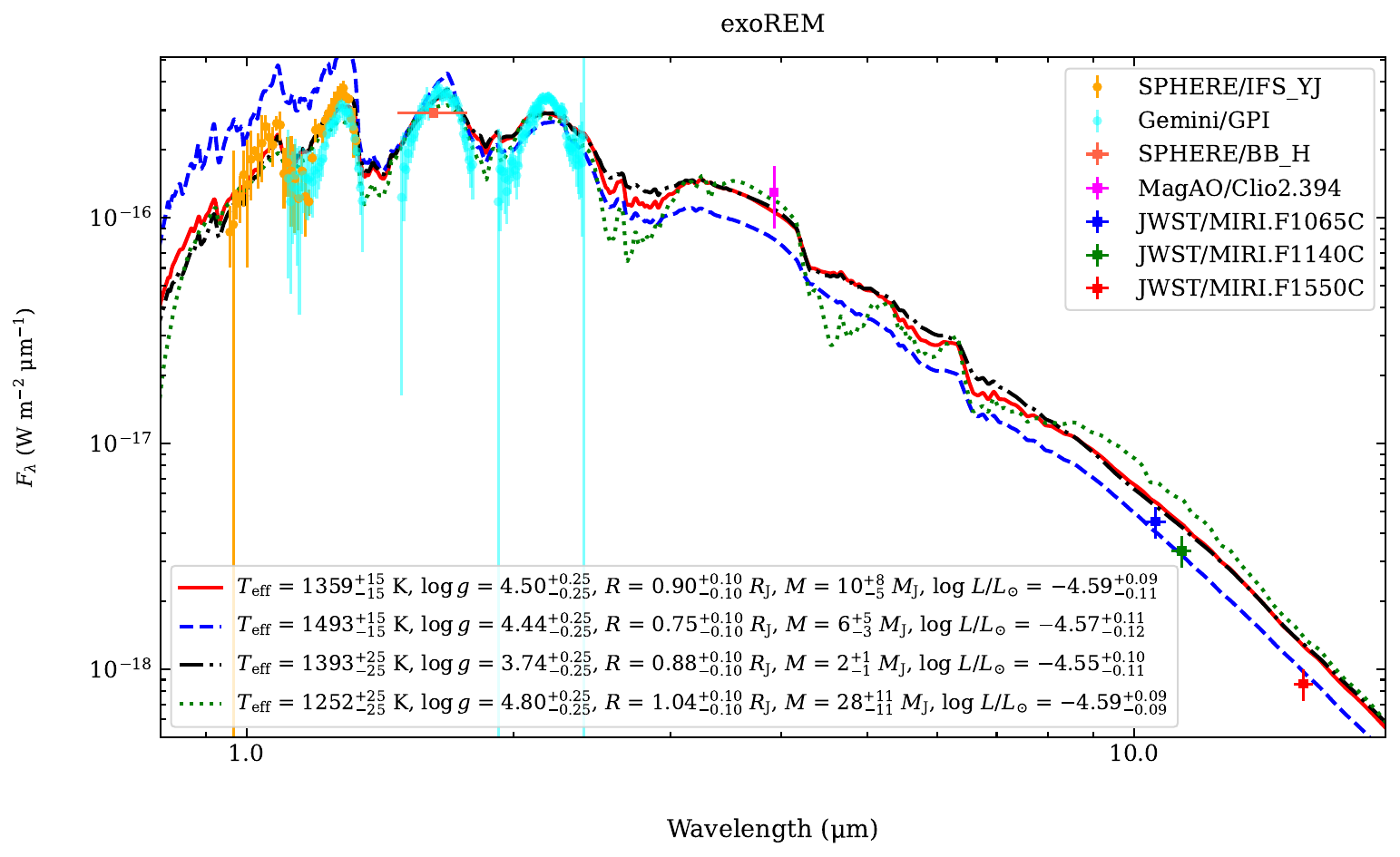}
\caption{Best-fit models using the \texttt{Exo-REM} grid via $\chi^2$ mapping. The different colored lines correspond to the different combinations shown in Table\,\ref{table:Phys_constr_exoREM}. The red solid line uses the NIR data; the dashed blue line the MIRI data; the dashed-dotted black line the MIRI, SPHERE, and MagAO data; the dotted green line the full dataset. The colored symbols correspond to the different observational data.  }
\label{fig:exoREM_all_models}%
\end{figure*}

\textit{\texttt{Exo-REM} fit procedure}:\ In order to investigate the role of clouds in the atmosphere, we fit the observed data with the \texttt{Exo-REM} models. We used a new \texttt{Exo-REM} grid recently incorporated into \texttt{species} which covers the MIR wavelengths. Since the changes in $\mathrm{T_{eff}}$, log(g), and other parameters are not linear in this grid, we determined the best-fit parameters from the model that minimizes the $\chi^2$ with respect to the data. Since the grid presents degeneracies, we used the minimum value in $\chi^2$ (but not caring about this value) and we took all the surrounding values within $3\sigma$ deviation\footnote{Here we normalized the $\chi^2$ map from 100\% at the minimum and 0\% at the maximum, and we took all the values within 68\%. This value is strongly affected by the sampling more than the underestimated uncertainties related to each $\chi^2$ estimation.} as an estimator for the representative parameter values and uncertainties, which are strongly affected by the grid sampling meaning greater uncertainties. We used this approach to obtain the uncertainties that would otherwise be underestimated by performing an MCMC analysis. Furthermore, we have not explored the entire possible space of dataset combinations to analyze the effect of including or rejecting some of them in the fit (as was the case for \texttt{ATMO}). Instead, we focused on those combinations of datasets that could have a major impact on the fit and cover different ranges of wavelengths. Table\,\ref{table:Phys_constr_exoREM} shows the different combinations we investigated, together with the respective best-fit values and their associated uncertainties. In the case of the uncertainties, in some cases we have been conservative in assuming larger uncertainties given the low-sampling space of the grid ($\Delta T=50$K and $\Delta$log(g)=0.5) at the best-fit values. In the same way as for \texttt{ATMO}, we did not focus on some free parameters such as $\mathrm{[Fe/H]}$ and $\mathrm{C/O}$ ratio, but those were set as free in the modeling. Figure\,\ref{fig:exoREM_all_models} shows all combinations and sets of best parameters that reproduce the SED. Some of these combinations give more unrealistic masses ($2$ to $6\,\mathrm{M_{Jup}}$), while when using the full SED we obtained a mass in good agreement with previous studies (e.g., \citealt{Sutlieff+2021}). Note that, as for \texttt{ATMO}, the mass is calculated using the best-fit values of log(g) and radius. Unlike the \texttt{ATMO} procedure, we did not investigate the case of different mass priors for \texttt{Exo-REM}.

\begin{table*}[]
\centering
\caption{Best-fit parameters using the \texttt{Exo-REM} models for each combination of datasets.}
  \begin{tabular}{ l c c c c c c c c c c }
    \hline \hline
  Data & Teff & Mass & Radius & log\,(g) & log(L/$L_{\odot}$) \\
       &  [K] & [M$_\mathrm{Jup}$] & [R$_\mathrm{Jup}$] &  &  \\ 
    \hline
MIRI               & $1493^{+15}_{-15}$ & $6^{+5}_{-3}$ & $0.75^{+0.10}_{-0.10}$ & $4.44^{+0.25}_{-0.25}$ & $-4.57^{+0.11}_{-0.12}$  \\ 
MIRI+SPHERE+MagAO  & $1393^{+25}_{-25}$ & $2^{+1}_{-1}$ & $0.88^{+0.10}_{-0.10}$ & $3.74^{+0.25}_{-0.25}$ & $-4.55^{+0.10}_{-0.11}$  \\ 
SPHERE+GPI+MagAO   & $1359^{+15}_{-15}$ & $10^{+8}_{-5}$ & $0.90^{+0.10}_{-0.10}$ & $4.50^{+0.25}_{-0.25}$ & $-4.59^{+0.09}_{-0.11}$  \\ 
\hline
Full SED           & $1252^{+25}_{-25}$ & $28^{+11}_{-11}$ & $1.04^{+0.10}_{-0.10}$ & $4.80^{+0.25}_{-0.25}$ & $-4.59^{+0.09}_{-0.09}$  \\ 
\hline
  \end{tabular}
\tablefoot{No priors in mass have been assumed. }
\label{table:Phys_constr_exoREM}
\end{table*}

\begin{table*}[]
\centering
\caption{Summary of the atmospheric and physical properties of HR\,2562\,B.}
  \begin{tabular}{ l c c c c c c c c c c }
    \hline \hline
  Data & T$_{\mathrm{eff}}$ & Mass & Radius & log\,(g) & log(L/$L_{\odot}$) \\
       &  [K] & [M$_{\mathrm{Jup}}$] & [R$_{\mathrm{Jup}}$] &  &  \\ 
    \hline
\cite{Konopacky+2016} &$1200\pm100$  &  $30\pm15$  &  $1.11\pm0.11$  &  $4.70\pm0.32$  &  $-4.62\pm0.12$  \\
\cite{Mesa+2018}      & $1100\pm200$  & $32\pm14$  &   --  &  $4.75\pm0.41$ & --  \\
\cite{Sutlieff+2021}  & $[1200-1700]$  &  $29\pm15$  &  $[0.56-0.89]$  &  $[4-5]$  &  $[-4.87 - -4.60]$\\
\cite{Zhang+2023}  & --  &  $\leq18.5$  &  --  &  --  &  -- \\
This work - ATMO $10\,\mathrm{M_{Jup}}$ prior   & $1284^{+12}_{-11}$ & $10^{+1}_{-1}$ & $0.90^{+0.02}_{-0.02}$ &  $4.47^{+0.03}_{-0.08}$ & $-4.68^{+0.01}_{-0.01}$ \\
This work - ATMO no prior   & $1255^{+14}_{-13}$ & $14^{+2}_{-2}$ & $0.92^{+0.02}_{-0.02}$ & $4.59^{+0.05}_{-0.05}$ & $-4.69^{+0.01}_{-0.01}$  \\
This work - Exo-REM no prior   & $1252^{+25}_{-25}$ & $28^{+11}_{-11}$ & $1.04^{+0.10}_{-0.10}$ & $4.80^{+0.25}_{-0.25}$ & $-4.59^{+0.09}_{-0.09}$ \\
\hline
  \end{tabular}
\tablefoot{In both cases, \texttt{ATMO} and \texttt{Exo-REM}, the full SED were used in the fitting. Note that in the case of \cite{Zhang+2023}, we also have a most probable mass of $10\,\mathrm{M_{Jup}}$.}
\label{table:Phys_constr_summary}
\end{table*}

Comparing our fits done with \texttt{Exo-REM} and \texttt{ATMO}, we can highlight the related uncertainties, which are notoriously at a different order of magnitudes because of the different fit methods used. In Appendix\,\ref{apx:SED_err} we present some examples of the fits, and their respective associated error curves for the \texttt{Exo-REM} (Figure\,\ref{fig:exoREM_erros}) and \texttt{ATMO} (Figure\,\ref{fig:ATMO_some_err}) cases. Considering the uncertainties presented in \texttt{Exo-REM}, the model fits well with the data at $5$-$\sigma$, and is in good agreement with the estimates using \texttt{ATMO} at $3$-$\sigma$. However  (as we can see in Table\,\ref{table:Phys_constr_summary}), the \texttt{Exo-REM} fits fails in narrowing down some of the parameters, especially for the mass of HR\,2562\,B considering the values from the astrometric study of \cite{Zhang+2023}. Figure\,\ref{fig:ATMO_some_err} shows a good agreement when fitting the MIRI data. This could indicate, as mentioned in Section\,\ref{sec:bands}, that HR\,2562\,B may not have prominent silicate absorptions. It is worth noting that these models are 1D, which means that we are obtaining average models of the atmosphere that do not take into account inhomogeneities across the surface. These fits could be refined by implementing more complex models, such as considering different sedimentations depending on the atmospheric layer (e.g., \citealt{Mukherjee+2023}), including the effect of refractory cloud species and including silicates clouds (\citealt{Morley+2024}; Sonora Diamondback), and even considering 3D models (e.g., \citealt{Showman+Kaspi-2013}; \citealt{Morley+2014}; \citealt{Zhang+Showman-2014}; \citealt{Tan+Showman-2021}). However, such improvements in SED modeling are beyond the purpose of this work.

\section{Discussion}\label{sec:dis}

In this section, we discuss our results on the mass estimates of HR\,2562\,B, and on the comparison with a companion with similar physical characteristics, namely, VHS\,1256\,b.

\subsection{The mass of HR\,2562\,B}\label{subsec:mass}

Previous studies based on SED analysis (see Table\,\ref{table:Phys_constr_summary}) have inferred a mass for HR\,2562\,B that varies over a wide range of values (from $15\mathrm{M_{Jup}}$ to $45\mathrm{M_{Jup}}$ at 1-$\sigma$). This range of possible masses is mostly inconsistent with the upper limit and most probable mass obtained from astrometric studies ($\leq 18.5\mathrm{M_{Jup}}$ and $10\mathrm{M_{Jup}}$, \citealt{Zhang+2023}, for which this object could have planetary mass). From our results using \texttt{ATMO} with different mass priors, we can notice that currently, the observations are not enough sensitive to log(g) (and mass, see Table\,\ref{table:Phys_constr_ATMO_prior} and Fig.\,\ref{fig:ATMO_best_model_priors}), as we demonstrated using different mass priors. Assuming that \texttt{ATMO} is the best atmospheric model that reproduces the spectrum of HR\,2562\,B, it is possible to better delimit the value of the mass by obtaining new observations. For example, at $3.1-3.4\mu m$ the $\mathrm{CH_4}$ absorption feature is quite associated with log(g) (and then with the mass), making this range of wavelengths more suitable to better constrain the mass of HR\,2562\,B (among other features that can help to delimit the mass as well, see Fig.\,\ref{fig:ATMO_best_model_priors}). Surface gravity has a direct impact on the formation of clouds in the upper layers of the atmosphere and, in turn, they affect the absorption features and depth of some molecules. For this reason, narrowing down the surface gravity helps us to understand atmospheric physics and compositions in HR\,2562\,B high altitude atmosphere and to better understand the mismatch at \texttt{F1140C} filter. Despite this, from our best-fit using \texttt{ATMO} joined with astrometric mass limits, we can narrow down the mass of HR\,2562\,B to between $8\mathrm{M_{Jup}}$ (3-$\sigma$ lower limit from the non-prior mass fit, see Table\,\ref{table:Phys_constr_summary}) and $18.5\mathrm{M_{Jup}}$ (from \citealt{Zhang+2023}). We note that, with these mass limits, HR\,2562\,B could be considered a massive planet/planetary-mass companion.

In the case of \texttt{Exo-REM}, we obtained a mass in good agreement with the previous estimates when using a combination of photometric datasets in both NIR and MIR. However, our estimate is not compatible with the astrometric mass. The \texttt{Exo-REM} models are based, fundamentally, on the presence of clouds at different altitudes and compositions. From Figure\,\ref{fig:exoREM_all_models} we see that models fail to replicate some shapes of the SED at NIR, and in particular in the MIR regime. When we fit the SED using only the NIR or MIR, the model reproduces most of the shapes presented in those wavelengths. However, this is not the case when we fit the SED combining both NIR and MIR observations. Following the analysis by \cite{Miles+2023}, this can be explained if we have different cloud layers that are affected by different physicochemical processes (e.g., species in equilibrium and disequilibrium chemistry), which requires more than one model to simultaneously fit the contribution to the NIR that comes from more internal atmospheric layers, as well as the contribution of the MIR that comes from more external layers in the atmosphere. For example, the presence of clouds in most internal layers can contribute to the NIR features, and more advanced sedimentation of the upper cloud particles could explain better the mismatches at MIR wavelengths. Another explanation could be the fraction of atmospheric dust and particle sizes which could be different depending on the atmospheric altitudes, generating different flux contributions in NIR and MIR regimes. However, as it seems in Figure\,\ref{fig:exoREM_all_models}, the presence of clouds generates distinguishable shapes and absorptions in the spectrum (wider and more extended shapes with spread wings in H and K bands, for example), so we can deduce that the atmosphere does not have clouds that dominate the SED. The atmospheric models which include clouds, inflate the value of the log(g) and mass when the object has a more likely cloud-free atmosphere. This may explain the large values of masses that we, and previous studies, found using these models. We can conclude that the atmospheric model \texttt{Exo-REM} that uses clouds to model the reddening in the atmosphere of exoplanets, is not the most representative of the atmosphere of HR\,2562\,B. The atmosphere is more likely free of clouds or, at least, the clouds have a minor impact on the SED shape. An alternative option is to study the effect of adding complexity to the atmospheric models that include clouds, to reproduce the shapes of the SED satisfactorily (e.g., 3D models), which is beyond the objectives of this study.

We can conclude that the mass of HR\,2562\,B moves in a range between $8$ and $18.5$\,$\mathrm{M_{Jup}}$ obtained from \texttt{ATMO} atmospheric model that does not use clouds as the main atmospheric modeler. Atmospheric model that uses clouds as the main component can give us erroneous mass estimates when the object is not dominated by clouds, which is the case of HR\,2562\,B when using \texttt{Exo-REM}. The L/T transitional objects are a group of sources that move from cloud-dominant atmospheres to cloud-free ones, so the use of different models with different approaches can help us to obtain a broader view of the physical properties and, in general, the characteristics of the atmosphere.

\subsection{Comparison with VHS\,1256\,b}

\begin{figure*}
\centering
\includegraphics[width=17cm]{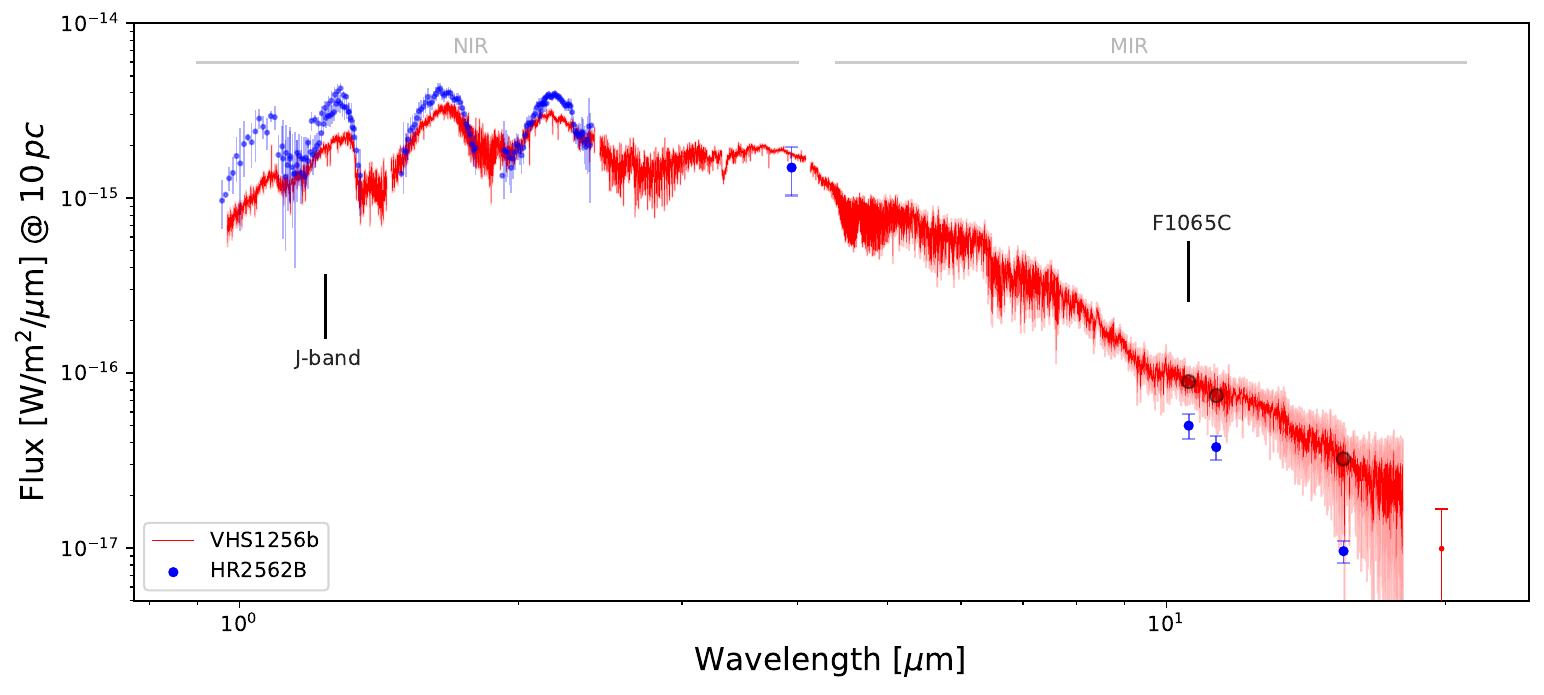}
\caption{Spectral energy distribution of both HR\,2562\,B (blue dots) and VHS\,1256\,b (red line and dot). The vertical lines highlight the \texttt{J} and \texttt{F1065C} bands for comparison purposes (see main text). The fluxes were normalized at $10$\,pc. The filled red area and errorbars correspond to 1-$\sigma$ uncertainties. The error bars of the first and last points in the \texttt{Y-}, \texttt{J-}, \texttt{H-}, and \texttt{K-}bands of HR\,2562\,B IFS data were set to zero for aesthetic purposes. The dark red dots are the synthetic MIRI photometry of VHS\,1256\,b.
  }
\label{fig:SED_comparison}%
\end{figure*}

The studies of VHS\,1256\,b (e.g., \citealt{Miles+2023}) show a younger planetary-mass companion than HD\,2562\,B ($140\pm20$\,Myr vs. [$200$-$700$]\,Myr), with a surface gravity of 4.5 dex, effective temperature around $1100\,K$, a radius of $1.25\,\mathrm{R_{Jup}}$, log(L/L$_\odot$) of $-4.55$, a mass less than $20\,\mathrm{M_{Jup}}$, with a spectral type very similar to L4.5 (from a comparison with the silicate absorption feature, \citealt{Miles+2023}). From Table\,\ref{table:Phys_constr_summary}, HR\,2562\,B presents a very similar range of physical parameters, even at the upper bound of mass. Figure\,\ref{fig:SED_comparison} shows the SED of both HR\,2562\,B (blue dots) and VHS\,1256\,b (red line and dot) from \cite{Miles+2023}, in which we highlight in the NIR and MIR regime and the location of \texttt{J-band} and \texttt{F1065C} for easy visualization and comparison (see text below). HD\,2562\,B being older than VHS\,1256\,b, we have a unique opportunity to study young objects in different stages of the T/L transition. Indeed, from Fig.\,\ref{fig:CMD_JK} we note these similarities at NIR wavelengths, in which the \texttt{J-K} color are similar, but HR\,2562\,B being brighter than VHS\,1256\,b (see Table\,\ref{table:comp} and Fig.\,\ref{fig:SED_comparison}). However, other differences between both companions are observed in the \texttt{F1065C} and \texttt{F1140C} color-magnitude diagram (see Fig.\,\ref{fig:CMD_index} and Table\,\ref{table:comp}), where VHS\,1256\,b and HR\,2562\,B have not only different absolute magnitudes in \texttt{F1065C} filter but also different colors. The first difference we note is that VHS\,1256\,b is brighter than HR\,2562\,B at $10\mu m$, $11\mu m$, and $15\mu m$ (see the dark-red and blue dots in Fig.\,\ref{fig:SED_comparison}) with a slightly redder color. The reduced luminosity of VHS\,1256\,b in the NIR can be due to the presence of more clouds at medium/higher atmospheric altitudes which decrease (e.g., absorption, scattering) the emitted flux. The flux absorbed by the clouds is thermally reemitted at longer wavelengths (MIR), which may explain the enhanced luminosity of VHS\,1256\,b compared to HR\,2562\,b. VHS\,1256\,b is known as a variable brown dwarf (e.g., \citealt{Zhou+2022}), whose variability is attributable to a heterogeneous distribution of clouds. From \cite{Miles+2023} we know that VHS\,1256\,b has silicates at MIR wavelengths, which is reflected also in our color-magnitude diagram (Fig.\,\ref{fig:CMD_index}). The fact that HR\,2562\,B emits less flux at $10\mu m$ and more at \texttt{J}-band (compared to VHS\,1256\,b, see Fig.\,\ref{fig:SED_comparison}) could be interpreted as a lack of clouds. The presence of clouds at lower/medium altitudes in the atmosphere dampens the NIR emission, scattering and re-emitting the light at longer wavelengths (see, for example, Fig.\,10 in \citealt{Currie+2023}). Indeed, the much bluer \texttt{F1065C-F1140C} color of HR\,2562\,B could also be attributable to a lesser effect of silicate absorption (\citealt{Wang-2023}), which means more advanced sedimentation of the upper cloud particles.

\begin{table}[]
\centering
\caption{Colors and absolute magnitudes in the \texttt{2MASS} and \texttt{MIRI} filters of HR\,2562\,B and VHS\,1256\,b. }
  \begin{tabular}{ l c c c c c c c c c c }
    \hline \hline
  Target & $\mathrm{M_{Jup}}$ & $\mathrm{J-K_{S}}$ & $\mathrm{M_{F1065C}}$ & $\mathrm{F1065C}$-$\mathrm{F1140C}$  \\
         &  [mag] & [mag] & [mag] & [mag]   \\ 
    \hline
HR\,2562\,B  &  $15.3$ &  $2.5$ & $10.64$ & $-0.03$  \\
VHS\,1256\,b & $16.37$ &  $2.6$ & $10.07$ & $0.15$ \\
\hline
  \end{tabular}
\label{table:comp}
\end{table}

\cite{Miles+2023} show that to model the spectrum of VHS\,1256\,b two thin cloud decks with a $\mathrm{f_{sed}}$ of $0.6$ and $1.0$ (with contributions of $90\%$ and $10\%$, respectively) are needed. The current atmospheric models do not match very well the observed spectrum of VHS\,1256\,b, even when the complexity of the model is increased. The presence of heterogeneous clouds with different states (particle sedimentation,  species in (dis)equilibrium, amongst others) results in a great challenge at the moment of retrieving some of the physical parameters such as the mass. On the other hand, from our modeling and analysis, HR\,2562\,B has an atmosphere most likely free of clouds, which could be beneficial to narrow down its mass. The main challenge is to identify absorption lines that are sensitive to surface gravity and mass. However, when we use atmospheric models that do not consider clouds as the main factor in the model, we can be biased when retrieving the radius of the companion. The presence of clouds (even a small amount), generates a slight re-distribution of the emitted flux, changing the SED shape compared to an atmosphere free of clouds. For this reason, the cloud-free model will compact the planet to fit this excess of flux. This can explain that VHS\,1256\,b and HR\,2562\,B, even having similar log(g) and mass, the radius of HR\,2562\,B is significantly smaller than the one of VHS\,1256\,b ($0.92\mathrm{R_{Jup}}$ vs. $1.25\mathrm{R_{Jup}}$). Indeed, from Sections\,\ref{sec:bands} and \ref{subsec:mass} we know that exists a low amount of clouds in the atmosphere of HR\,2562\,B (most probably of silicates), that produces a slight absorption at $11\mu m$.

The chemical composition is another key factor that needs to be investigated. The metals in the atmosphere are related directly to the concentration of specific chemical species, which influence the chemical equilibrium and absorption features. The different species can alter the pressure and temperature at different atmospheric altitudes, which generates different SEDs even for objects with the same physical properties. For example, the presence of iron could be necessary to explain the L/T transitional objects observations. These grains can be at different pressure layers (\citealt{Vos+2023}) or mixed with silicates (\citealt{Luna+Morley-2021}; \citealt{Suarez+Metchev-2023}), affecting the sedimentation of dust in low-gravity atmospheres. The [C/O] ratio, along with the presence and amount of silicates, ammonia, and water, among other chemical species, are fundamental to understanding better the main differences between VHS\,1256\,b and HR\,2562\,B. Currently, we do not have the observations to provide this comparison. The chemical composition is fundamental to improving atmospheric models (currently in tension with the observations, for example, \citealt{Miles+2023}), which can be tested directly in these two planetary-mass companions which are at different stages in the L/T transition but have similar physical properties.

\section{Summary and conclusions}\label{sec:sac}

We analyzed the spectral energy distribution of HR\,2562\,B with observations acquired with JWST/MIRI with \texttt{F1065C}, \texttt{F1140C}, and \texttt{F1550C} bandpasses. HR\,2562\,B is characterized by being a young companion ($450$\,Myrs) in the L/T transition, whose mass had not been well delimited by previous studies using atmospheric models ($30\pm15\,\mathrm{M_{Jup}}$), but narrowed down  to $18.5\,\mathrm{M_{Jup}}$ upper limit by astrometric measurements. Our research addresses the characterization of HR\,2562\,B in three aspects: a study of the color-magnitude diagram in the MIRI filters, the analysis of the SED with two atmospheric models based on different physics, and a comparison with another target in the L/T transition with similar physical characteristics (VHS\,1256\,b, mass$\leq20\mathrm{M_{Jup}}$, $\mathrm{T_{eff}}$=$1100$\,K).

We used a field of sub-stellar objects for which a spectral index is provided for some chemical species based on spectroscopic analysis (\citealt{Suarez+Metchev-2022}). We used this information to place HR\,2562\,B in the context of chemical species based on color-magnitude diagrams. Regarding the SED analysis, we used two atmospheric models to fit and study the atmosphere of HR\,2562\,B: \texttt{ATMO}, a free-cloud model, and \texttt{Exo-REM} a model that includes clouds. Comparing the SED fitting results with the values found in previous studies, our analysis using color-magnitude diagram, and comparison with VHS\,1256\,b, we conclude the following: \textbf{1)} the degeneracy in solution space for $\mathrm{T_{eff}}$ has been solved, placing HD\,2562\,B in the $\mathrm{T_{eff}}$ range of between $1200$\,K and $1300$\,K, our best-fit value being $1255\pm14$\,K (\texttt{ATMO} model without mass prior). On the other hand, we were able to narrow down the log(g), whose values range between $4.4$ and $4.8$\,dex.; \textbf{2)} although the SED adjusts reasonably with a wide range of masses, the bounded mass from the dynamic study ($\leq18.5\mathrm{M_{Jup}}$, \citealt{Zhang+2023}) gives us a range that is much closer to that predicted in the case of \texttt{ATMO} ($14\,\mathrm{M_{Jup}}$, without mass prior). Considering the range of possible masses, this companion should be considered as a planetary-mass companion instead of BD, whose name should be changed to \textit{HR\,2562\,b}. This if we consider the planetary-mass threshold of $\sim 16\mathrm{M_{Jup}}$ defined by the deuterium-burning mass limit (\citealt{Spiegel+2011}), or more recently the suggested definition given by \citealt{Currie+2023} (mass$<25\mathrm{M_{Jup}}$, $q<0.025$, a$<300$au), with $10\mathrm{M_{Jup}}$ as the most probable mass (and $18.5\mathrm{M_{Jup}}$, upper-limit), $q<0.013$, and a$\sim23$au for HR\,2562\,B; \textbf{3)} \texttt{ATMO} fits the data well, but still needs more chemical species and dust particles in atmospheric upper layers (for example, the silicates at $11$ microns) to replicate small mismatches with the observations. This mismatch at 11 microns could come from a low amount of silicates, for example, meaning a low amount of clouds/dust in the atmosphere, or the observed rotational axis inclination in our line of sight that affects the silicates depth absorption; \textbf{4)} \texttt{Exo-REM} does not fit satisfactorily the observations, and fails in obtaining a representative mass for HR\,2562\,B around $30\mathrm{M_{Jup}}$ (well above the astrometric upper limit mass, $18.5\mathrm{M_{Jup}}$); \textbf{5)} with our MIRI observations, we can not conclude that a specific chemical species is present in the atmosphere of HR\,2562\,B, however, and comparing the color-magnitude diagram with chemical abundance, the SED analysis, and comparison with VHS\,1256\,b, the silicates could be the most favorable species (but with a weak absorption); \textbf{6)} HR\,2562\,B, and VHS\,1256\,b have similar physical properties, however, HR\,2562\,B is in a late stage in the L/T transition, reflected in a more advanced sedimentation of dust and particles and a lack of clouds, as reflected in our SED fit with atmospheric models.

To precisely obtain information about the chemical composition it is necessary to obtain spectroscopic observations covering the MIR regime ($4$-$20\mu m$). The presence of silicates and other chemical species can be directly measured from the spectrum and can help us to better understand the actual state of the atmosphere of HR\,2562\,B (e.g., chemical dis-equilibrium, P-T profiles). To narrow down the mass of HR\,2562\,B, it is necessary to obtain new photometric observations at wavelengths sensitive to the surface gravity. From our study, we suggest using the $\sim3.3\mu m$ $\mathrm{CH_{4}}$ absorption feature, which is sensitive to the surface gravity (see Fig.\,\ref{fig:ATMO_best_model_priors}). Since the $\mathrm{T_{eff}}$ is well-constrained, atmospheric models can better narrow down the log(g). The new observations could be beneficial in terms of improving our knowledge about the L/T transitional objects, serving as a complement to the younger and clouded counterpart VHS\,1256\,b.

\begin{acknowledgements}
We thank B. Miles et al. for providing the VHS\,1256\,b spectrum and B. Miles for the discussions about chemistry on VHS\,1256\,b relative to HR\,2562\,B. We thank A. Carter and Y. Zhou for providing the photometric data of the population of low-mass stars, brown dwarfs, and direct imaged planets/planetary-mass companions. We thank the authors Konopacky et al., Mesa et al., and Sutlieff et al. for providing and/or publishing the observational data used in this study. This project is funded/Co-funded by the European Union (ERC, ESCAPE, project No 101044152). Views and opinions expressed are however those of the author(s) only and do not necessarily reflect those of the European Union or the European Research Council Executive Agency. Neither the European Union nor the granting authority can be held responsible for them. Part of this work was carried out at the Jet Propulsion Laboratory, California Insitute of Technology, under contract with NASA (80NM00018D0004). C.D. acknowledges financial support from the INAF initiative ``IAF Astronomy Fellowships in Italy'', grant name \textit{GExoLife}. This work is based on observations made with the NASA/ESA/CSA James Webb Space Telescope. The data were obtained from the Mikulski Archive for Space Telescopes at the Space Telescope Science Institute, which is operated by the Association of Universities for Research in Astronomy, Inc., under NASA contract NAS 5-03127 for JWST. These observations are associated with program \#1241. This work has made use of data from the European Space Agency (ESA) mission {\it Gaia} (\url{https://www.cosmos.esa.int/gaia}), processed by the {\it Gaia} Data Processing and Analysis Consortium (DPAC, \url{https://www.cosmos.esa.int/web/gaia/dpac/consortium}). Funding for the DPAC has been provided by national institutions, in particular, the institutions participating in the {\it Gaia} Multilateral Agreement. This research has made use of the Spanish Virtual Observatory (\url{https://svo.cab.inta-csic.es}) project funded by MCIN/AEI/10.13039/501100011033/ through grant PID2020-112949GB-I00. This publication makes use of VOSA, developed under the Spanish Virtual Observatory (https://svo.cab.inta-csic.es) project funded by MCIN/AEI/10.13039/501100011033/ through grant PID2020-112949GB-I00. VOSA has been partially updated by using funding from the European Union's Horizon 2020 Research and Innovation Programme, under Grant Agreement nº 776403 (EXOPLANETS-A). This publication makes use of data products from the Two Micron All Sky Survey, which is a joint project of the University of Massachusetts and the Infrared Processing and Analysis Center/California Institute of Technology, funded by the National Aeronautics and Space Administration and the National Science Foundation. This publication makes use of data products from the Wide-field Infrared Survey Explorer, which is a joint project of the University of California, Los Angeles, and the Jet Propulsion Laboratory/California Institute of Technology, funded by the National Aeronautics and Space Administration. This work has made use of data from the European Space Agency (ESA) mission Gaia (https://www.cosmos.esa.int/gaia), processed by the Gaia Data Processing and Analysis Consortium (DPAC, https://www.cosmos.esa.int/web/gaia/dpac/consortium). Funding for the DPAC has been provided by national institutions, in particular the institutions participating in the Gaia Multilateral Agreement. This work has benefitted from The UltracoolSheet, maintained by Will Best, Trent Dupuy, Michael Liu, Rob Siverd, and Zhoujian Zhang, and developed from compilations by Dupuy \& Liu (2012, ApJS, 201, 19), Dupuy \& Kraus (2013, Science, 341, 1492), Liu et al. (2016, ApJ, 833, 96), Best et al. (2018, ApJS, 234, 1), and Best et al. (2020b, AJ, in press). This research has made use of the Washington Double Star Catalog maintained at the U.S. Naval Observatory.
\end{acknowledgements}

\appendix

\section{Best extraction companion}\label{Apx:comp_stat}

\begin{figure*}
\centering
\includegraphics[width=18cm]{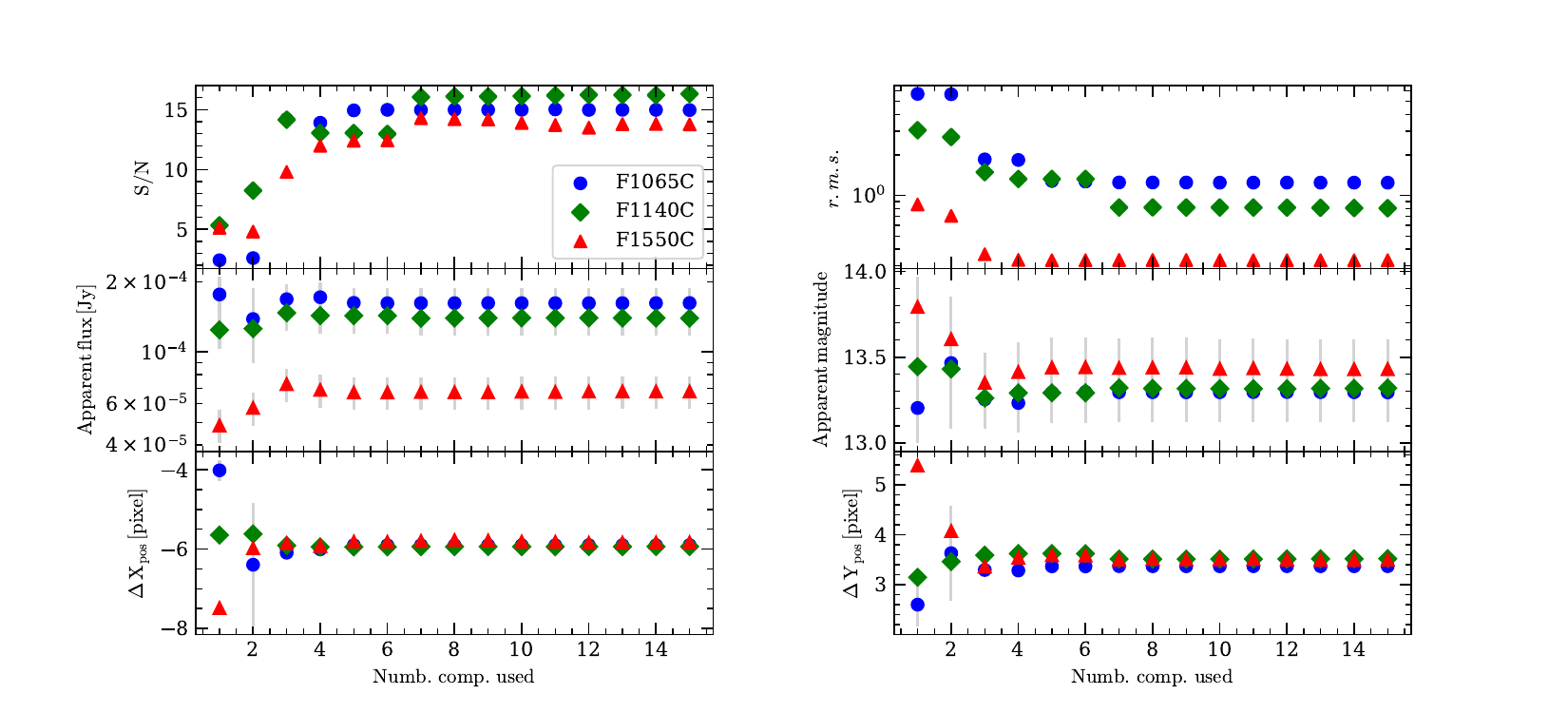}
\caption{General extraction quality for HR\,2562\,B. \textit{Left-top}: Signal-to-noise ratio as a function of the number of components used (NPC). \textit{Left-middle}: Apparent flux vs. NPC. \textit{Left-bottom}: Relative position (with respect to the coronograph) along the X-axis vs. NPC (in pixels). \textit{Right-top}: Root-mean-square in the residual image after subtracting the companion, vs. NPC. \textit{Right-middle}: Apparent magnitude in Vega system vs. NPC. \textit{Right-bottom}: Relative position along the Y-axis vs. NPC (in pixels). Blue circles, red triangles, and green diamonds correspond to the bandpasses \texttt{F1065C}, \texttt{F1140C}, and \texttt{F1550C}, respectively. Vertical gray lines are the uncertainties. }
\label{fig:Comp_stat}%
\end{figure*}

To obtain the flux and astrometric position of HR\,2562\,B, we performed different tests to obtain the lowest possible uncertainty and less bias effects. The post-processing of the images is carried out with different numbers of principal components, which have different performance when removing the coronagraphic PSF and the speckles. The residuals have an impact on the companion signal, flux measurement, and astrometric position, as well as their uncertainties. We analyzed the extraction of the companion using the fake planet injection method from \texttt{spaceKLIP} using from $1$ to $15$ components. In this process, we also calculated the S/N using \texttt{VIP}, and measured the fluctuation of the post-extraction residuals without the companion contribution (including rms), flux, and magnitude, as well as the relative position in pixels. Figure\,\ref{fig:Comp_stat} shows these values as a function of the number of components subtracted for each of the three filters. We obtained the best results with $6$, $7$, and $7$ components, respectively, for \texttt{F1065C}, \texttt{F1140C}, and \texttt{F1550C}. The best values were selected based on the highest S/N, a relative position that reaches a value that does not fluctuate, and fluxes that reach a fixed value with an uncertainty of the same order of magnitude as their counterparts using larger values of components. Furthermore, the rms stabilizes with the lowest values at these number of components, meaning these components removed efficiently the starlight.

\section{The source ``speckle''}\label{Apx:speckle}

\begin{figure}[htb]
    \centering 
 \begin{subfigure}{}
  \includegraphics[width=2cm]{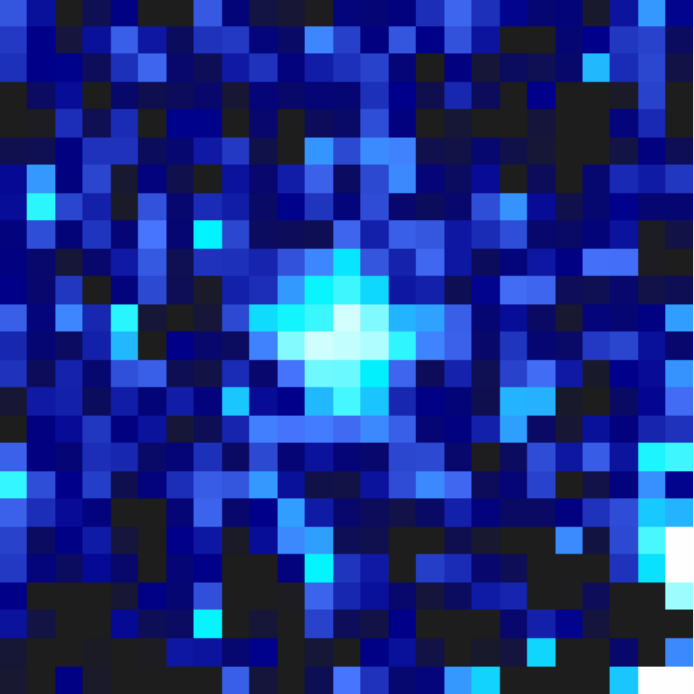}
 \end{subfigure}\hfil 
 \begin{subfigure}{}
  \includegraphics[width=2cm]{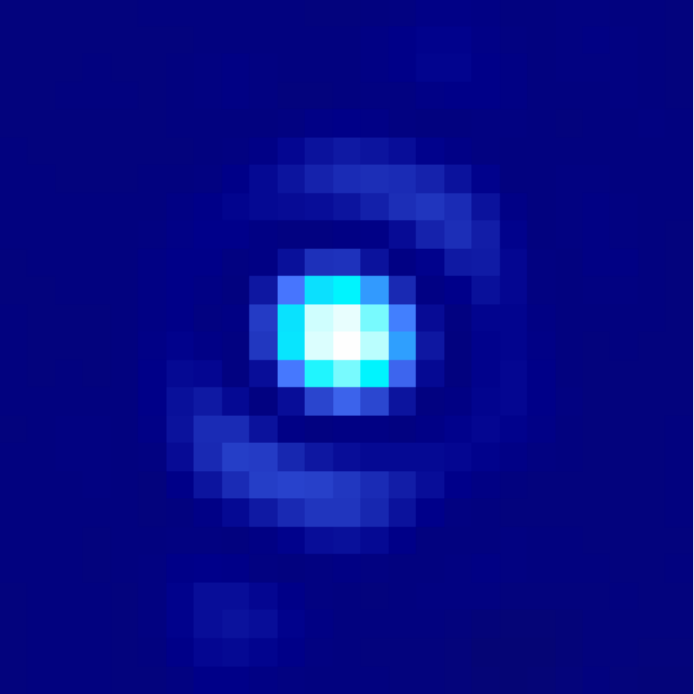}
 \end{subfigure}\hfil 
 \begin{subfigure}{}
  \includegraphics[width=2.6cm]{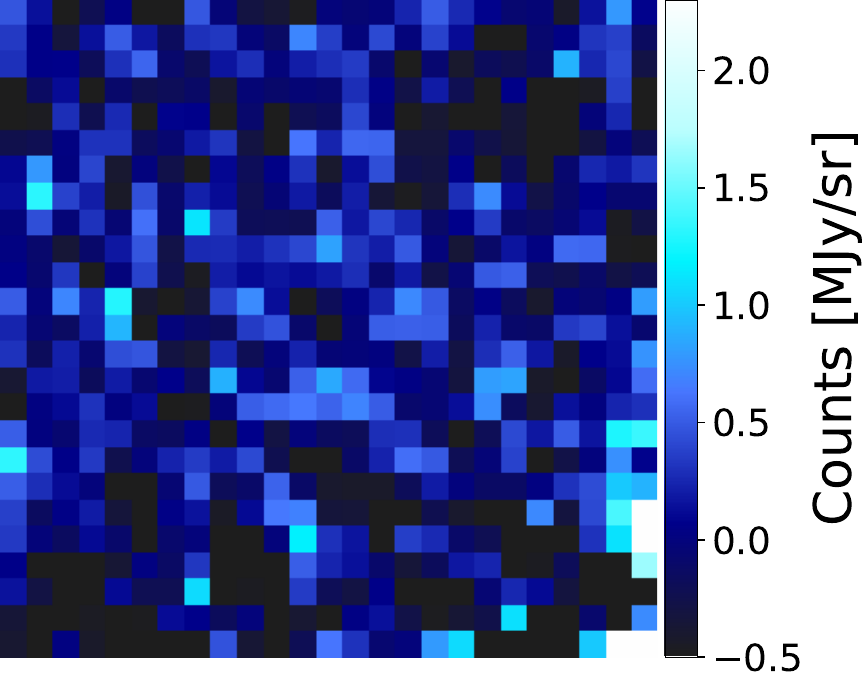}
 \end{subfigure}
 \medskip
     \centering 
 \begin{subfigure}{}
  \includegraphics[width=2cm]{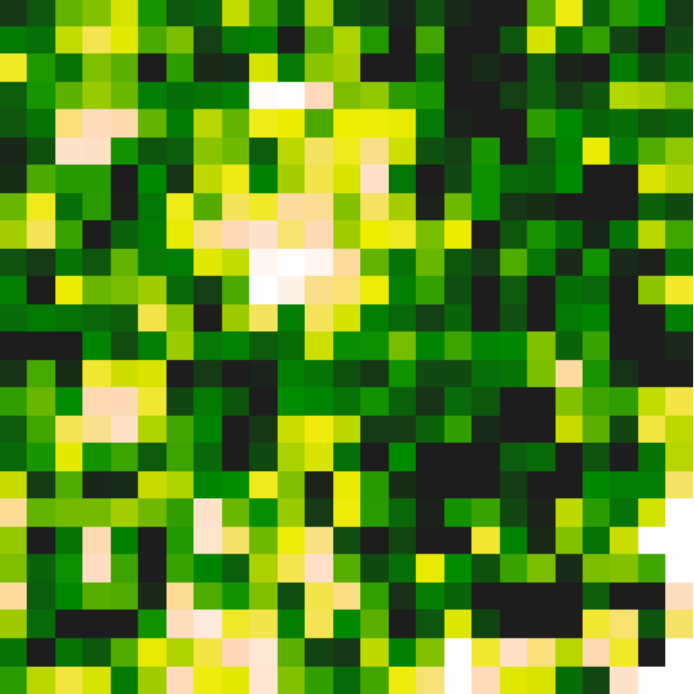}
 \end{subfigure}\hfil 
 \begin{subfigure}{}
  \includegraphics[width=2cm]{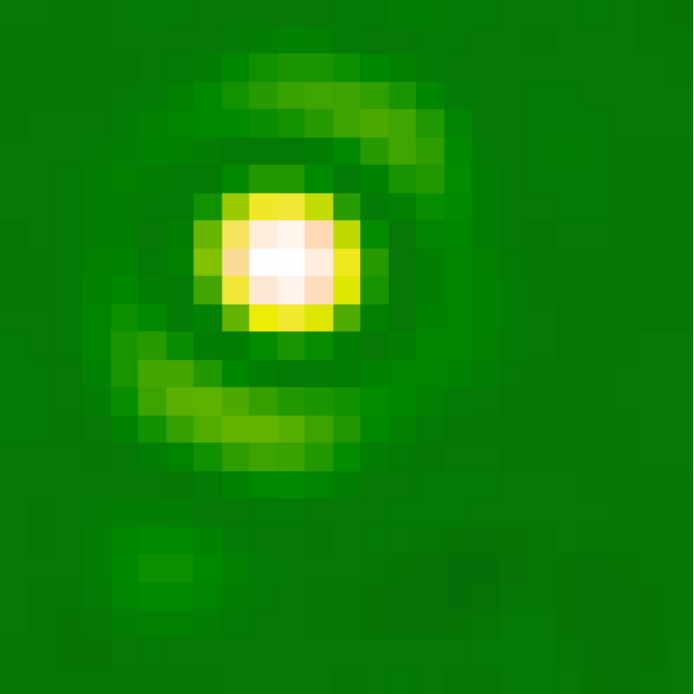}
 \end{subfigure}\hfil 
 \begin{subfigure}{}
  \includegraphics[width=2.6cm]{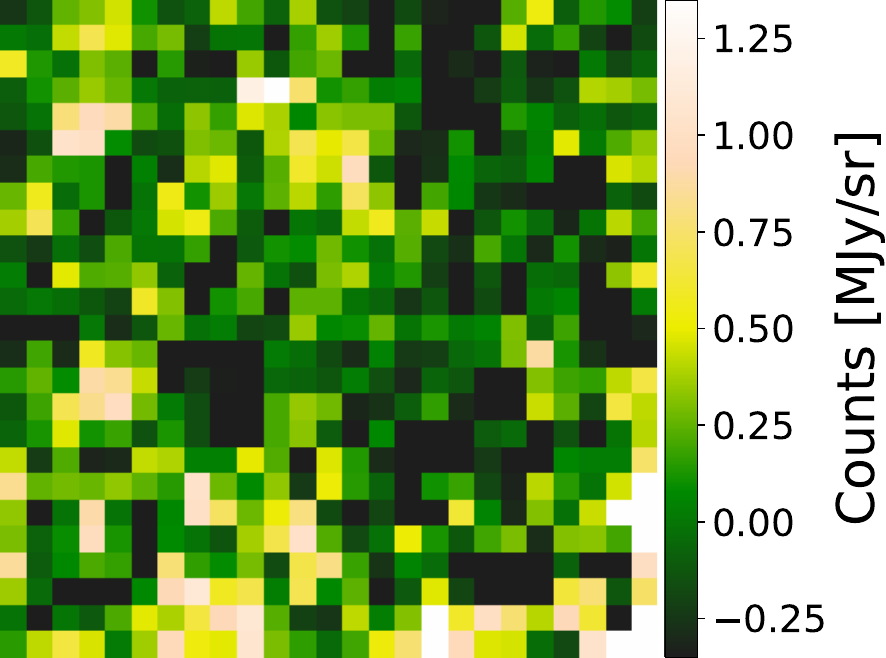}
 \end{subfigure}
 \medskip
     \centering 
 \begin{subfigure}{}
  \includegraphics[width=2cm]{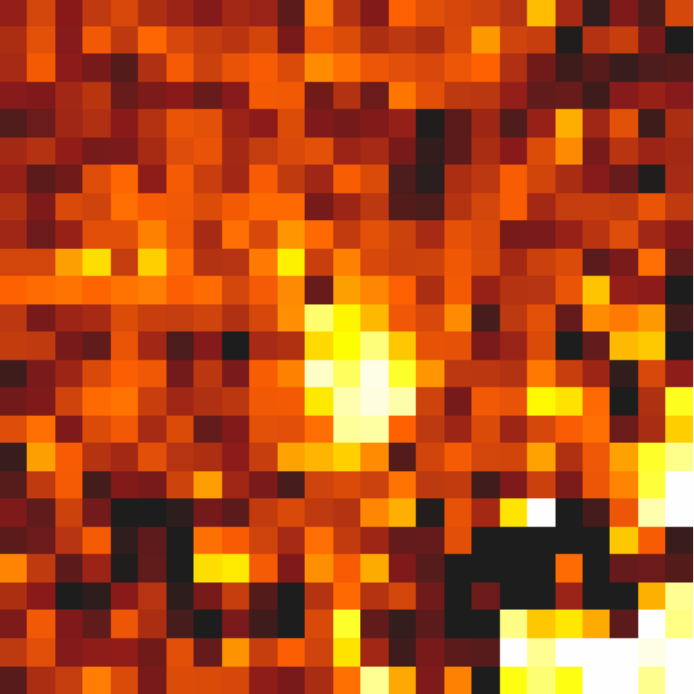}
 \end{subfigure}\hfil 
 \begin{subfigure}{}
  \includegraphics[width=2cm]{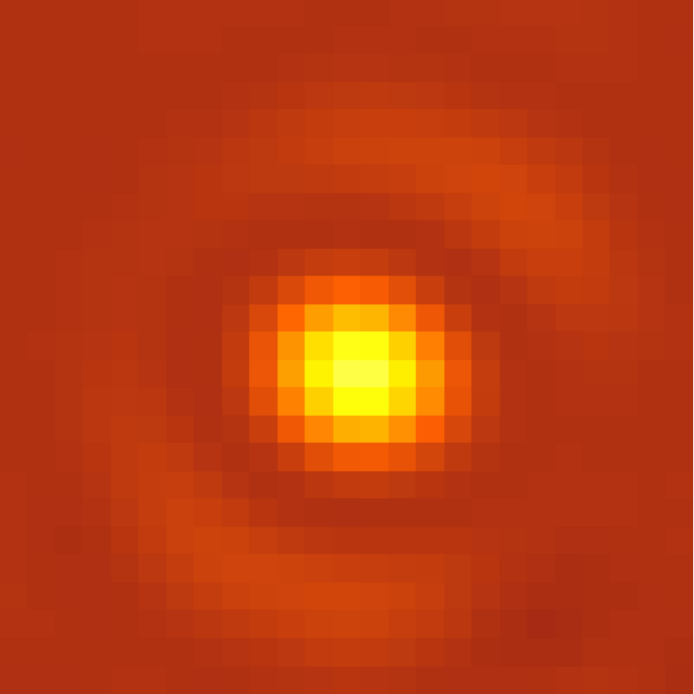}
 \end{subfigure}\hfil
 \begin{subfigure}{}
  \includegraphics[width=2.6cm]{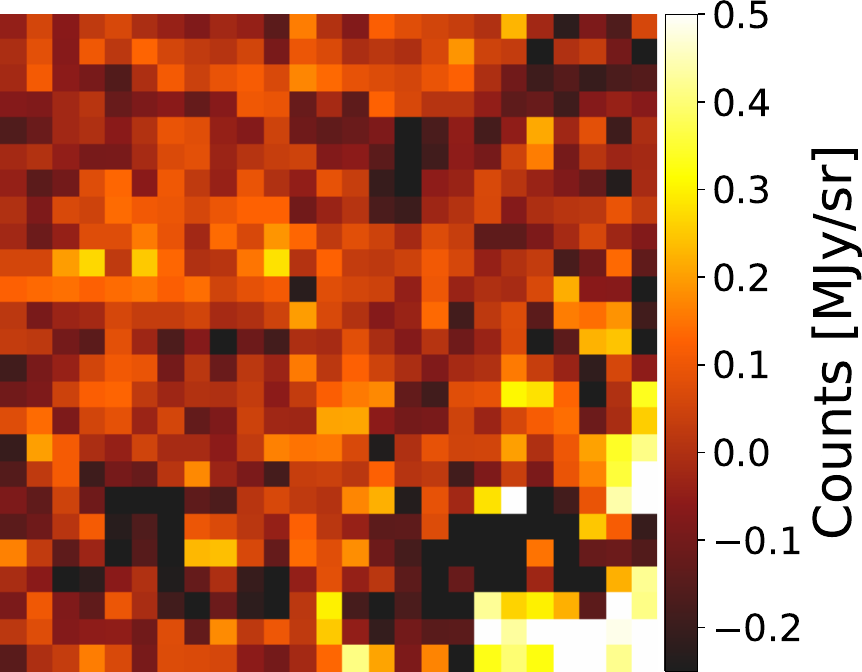}
 \end{subfigure}
\caption{Same as Figure\,\ref{fig:Extraction_comp} but for the inner residual speckle. From \textit{top} to \textit{bottom}: Filters \texttt{F1065C}, \texttt{F1140C}, and \texttt{F1550C}. }
\label{fig:Extraction_c3}
\end{figure}

\begin{figure}
\centering
\includegraphics[width=8cm]{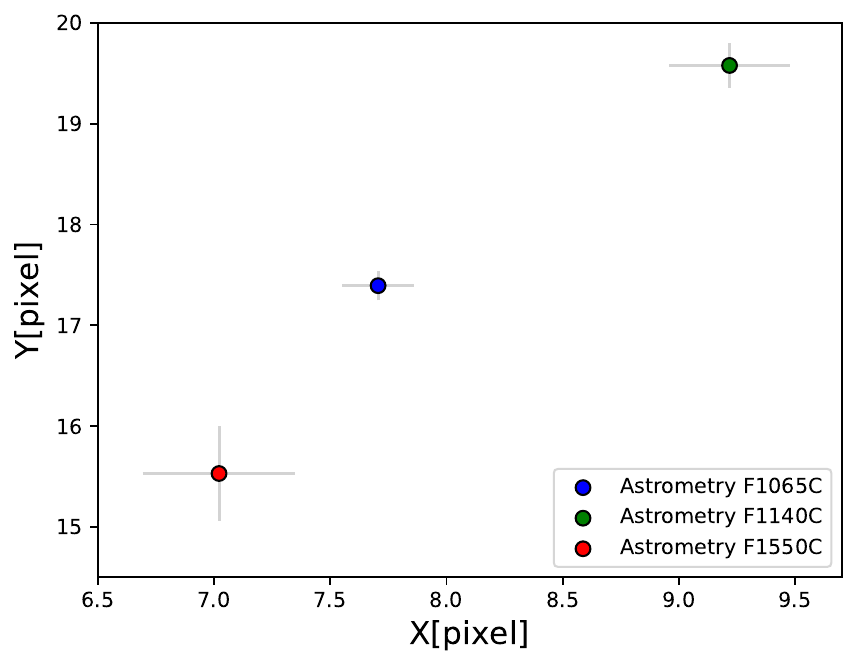}
\caption{Relative position (with respect to the star) of the inner source.}
\label{fig:Astr_c3}%
\end{figure}

We studied the source that appears in the inner region labeled ``speckle'' (highlighted with an arrow in Figure\,\ref{fig:images}) to investigate if this is an astronomical source or a residual speckle. We extracted the photometry and astrometry using \texttt{spaceKLIP}, being the astrometric position the most important for our analysis. Figure\,\ref{fig:Extraction_c3} shows the modeling and extraction of this source at each filter using the same number of components used for HR\,2562\,B. It is relevant to note that the center of each figure corresponds to the same coordinate, so the shifts are real. We corroborated the last point by plotting the astrometry of the source. Figure\,\ref{fig:Astr_c3} shows the relative position of the source with respect to the star, in pixels. Compared to HR\,2562\,B (see Figure\,\ref{fig:Comp_stat}), this source moves considerably between filters (maximum separation of $4.6\pm0.5$  pixels or $0.51\pm0.05\arcsec$). Therefore, we concluded the source is a residual speckle.

\section{The source ``bkg''}\label{Apx:c2_source}

\begin{figure}
\centering
\includegraphics[width=2.75cm]{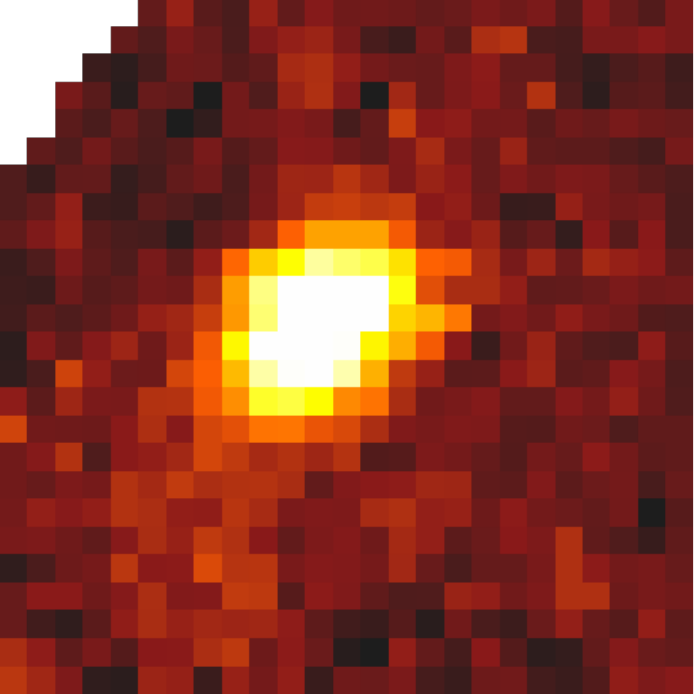}
\includegraphics[width=2.75cm]{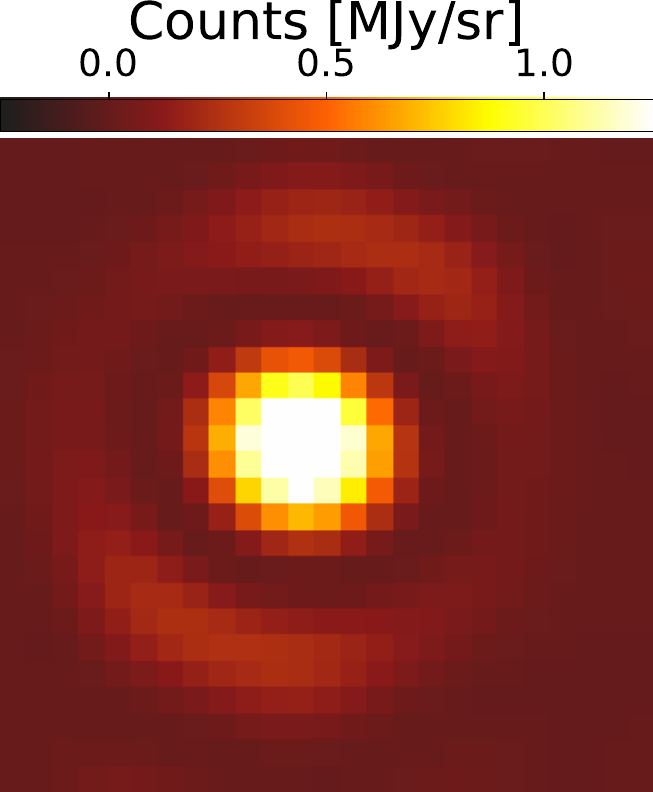}
\includegraphics[width=2.75cm]{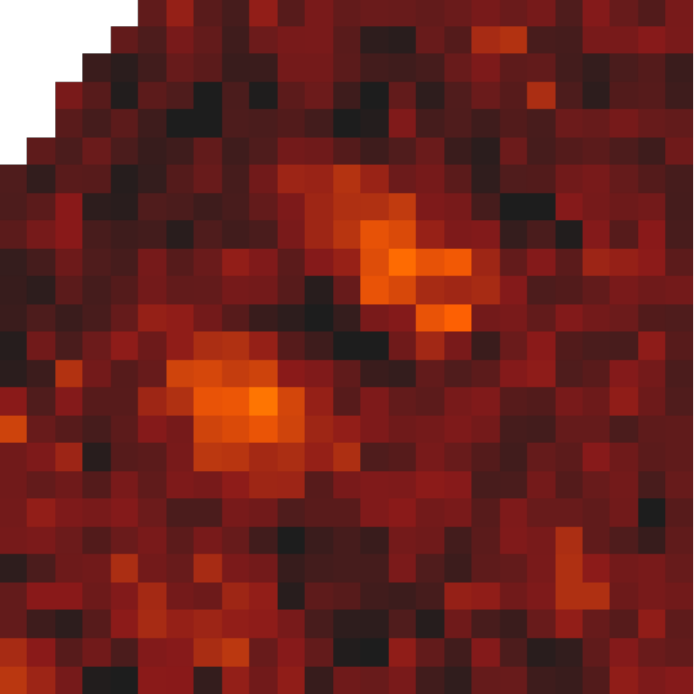}
\caption{Example of modeling and extraction of the first identified source at $15\mu m$ upper-left side, post-processing image. \textit{Left}: Science image at $15\mu m$ (\texttt{F1550C} filter), using one component in the post-processing. \textit{Middle}: Best \texttt{spaceKLIP} model of the source using \texttt{webb\_psf}. \textit{Right}: Residuals after subtracting the model in the science data. The field of view corresponds to $25\times25$ pixels ($\sim2.75\arcsec\times2.75\arcsec$). The three images have the same color scale.}
\label{fig:Extraction_c2}%
\end{figure}

The first source identified in the field of view and labeled ``bkg'' in Figure\,\ref{fig:images} appears only at $15\mu m$. We used \texttt{spaceKLIP} to extract the source photometry and astrometry as is shown in Figure\,\ref{fig:Extraction_c2}. From the residuals, we can see evidence of an extended source due to poor extraction of the companion (assuming a point source model). This is also slightly evident from the image at $15\mu m$. We proceed to estimate the photometry using aperture photometry since the source is not affected by the starlight contamination. For the uncertainties, we assumed a value of $0.005$\,mJy conservatively at $15\mu m$ (a larger value than the mean standard deviation around the source). We also estimated the upper limits from the contrast curves at $10\mu m$ and $11\mu m$. These values are presented in Table\,\ref{table:Fluxes}. Based on the extended source emission, we can conclude we are observing a galaxy or a young star with a disk. However, the last one is not consistent with the stellar population and age of the young association of HR\,2562. We also note that the observation at $15\mu m$ is deeper than the ones at $10\mu m$ and $11\mu m$, so the disk could not appear necessarily at \texttt{F1065C} and \texttt{F1140C}. However, it should be expected to observe the stellar emission at shorter wavelengths. Given the small field of view of SPHERE and GPI, it is not possible to check the stellar emission at NIR wavelengths. The angular separation of the source ``bkg'' is $10.1\arcsec$ vs the field-of-view of SPHERE which is $11\arcsec\times11\arcsec$, meaning a maximum separation from the center of below $<8\arcsec$. In addition, we did not identify any source at the angular separation and position angle of ``bkg'' using the Gaia EDR3 catalog within $50\arcsec$. 

\begin{figure}
\centering
\includegraphics[width=8cm]{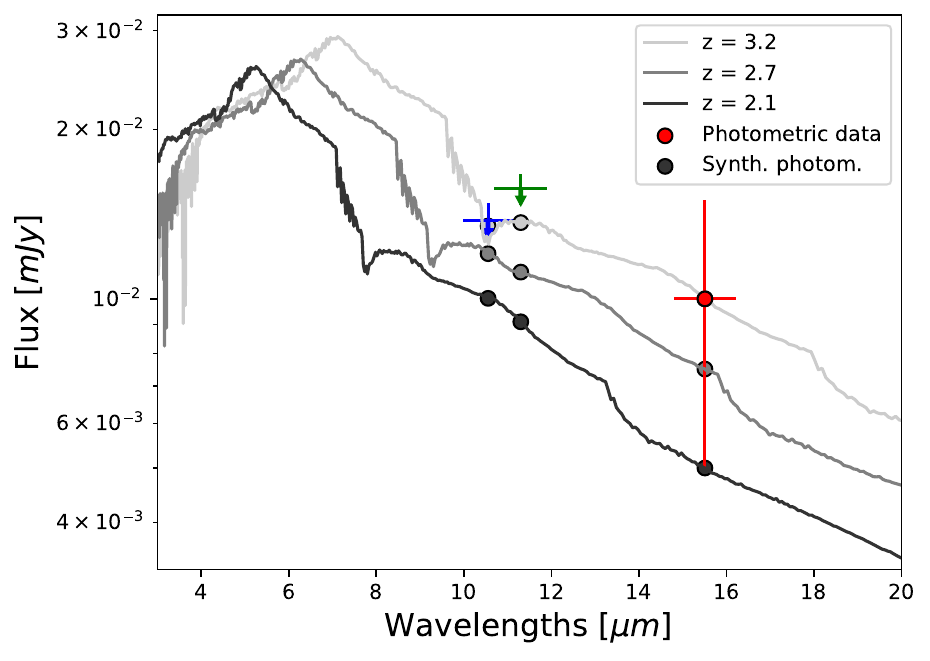}
\caption{Spectral energy distribution of the source labeled ``bkg,'' assuming a galaxy nature. The red dot corresponds to the photometry at $15\mu m$, while the arrows are upper limits at $10\mu m$ and $11\mu m$. The vertical lines denote the uncertainties, and the horizontal lines the $\mathrm{W_{eff}}$ of each filter. The gray lines are the galaxy models from \cite{Bruzual+Charlot-2003} at three different redshifts that better represent the actual data. The black dots are the synthetic photometry of each galaxy model.}
\label{fig:SED_c2}%
\end{figure}

On the other hand, the source could be a background galaxy. Since the source is quite elongated, it could be an elliptical galaxy or a spiral galaxy\footnote{Other alternatives are the Active Galactic Nuclei (AGN) and quasi-stellar objects. However, for those, it is expected a point-like source or an extended source with a bright bulge.}. To investigate the last point, we used the spectra galaxy models from \cite{Bruzual+Charlot-2003} to constrain the redshift. We used the spectra of a ``normal galaxy'' (S0-Sa spiral galaxy). At first order, we can discard high-redshift ($z>5$) galaxies since they look more like point sources instead of elongated ones. Also, with our integration time ($<5\,\mathrm{hours}$), it is not possible to detect those types of galaxies since they are too faint. The fact that the source is not visible at $10\mu m$ and $11\mu m$ could be due to the integration time (which is $\sim12$ times shorter than for $15\mu m$). We can discard some galaxy features such as the Balmer break because that can mean an extremely high redshift. We focused on the $\sim2.5\mu m$ absorption feature that can fit better the observations. Figure\,\ref{fig:SED_c2} shows our first approach to constrain the redshift given our upper limits and photometry plus uncertainties. If this source is a galaxy, the best (and maximum) redshift is $z\approxeq 3.2$, which could be as low as $z=2.1$ considering the uncertainties, so $z$ can lie between $2.1$ and $3.2$. If we consider galaxies of type Sb, Sc, Sd, which present emission lines, we can get a different range of values for the redshift, between $0.5$ and $1.5$ (depending on the emission line we use). In addition, from ALMA observations (\citealt{Zhang+2023}) we detected a point source at the coordinates of this possible galaxy but with a signal comparable to the noise. However, the position of the identified signal is inconsistent with the stellar proper motion for a background object (appears at a slightly greater distance than expected), so we concluded that the point source we identified in the ALMA data is more likely noise.

\section{Sensitivity for \texttt{ATMO} chemical equilibrium}\label{apx:PMD}

\begin{figure*}
\centering
\includegraphics[width=5.75cm]{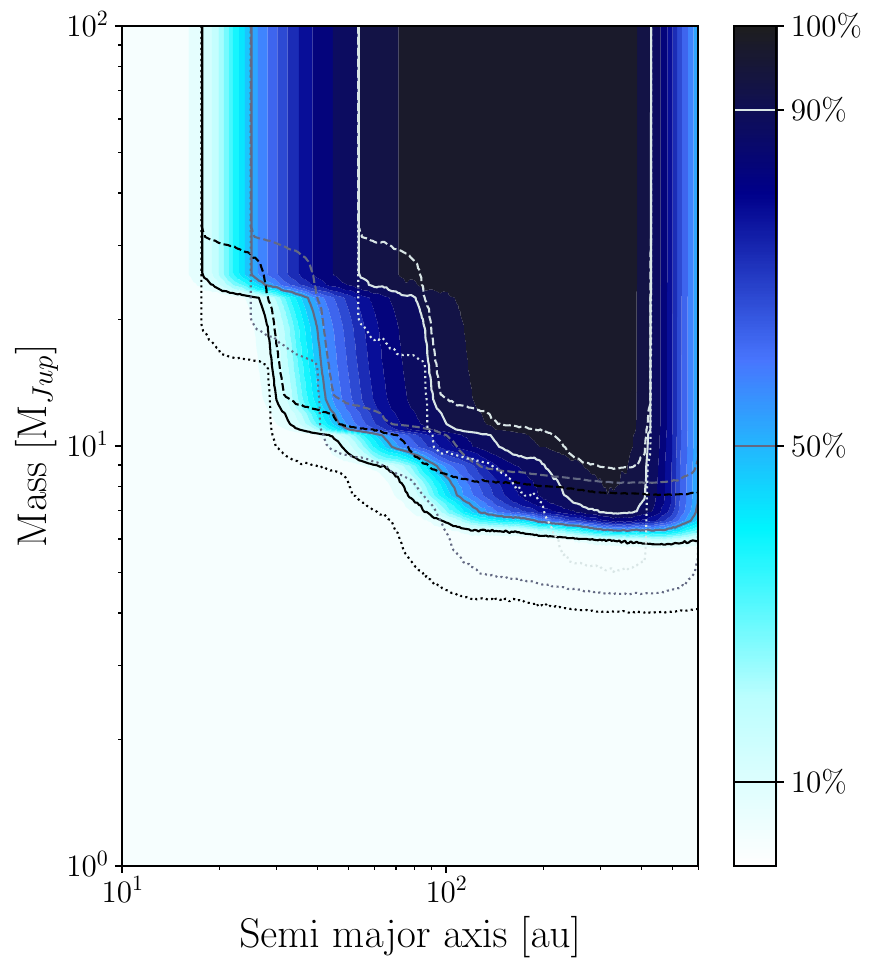}
\includegraphics[width=5.75cm]{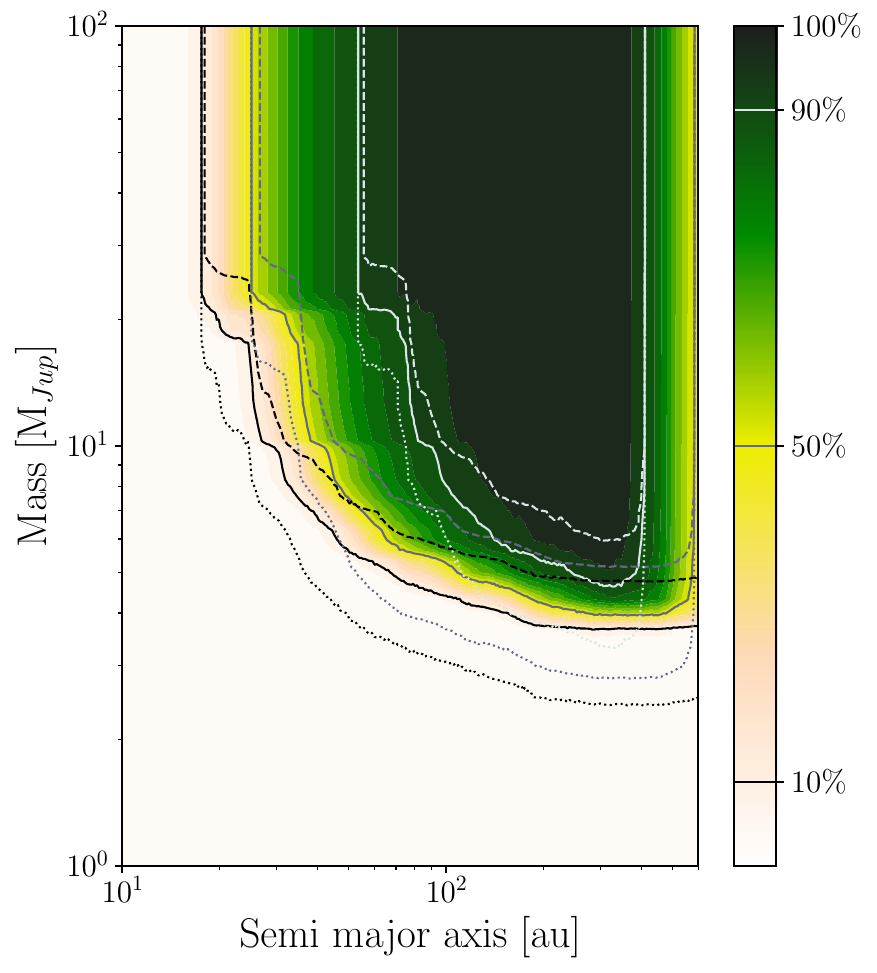}
\includegraphics[width=5.75cm]{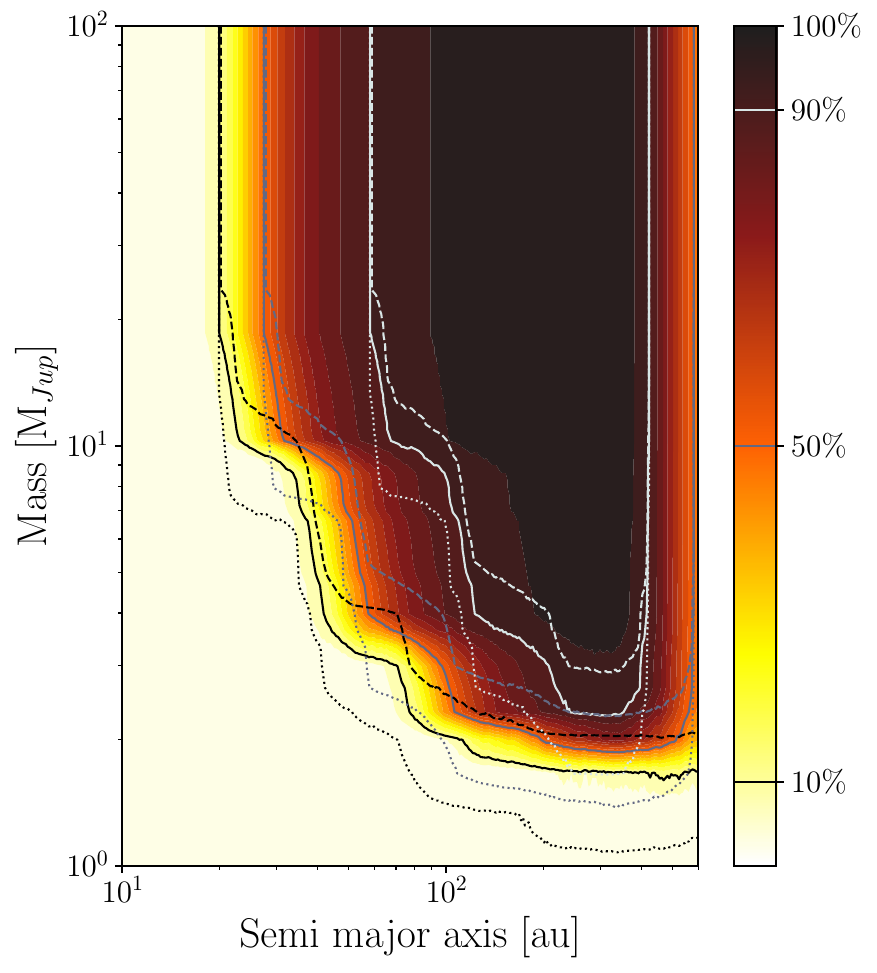}
\caption{Same as Figure\,\ref{fig:PMD-ATMOceq} but using \texttt{ATMO} chemical equilibrium. }
\label{fig:PMD-ATMO_uncert_ceq}%
\end{figure*}

Figure\,\ref{fig:PMD-ATMO_uncert_ceq} shows the \texttt{ATMO} mass sensitivity for the filters \texttt{F1065C}, \texttt{F1140C}, and \texttt{F1550C} using the chemical equilibrium model. As for Figure\,\ref{fig:PMD-ATMOceq}, the solid lines highlight the detection probability at 10\%, 50\%, and 90\%. The dashed and dotted lines are the respective $1\sigma$ uncertainty.


\section{Color-magnitude diagram and water}\label{apx:CMD_w}

\begin{figure}
\centering
\includegraphics[width=5.8cm]{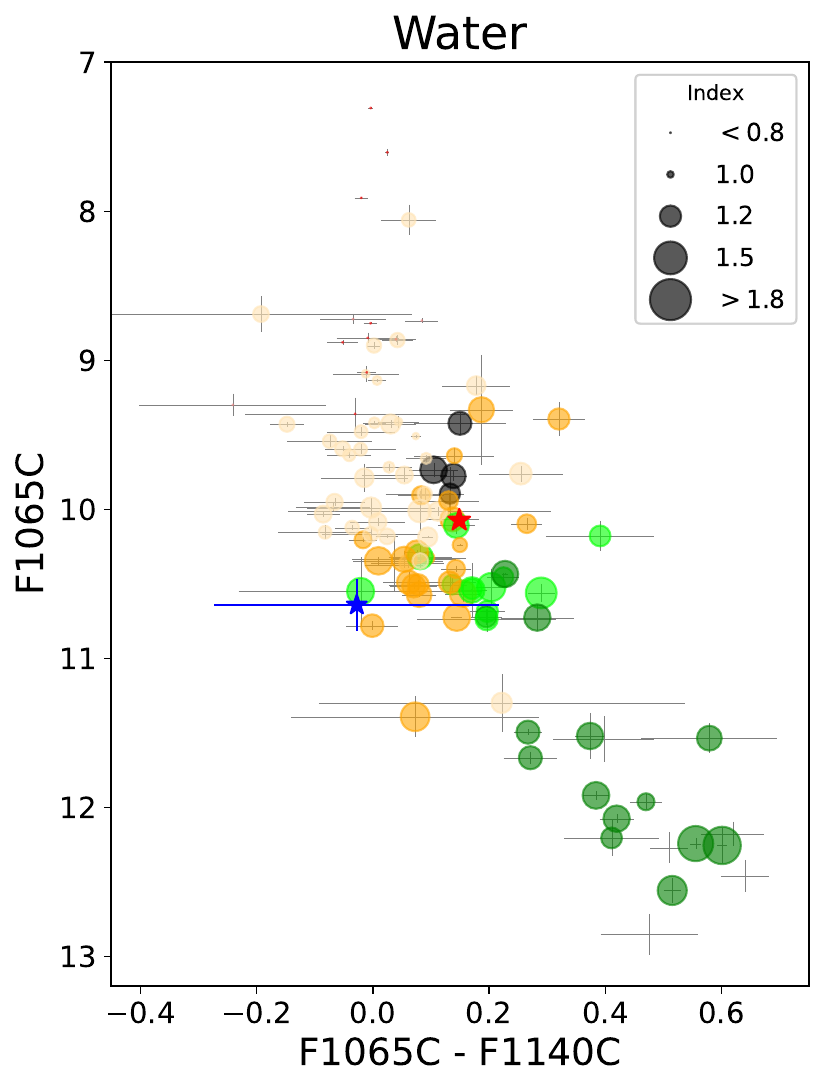}
\caption{Color magnitude diagram using $10$ and $11\,\mu m$. The circle size means the water index. Colors and symbols are the same as Figure\,\ref{fig:CMD_index}.}
\label{fig:CMD_water}%
\end{figure}

From the color-magnitude diagram constructed with filters \texttt{F1065C} and \texttt{F1140C}, it is possible to identify, with a certain degree of uncertainty, the different spectral types and their different trends and colors. However, the presence of water vapor is also probable, given that it has a strong absorption at $\geq11$ microns (e.g., \citealt{Miles+2023}). The presence of water is not unusual in brown dwarfs, as has been observed in different spectral types and temperature ranges (e.g., \citealt{Allers+2013}). Also, it is possible to detect water in all spectral types from L1 to T9 (with a progressively increasing spectral index) at MIR wavelengths at MIRI filters (\citealt{Suarez+Metchev-2022}). Figure\,\ref{fig:CMD_water} shows the color-magnitude diagram in the same style as Figure\,\ref{fig:CMD_index} but for the water spectral index. From the color-magnitude diagram, we can see that almost all the sources present water absorption which affects the shape of the spectrum mostly in the same manner. This means the presence of water cannot be used to clearly distinguish between spectral types and objects at early or late stages in the L/T transition (but could roughly indicate the spectral type). The water could not have a dominant effect on the color of sources with temperatures between 1000K and 1300K (e.g., VHS\,1256\,b, \citealt{Miles+2023}) at \texttt{F1065C} and \texttt{F1140C}. However, since the water absorption appears in all the L/T transitional objects (see Fig.\,\ref{fig:CMD_water}), it could contribute (to a medium/small degree) to the features created by other species. We can expect to have water vapor absorption in the spectrum of HR\,2562\,B, but this is not the mean dominant chemical species at these MIRI filters.

\section{Dataset summary}\label{apx:data}

Table\,\ref{table:Obs_summary} summarize the observing mode, filters, type of observation (imaging of IFS), and observing dates for each of the datasets used in this study.

\begin{table*}[]
\centering
\caption{Summary of the dataset and observations of HR\,2562 used in this work for atmospheric modeling. }
  \begin{tabular}{ l c c c c c c c c c c }
    \hline \hline
  Data & Instrument & Observing mode & Band/filter & Obs. date   \\
    \hline
\cite{Konopacky+2016} & Gemini/GPI   & IFU     & JHK    &  January/February 2016   \\
\cite{Mesa+2018}      & SPHERE/IFS   & IFU     & YJ     &  February 06 2017   \\
                      & SPHERE/IRDIS & imaging & BB\_H  &  February 06 2017   \\
\cite{Sutlieff+2021}  & MagAO/Clio2  & imaging & 3.94   &  February 06/07 2017   \\
This work             & JWST/MIRI    & imaging & F1065C &  March 10 2023   \\
                      & JWST/MIRI    & imaging & F1140C &  March 10 2023    \\
                      & JWST/MIRI    & imaging & F1550C &  March 10 2023    \\
\hline
  \end{tabular}
\label{table:Obs_summary}
\end{table*}

\section{The posteriors of the atmospheric fits}\label{apx:SED_err}

\begin{figure*}
\centering
\includegraphics[width=9.75cm]{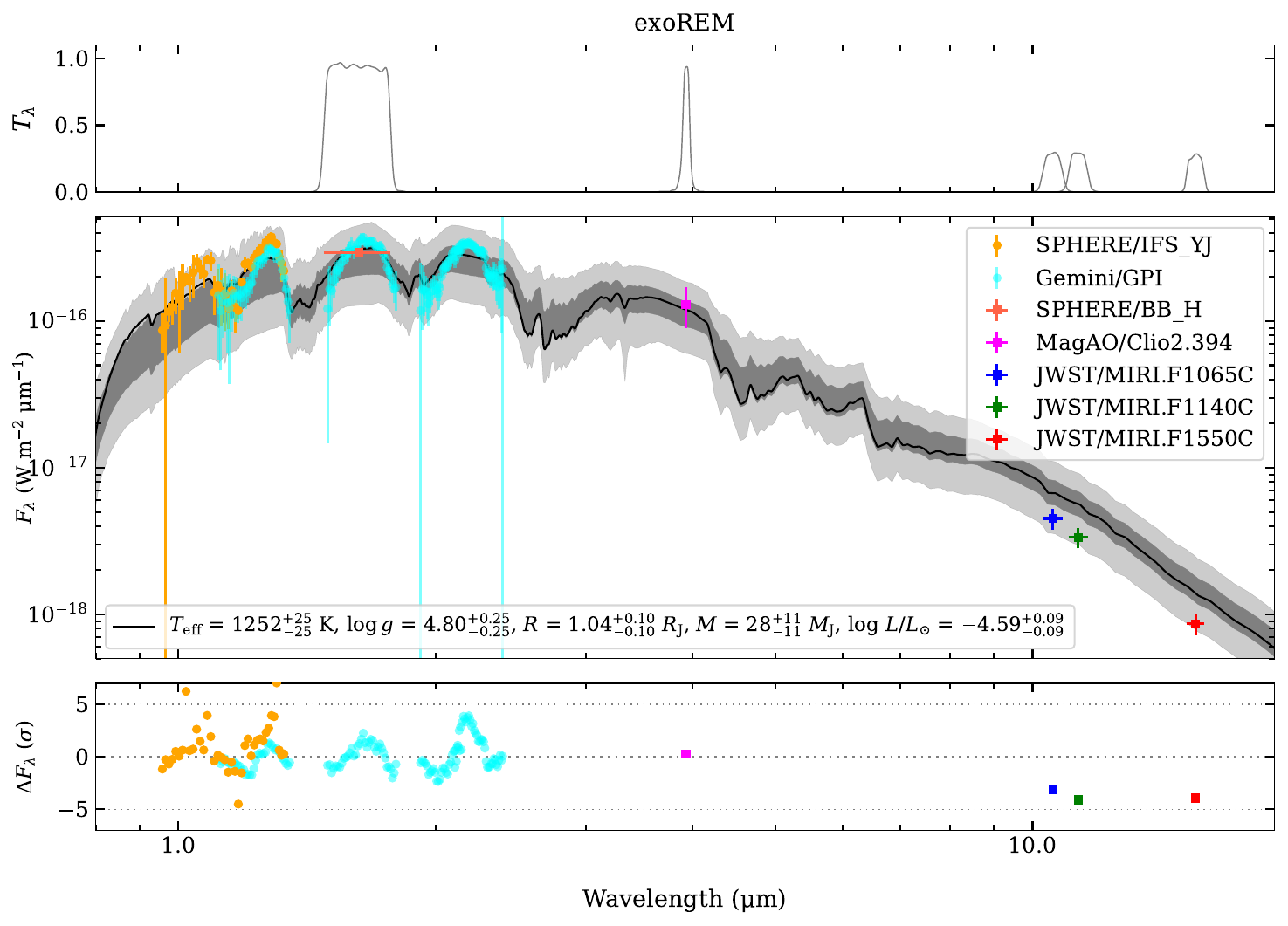}
\caption{ Same as Figure\,\ref{fig:ATMO_best_model} but for the best-fit model using \texttt{Exo-REM} grid via $\chi^2$ mapping with all the available datasets. The model correspond to the dotted green line in Figure\,\ref{fig:exoREM_all_models}. Regions colored with different shades of gray correspond to $2-$ and $3-\sigma$ uncertainties, and the black line to the best-fit model. The different colored symbols correspond to the observations.}
\label{fig:exoREM_erros}%
\end{figure*}

\begin{figure*}
\centering
\includegraphics[width=9.75cm]{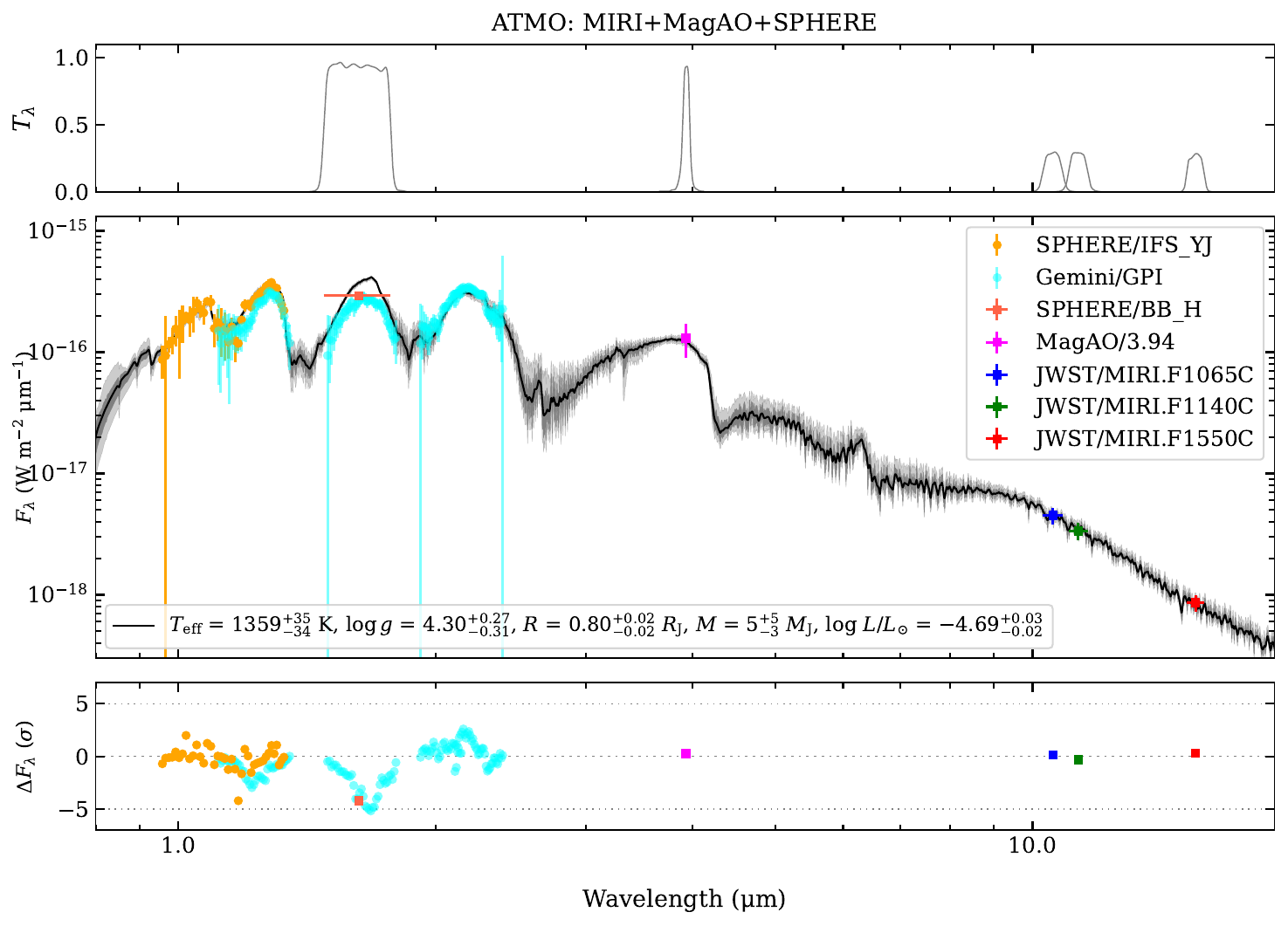}
\caption{Same style as Figure\,\ref{fig:exoREM_erros} but using \texttt{ATMO} and the SPHERE, MagAO, and MIRI observations.}
\label{fig:ATMO_some_err}%
\end{figure*}

We did the analysis of the SED of HR\,2562\,B using different combinations of datasets. Furthermore, in the case of \texttt{ATMO}, we used different priors for the mass. For \texttt{Exo-REM}, the uncertainties are based on covering the range of best solutions using $\chi^2$, which results in higher uncertainties. Although we noted that \texttt{Exo-REM} does not satisfactorily fit the data, the models are within the associated uncertainties. Figure\,\ref{fig:exoREM_erros} shows a clear example of the effect of these uncertainties on the variables, where the gray areas correspond to $2$ and $3$-$\sigma$, and the black line to the best fit. For this fit, we used all the datasets. Figure\,\ref{fig:ATMO_some_err} shows, on the other hand, the same figure but using the \texttt{ATMO} models for a specific combination of datasets (MIRI+MagAO+SPHERE), assuming no priors for the mass.

\begin{figure*}
\centering
\includegraphics[width=14cm]{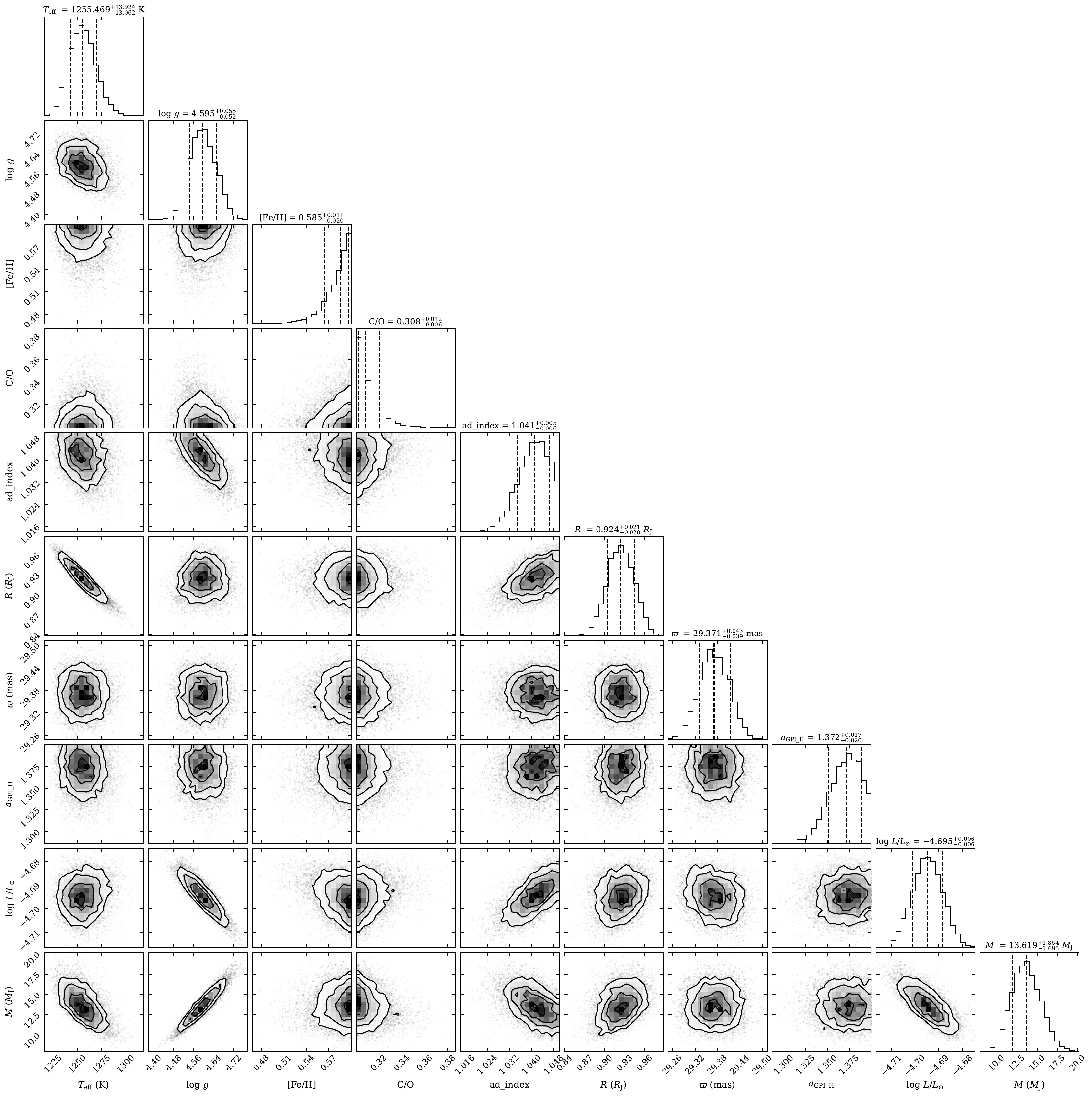}
\caption{The posteriors distributions and correlation diagrams from the \texttt{ATMO} fit assuming no-priors in the mass. The \texttt{ATMO} models incorporate other parameters presented in the Figure: metalicity [Fe/H], C/O ratio, adiabatic index $mathrm{ad\_index}$. None of these parameters were analyzed in this study. Our fitting procedure also considers the correction factor $\mathrm{a_{GPI\_H}}$ for the GPI H-band.}
\label{fig:ATMO_posterior}%
\end{figure*}

As an example, Figure\,\ref{fig:ATMO_posterior} shows the posterior distributions of the different variables fitted using \texttt{ATMO}. In this case, we use all datasets, and no priors for the mass. In the case of \texttt{Exo-REM}, the distributions are more centered on the ranges of models that best fit the data, so obtaining a posterior distribution would generate a bias in the uncertainties (given the nonlinearity when moving in the space of the variables). As we mentioned before, the uncertainties are more associated with the range of best fits than a model on the residuals.

%
%

\bibliographystyle{aa} 

\end{document}